\documentclass[referee,a4paper,12pt,traditabstract]{jswsc}

\usepackage{graphicx}
\usepackage{epstopdf}

\usepackage[authoryear,round]{natbib}
\usepackage[backref]{hyperref}
\usepackage{url}

\usepackage{dcolumn}   % needed for some tables
\usepackage{amsmath,amssymb,amssymb}   % for math

\usepackage[british]{babel}
\usepackage[normalem]{ulem} % Strike text

\usepackage{graphics, setspace}
\usepackage[section]{placeins}

\DeclareMathOperator*{\argmin}{arg\,min}

\usepackage{float}
\usepackage{color}
\usepackage[normalem]{ulem}

\newcommand{\mathsym}[1]{{}}
\newcommand{\unicode}[1]{{}}

\usepackage{soul}               % to striketrhourhg text

\usepackage[usenames,dvipsnames]{xcolor}

\newcommand{\red}{\color{red}}

\definecolor{dgreen}{rgb}{0.0, 0.5, 0.0}
\newcommand{\dgreen}{\color{dgreen}} 
\begin{document}

%\begin{linenumbers}  

   %\title{Interplanetary Medium Monitoring with LISA}
   %\subtitle{I. Overviewing the $\kappa$-mechanism}
   \title{Interplanetary medium monitoring with LISA: lessons from LISA Pathfinder}
      
   \titlerunning{Interplanetary Medium Monitoring with LISA}

   \authorrunning{A.~Cesarini et al.}

   \author{A.~Cesarini
          \inst{1}\fnmsep\thanks{Corresponding author : \email{\href{mailto:andrea.cesarini@fi.infn.it}{andrea.cesarini@fi.infn.it}}},
          C.~Grimani\inst{1}\fnmsep\inst{2},
          S.~Benella\inst{3},
          M.~Fabi\inst{1}\fnmsep\inst{2},
          F.~Sabbatini\inst{1}\fnmsep\inst{2},
          M.~Villani\inst{1}\fnmsep\inst{2}       
          \and
          D.~Telloni\inst{4}
          }

   \institute{INFN - Sezione di Firenze,
              Via B. Rossi, 1, 50019, Sesto Fiorentino, Florence, Italy\\
              %\email{\href{mailto:andrea.cesarini@fi.infn.it}{andrea.cesarini@fi.infn.it}}
         \and
             DISPEA, Universit\`a di Urbino ``Carlo Bo'', Via S. Chiara, 27, 61029, Urbino, Italy\\
             %\email{\href{mailto:catia.grimani@uniurb.it}{catia.grimani@uniurb.it}}
             %\thanks{The university of heaven temporarily does not accept e-mails}
         \and
             INAF, Istituto di Astrofisica e Planetologia Spaziali,
             Via del Fosso del Cavaliere, 100, 00133 Roma, Italy\\
         \and
             Istituto Nazionale di Astrofisica (INAF), Osservatorio Astronomico di Torino, Via Osservatorio 20, 10025 Pino Torinese, and INFN Sezione di Firenze, Italy\\
             }

%presently at Istituto Tecnico Tecnologico (I.T.T.) ``Guglielmo Marconi''

%\def\addressGargiulo{INAF, Instituto di Astrofisica e Planetologia Spaziali, Via del Fosso del Cavaliere, 100, 00133 Roma, Italy}
%\def\addresspp{Istituto Nazionale di Astrofisica (INAF), Osservatorio Astronomico di Torino, Via Osservatorio 20, 10025 Pino Torinese, and INFN Sezione di Firenze, Italy}

% %
%\date{\today}
% %

%%   \date{Received September 15, 1996; accepted March 16, 1997}

  % \abstract{}{}{}{}{}        %% uncomment if structured abstract is desired
 %% 5 {} token are mandatory
 
  \abstract
 %% context heading (optional). leave {} empty if necessary  
   {The Laser Interferometer Space Antenna (LISA) of the European Space Agency (ESA) will be the first low-frequency gravitational-wave observatory orbiting the Sun at 1 AU. 
   %It will consist of a constellation of three spacecraft (S/C) placed in a triangular formation of $2.5\times 10^6\,\mathrm{km}$ on each side trailing Earth at about fifty million kilometers. 
   The LISA Pathfinder (LPF) mission, aiming at testing of the instruments to be located on board the LISA spacecraft (S/C), hosted, among the others, fluxgate magnetometers and a particle detector as parts of a diagnostics subsystem. 
   These instruments allowed us for the estimate of the magnetic and Coulomb spurious forces acting on the test masses that constitute the  mirrors of the interferometer. 
   With these instruments we also had the possibility to study the galactic cosmic-ray short term-term variations as a function of the particle energy and the associated interplanetary disturbances.
   Platform magnetometers and particle detectors will be also placed on board each LISA S/C.
   This work reports about an empirical method that allowed us to disentangle the interplanetary and onboard-generated components of the magnetic field by using the LPF magnetometer measurements.
   Moreover, we estimate the number and fluence of solar energetic particle events expected to be observed with the ESA Next Generation Radiation Monitor during the  mission lifetime. An additional cosmic-ray detector, similar to that designed for LPF, in combination with magnetometers, would permit to observe the evolution of recurrent and non-recurrent galactic cosmic-ray variations and associated increases of the interplanetary magnetic field at the transit of high-speed solar wind streams and interplanetary counterparts of coronal mass ejections.
   %, precious clues will be provided about the passage of interplanetary magnetic structures, high-speed solar wind streams and S/C heliospheric current sheet crossing. 
   %This instrument is optimized  for solar energetic particle (SEP) event detection. 
   %It is worthwhile to point out that an additional detector similar to that hosted on the LPF S/C devoted to cosmic rays may   
   %The interplanetary medium monitoring found as necessary for the estimate of the spurious forces acting on the 
   The diagnostics subsystem of LISA makes this mission also a natural multi-point observatory for space weather science investigations.
   }        %% replace by pair of curly brackets, {}, if structured abstract is selected

   \keywords{gravitational wave interferometers --
             magnetometers --
             interplanetary magnetic field --
             galactic cosmic rays
            }

   \maketitle
%%
%%________________________________________________________________

\section{Introduction}
% % %
\vspace{-3mm}

% The Laser Interferometer Space Antenna (LISA) devoted to low-frequency gravitational wave detection is a European Space Agency (ESA) mission scheduled to launch in 2035 \cite{2017arXiv170200786A}. It will consist of three spacecraft (S/C) orbiting on a triangular formation of $2.5\times 10^6\,\mathrm{km}$ on each side around the Sun near the ecliptic, trailing the Earth at about $20^\circ$ in longitude.
% towards the fifth Sun-Earth Lagrangian Libration point (L5). 
%
%Similarly to the case of ground-based gravitational wave detectors, the LISA space-based observatory will experience noise produced by the space environment. The monitoring of the space environment will become fundamental in order to guarantee the interferometer performance since plasma, 
The  European Space Agency (ESA) Laser Interferometer Space Antenna (LISA) is the first mission designed for gravitational wave detection in space in the frequency range 2$\times$10$^{-4}$ - 10$^{-1}$ Hz \citep{2017arXiv170200786A}.
LISA consists of three identical spacecraft (S/C) placed at the corners of an equilater triangle of 2.5 million km of side length corresponding to the arm of the interferometer. The LISA satellite constellation will be inclined at an angle of 60 degrees on the ecliptic plane trailing Earth in its orbit around the Sun at fifty million kilometer distance.
%,  at 20 degrees in heliolongitude towards  the fifth Lagrange point with respect to Earth.                                                      
The S/C triangular formation will rotate yearly clockwise with the center of mass of the formation remaining on the ecliptic. Free-falling cubic test masses (TMs)  of gold and platinum constitute the  mirrors of the interferometer. LISA is scheduled for launch in 2035 at the maximum of the solar cycle 26 during a positive polarity epoch of the global solar magnetic field (GSMF; when the magnetic field lines exit from the Sun North Pole).
The  LISA Pathfinder (LPF)  \citep{{Antonucci:2012:12/124014},{PhysRevLett.116.231101},{2018PhRvL.120f1101A}} mission designed for the testing of the LISA instrumentation  was launched on December 3, 2015 at 4:04 UTC from the near equatorial cosmodrome in Kourou (French Guiana). The LPF S/C entered its final Lissajous orbit around the Sun-Earth first Lagrangian libration point (L1) at the end of January 2016. 
The S/C orbit was quasi-elliptic with minor and major axes of $5\times 10^{5}\,\mathrm{km}$ and $8\times 10^{5}\,\mathrm{km}$, respectively. On July 18, 2017, LPF received its last command.
%{\red \sout{During S/C operations, several in-flight experiments were carried out with LPF to demonstrate the technological feasibility of operating a gravitational wave observatory in space. After three months of science mode operation, the free-fall acceleration noise between two proof-test masses (TMs) placed on board LPF was measured to be $5.2\pm0.1 \,\mathrm{fm\,s^{-2}\, Hz^{-1/2}}$ in the $0.7-20\,\mathrm{mHz}$ frequency range \cite{PhysRevLett.116.231101}. 
%Before the mission end, LPF succeeded in outperforming the LISA sensitivity requirements by measuring a residual differential acceleration noise of $1.74 \pm 0.05\,\mathrm{fm\,s^{-2}\, Hz^{-1/2}}$ between $2$ and $8\,\mathrm{mHz}$ and $60\pm10 \,\mathrm{fm\,s^{-2}\, Hz^{-1/2}}$ at $20\,\mathrm{\mu Hz}$ \cite{PhysRevLett.120.061101}.
%Enabling and cutting-edge technology subsystems carried on board LPF were specifically developed to perform in-flight dedicated experiments \cite{McNamara_2008} to study different components of the TM differential acceleration noise in order to achieve the mentioned unprecedented free-fall sensitivity \cite{2018PhRvD..97l2002A}.}}
The Pathfinder mission of LISA hosted a diagnostics subsystem of instruments \citep{2009CQGra..26i4005C} to carry out a continuous monitoring of 1) the magnetic field near the TMs, 2) the overall high-energy particle flux incident on the S/C \citep{{2017JPhCS.840a2037G},{apj1},{10.1093/mnras/staa830}} and 3) the temperature stability and gradient \citep{10.1093/mnras/stz1017}. % \citep{2016PhRvL.116w1101A}.
As a result, important lessons were learned about spurious forces induced on the TMs by magnetic field variations and high-energy particles. The LPF platform magnetometers were located within the S/C hull to monitor  the magnetic field  generated  by wiring and instruments 
%Conversely, in missions dedicated to the study of the interplanetary medium, 
while science magnetometers are typically placed outside the hull, mounted on extensible booms. As a result, science magnetometers increase mission total costs \citep{Nishio_2007} and mass balance. 
%ESA Ulysses and Cluster \cite{1992A&AS...92..221B,2001AnGeo..19.1197E} and National Aeronautics and Space Administration (NASA) Wind and ACE \cite{1995SSRv...71..207L,2005GeoRL..3215S01B} are examples of missions devoted to measure interplanetary plasma parameters and the IMF intensity. 
To pursue mission cost reduction strategy, space agencies are carrying out many efforts to enable scientists to infer  the interplanetary magnetic field (IMF) data from platform magnetometers present in a wide variety of space missions like, {\it e.g.}, RHESSI \citep{2002SoPh..210....3L}, TRACE\footnote{https://directory.eoportal.org/web/eoportal/satellite-missions/t/trace}, PICARD\footnote{https://directory.eoportal.org/web/eoportal/satellite-missions/p/picard} and IRIS\footnote{https://directory.eoportal.org/web/eoportal/satellite-missions/i/iris}. % (\citeauthor{2002SoPh..210....3L}, \citeyear{2002SoPh..210....3L}, and the \citeauthor{eoportal_org_ALL}). 

Previous works on magnetic field observations gathered on board the LPF S/C is reported in \citet{Mateos2015Thesis} and \citet{10.1093/mnras/staa830}. %\citep{{tscharnuter},{balluch}}
In the present paper we illustrate an empirical method to disentangle the LPF on-board generated magnetic field from the IMF component.
The same approach may be considered for LISA and other missions for which the on-board dominant magnetic field would not result correlated with the variations of the IMF. The survey of the transit of interplanetary structures would contribute to both mission magnetic noise evaluation and space weather science investigations. 

In addition to the magnetic noise, a further source of spurious Coulomb forces on the LISA TMs is associated with the charging process due to particles of galactic and solar origin with energies $>$ 100 MeV(/n) penetrating or interacting in  approximately 15 g cm$^{-2}$ of material surrounding the interferometer mirrors.
A particle detector (PD) was hosted on board the LPF satellite to estimate these forces.
%to monitor the overall incident flux of particles of galactic or solar origin energetic enough to penetrate the S/C material surrounding the test-masses. This  detector 
The PD consisted of two $\sim 300$ $\mu$m thick silicon wafers
%300 $\mu$m thick,                                                                                                                        
of  1.40 x 1.05 cm$^2$ area,   separated by 2 cm  and placed in
a telescopic arrangement. For particles with  energies $>$ 100 MeV n$^{-1}$ with an isotropic incidence on each silicon layer,
the instrument geometrical factor  was  found  energy independent and equal to  9 cm$^2$ sr for a total of 17 cm$^2$ sr.
In coincidence mode (particles traversing both silicon wafers), the geometrical factor was  about one tenth
of this  value. The silicon wafers were placed inside a shielding copper  box of 6.4 mm thickness meant to  stop particles with energies smaller than 70 MeV
n$^{-1}$. This conservative choice was made in order not to underestimate the overall flux of particles charging the TMs. The PD  allowed for the counting of
 protons and helium nuclei
 traversing  each silicon layer and for the measurement of ionization energy losses of particles in coincidence mode.
%The PD  data are stored  in the form of histograms over periods of 600                                                                   
%seconds and then sent to the on-board computer.                                                                                          
The maximum allowed particle  counting rate
was 6500 counts s$^{-1}$ on both silicon wafers, corresponding to an event integrated  proton fluence of   10$^8$
protons cm$^{-2}$ at energies $>$ 100 MeV.
%An increasing counting rate starting from  more than 120 counts per sampling time of 15 seconds was observed over\                       
% the mission lifetime due to a decreasing solar activity intensity.                                                                      
In coincidence mode up to 5000 energy deposits per second could be stored on the on board computer.
%Platform magnetometers and particle detectors (PDs) will allow for a partial  monitoring of the  S/C environment.          
%The ESA Next Generation Radiation Monitor expected to be hosted on the LISA S/C is optimized for harsh radiation environments but present a geometrical factor for cosmic-ray particles of just 0.01 cm$^2$: this amounts to about 100 times less with respect to the LPF radiation monitor (ref). 
Hourly integrated cosmic-ray data gathered with LPF presented 1\% statistical uncertainty  thus allowing us for the study of the galactic cosmic-ray (GCR) short-term variations with periodicities typical of the LISA band of sensitivity \citep{{apj1},{apj2},{paper3}}. 
%However, the LPF radiation monitor is characterized by a geometrical factor of 17 cm$^2$ sr that allowed for cosmic-ray hourly  measurements with 1\% uncertainty. 
%The ESA Next Generation Radiation Monitor, expected to be hosted on the LISA S/C, is optimized for harsh radiation environments but presents 
%a geometrical factor of just 0.01 cm$^2$ for cosmic-ray
%particles, about hundred times smaller than that of the LPF PD (ref).
%In order to obtain a similar uncertainty for the LISA measurements, data gathered with the ESA Next Generation Radiation Monitor (NGRM; ref)  should be integrated over timescales of not less than 48 hours preventing us from the possibility of monitoring the short-term galactic cosmic-ray (GCR) variations. 
Unfortunately, no solar energetic particle (SEP)  events  with  fluences  overcoming that of GCRs  above a few tens of MeV n$^{-1}$ were observed during the LPF mission operations. 
%with  fluences  overcoming that of GCRs  above a few tens of MeV n$^{-1}$. 
Measurements gathered with LPF \citep{2017PhRvL.118q1101A} in the spring 2016 indicated that the net charging of the TMs generated by cosmic rays was of tens of charges per second while the charging noise was of the order of one thousand charges per second. Monte Carlo simulations resulted in agreement with the measured net TM charging while the estimated noise appeared three-four times smaller than observations \citep{2015CQGra..32c5001G}. In addition, the simulations of the TM charging noise returned 10730 e s$^{-1}$ and 68000 e s$^{-1}$ during SEP events characterized by proton fluences of 4.2$\times$10$^7$ cm$^{-2}$ and 10$^9$ cm$^{-2}$ above 30 MeV, respectively \citep{ara05,2005CQGra..22S.319V}.
Recently, we have shown that the mismatch between simulated and observed charging noise was associated with the lack of propagation of a few eV electrons in the simulations \citep{{2020ApSS..51245734V},{2021CQGra..38d5013G},{2021CQGra..38n5005V}}. 
%The noise associated with this high TM charging was discussed in \cite{2005CQGra..22S.319V}.
%As a result, ESA  found as necessary to study the effects of galactic and solar particles in modulating the LISA TM charging  with dedicated Monte Carlo simulations (ESA ITT2020/10081).
For reliable TM charging estimates during the LISA mission operations we will benefit of improved simulation toolkits but it will be also of primary importance to follow the evolution of SEP events and GCR variations on board each S/C of the constellation.
The ESA Next Generation Radiation Monitor (NGRM; \citeauthor{ngrm}, \citeyear{ngrm}) is supposed to be adopted to this purpose. The NGRM consists of an electron detector and a proton unit. This instrument is optimized for particle measurements in harsh radiation environments and on board the LISA S/C would play a precious role for SEP event short-term forecasting and monitoring. The number of expected SEP events during the LISA mission operations is estimated here with the Nymmik model \citep{{Nymmik1999a},{Nymmik1999b}} as a function of event fluence on the basis of the predicted solar activity during the solar cycle 26 \citep{sing19}. LISA will allow us to carry out for the first time observations of SEPs at about one and twenty degrees  in heliolongitude that separate the three LISA S/C and LISA from near-Earth detectors and neutron monitors, respectively.     
% characterized by particle acceleration above 50 MeV.
Unfortunately, the NGRM presents a small geometrical factor for particles with energies larger than 200 MeV  thus impeding to follow the dynamics of short-term variations of the particle flux of galactic origin in the LISA band of sensitivity with a precision similar to that of LPF.
It would be more than recommended to add a dedicated cosmic-ray detector to the NGRM on board each LISA S/C.
The possible contribution that LISA shall give to Space Weather science investigations has been illustrated in the roadmap of the Italian Space Agency \citep{refId0}.  

%Since LISA won't be provided of magnetometers devoted to the IMF probing, it will be required to carry out a detailed investigation to infer the IMF 
%starting from platform magnetometer measurements.                                                                                                
              
%Each of the three LISA S/C will rely upon the experience gained from the design and build of the original LISA Pathfinder (LPF) design because 
%it aimed to test the technology to be placed on board LISA. Therefore, investigations applied to the case of LPF are expected to be re-used in the 
%case of LISA.

%LISA is scheduled for launch in 2035 at the maximum of the solar cycle 26 during a positive polarity epoch of the global solar magnetic field (GS\
%MF; when the magnetic field lines exit from the Sun North Pole), under very similar conditions of LISA Pathfinder \cite{sing19}.

This manuscript is organized as follows: the LPF magnetometer characteristics are described in Section~II.
Details of the magnetic environment within the LPF S/C are provided in Section~III.
In Section~IV the LPF magnetic field empirical modeling is described. In Section~V the LPF IMF data are compared to those gathered contemporaneously with the Wind mission.
% dedicated to interplanetary medium parameter observations. 
%orbiting around L1 at the same time of LPF. 
In Section~VI the expected flux of cosmic rays and the occurrence and intensity of SEP events during the time the LISA mission will remain into orbit are estimated. Finally in Section VII we discuss the LISA contribution to Space Weather science investigations with the NGRM and, possibly, with a GCR detector.
%the performance and limits of the particle detector for LISA are described. 
%Finally, conclusions are reported in Section~VII. 

\section{THE LISA PATHFINDER MAGNETOMETERS}
\label{section_LPFMAGS}

\begin{figure}[t!]
	%\hspace{-2cm}
	\hspace{-0.3cm}
	\centering
	\includegraphics[width=7.5cm,height=7cm]{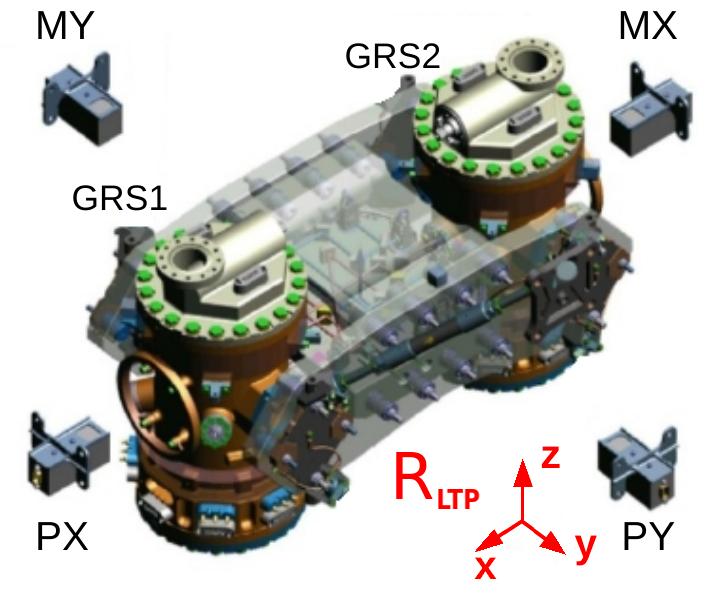} %COLOR
	\caption{Sketch of the position of the four 3-axis magnetometers placed on board LPF (MX, MY, PX and PY) with respect to the gravitational reference sensors (GRS1 and GRS2) enclosing the two LPF TMs. The magnetometers were arranged in a cross-shaped configuration. MX and PX were placed at the edges of the arm joining the TM centers ($x$ direction) where the negative (M) and the positive (P) sides are meant with respect to the middle point between the two GRSs, respectively. Similarly, MY and PY were placed at the edges of the second arm ($y$ direction) where M and P have the same meaning as for the $x$-axis. The $z$-axis was chosen to complete the right-handed tern of the LTP reference frame, $\pmb{\mathrm{R}}_\mathrm{LTP}$.}
	\label{fig_lpf_four_mags_map}
\end{figure}

The Lisa Technology Package (LTP) was placed at the centre of the LPF S/C platform \citep{Anza_2005} with four Billingsley TFM100G4-S 3-axis fluxgate magnetometers (MX, MY, PX and PY) as it is shown in Figure~\ref{fig_lpf_four_mags_map} \citep{Mateos2015Thesis}. 
Fluxgate magnetometers are considered optimum for space applications since they provide highly precise magnetic field measurements and coping with several design constraints ({\it e.g.}, mass, cost limitation, robustness and low-power consumption).
Each magnetometer was made of three distinct $1$-axis magnetic sensors aligned to form a Cartesian tern. Each sensor was provided with a primary inner drive consisting of a high permalloy magnetic core material and a secondary sensing (pickup) coil. The signal sensed by the pickup coil is amplified and demodulated \citep{Mateos2015Thesis}. The obtained constant part of the output signal is, then, integrated and sent back to the pickup coil as feedback of the control loop with the purpose of widening the measurement range.
When the sensor was in operation, a magnetic field  was produced by an exciting coil that periodically saturated the core in opposite directions.
%Therefore, magnetometer distances from the two TMs were carefully selected \cite{2007JPhCS..66a2003A}.
The magnetometers were properly placed at about $19\ \mathrm{cm}$ distance from each TM \citep{2007JPhCS..66a2003A}, and arranged in a cross-shaped configuration with two orthogonal arms of $\simeq 74\,\mathrm{cm}$ length \citep{2009CQGra..26i4005C,2010JPhCS.228a2038D}. 
The magnetometer sensing axes were aligned with the LTP reference frame {\red $\pmb{\mathrm{R}}_\mathrm{LTP}$} (Fig.~\ref{fig_lpf_four_mags_map}). The characteristics of the magnetometers are summarized in Table~\ref{tab0}.
\begin{table}[tbh]
	\centering
	\resizebox{6.5cm}{!}{ 
		\begin{tabular}{lc}\hline\hline
			Parameter & Value \\ \hline
			Field range & $\pm 60\ \mathrm{\mu T}$ \\
			Temperature coefficient & $1.2\, \mathrm{nT\, K^{-1}}$ \\
			Noise density & $<\! 100\ \mathrm{pT\, Hz^{-1/2}}\, @\, 1 \mathrm{Hz}$ \\
			Linearity & $0.015\%$ \\
			Sensitivity & $166.7\ \mathrm{\mu V\, nT^{-1}}$ \\
			Offset voltage & $25\ \mathrm{mV}$ ($150\ \mathrm{nT}$)\\
			Operating temperature & $-55\ \mathrm{^{\circ} C}$ to $+85\ \mathrm{^{\circ} C}$ \\
			Susceptibility & $\pm 20\,\mathrm{nT}$ with $0.5\ \mathrm{mT}$ \\\hline\hline
		\end{tabular}
	}
	\caption{LPF magnetometer characteristics \citep{Mateos2015Thesis}.}
	\label{tab0}
\end{table}

The magnetic subsystem, including the four magnetometers and the acquisition electronics, was tested before the mission launch at the Universidad Politecnica de Madrid \citep{Mateos2015Thesis}. The calibration of the magnetometers  was carried out with a 3-D Helmholtz coil \citep{6740059}.
By denoting with $\mathbf{B}^\mathrm{raw}_{m}[t]$ the raw magnetic field measurements of each magnetometer:
\begin{eqnarray}
\label{eq_Bcomponents}
\displaystyle \mathbf{B}^\mathrm{raw}_{m}[t] &=& \mathrm{B}^\mathrm{raw}_x [t] \, \hat{x} + \mathrm{B}^\mathrm{raw}_y [t] \, \hat{y} + \mathrm{B}^\mathrm{raw}_z [t] \, \hat{z},
\end{eqnarray}
%where $\mathbf{B}^\mathrm{raw}_{m}[t]$ are the raw measurements of the magnetic field amplitude carried out in the LTP reference frame with 
where $m$ ranges from $1$ to $4$ to indicate the four magnetometers MX, MY, PX and PY, respectively (Fig.~\ref{fig_lpf_four_mags_map}). 
%The calibration of the magnetometers  was carried out with a 3-D Helmholtz coil \cite{6740059}. 
This calibration procedure allowed us to determine for each magnetic field component of individual instruments the parameters associated with   gain,  mis-alignment and  bias.
The parameters relative to the gain and to the axis mis-alignment fill two    $3\times 3$ matrices: $\mathbf{s}_m$ and  $\mathbf{T}_m$, respectively.
%and the parameters representing the axis mis-alignment are also reported in a $3\times 3$ matrix $\mathbf{T}_m$. 
Finally, the measurement offset is represented by a $3\times 1$ vector $\mathbf{b}_m$ as indicated below:
\begin{eqnarray}
\label{eq_BINS_to_BMAG}
\displaystyle 
\mathbf{B}^\mathrm{MAG}_{m}[t] &=& \mathbf{s}_{m} \, \mathbf{T}_{m}\, \mathbf{B}^\mathrm{raw}_{m}[t] + \mathbf{b}_m,
\end{eqnarray}
where $\mathbf{B}^\mathrm{MAG}_{m}[t]$ indicates the calibrated magnetic field measurements.

\section{LPF SPACECRAFT MAGNETIC FIELD}

\label{section_magnetic_monitoring}
\begin{figure}[t!]
    \centering
	%\hspace{-2cm}
	\hspace{-0.3cm}
	\includegraphics[width=8.5cm,height=8cm]{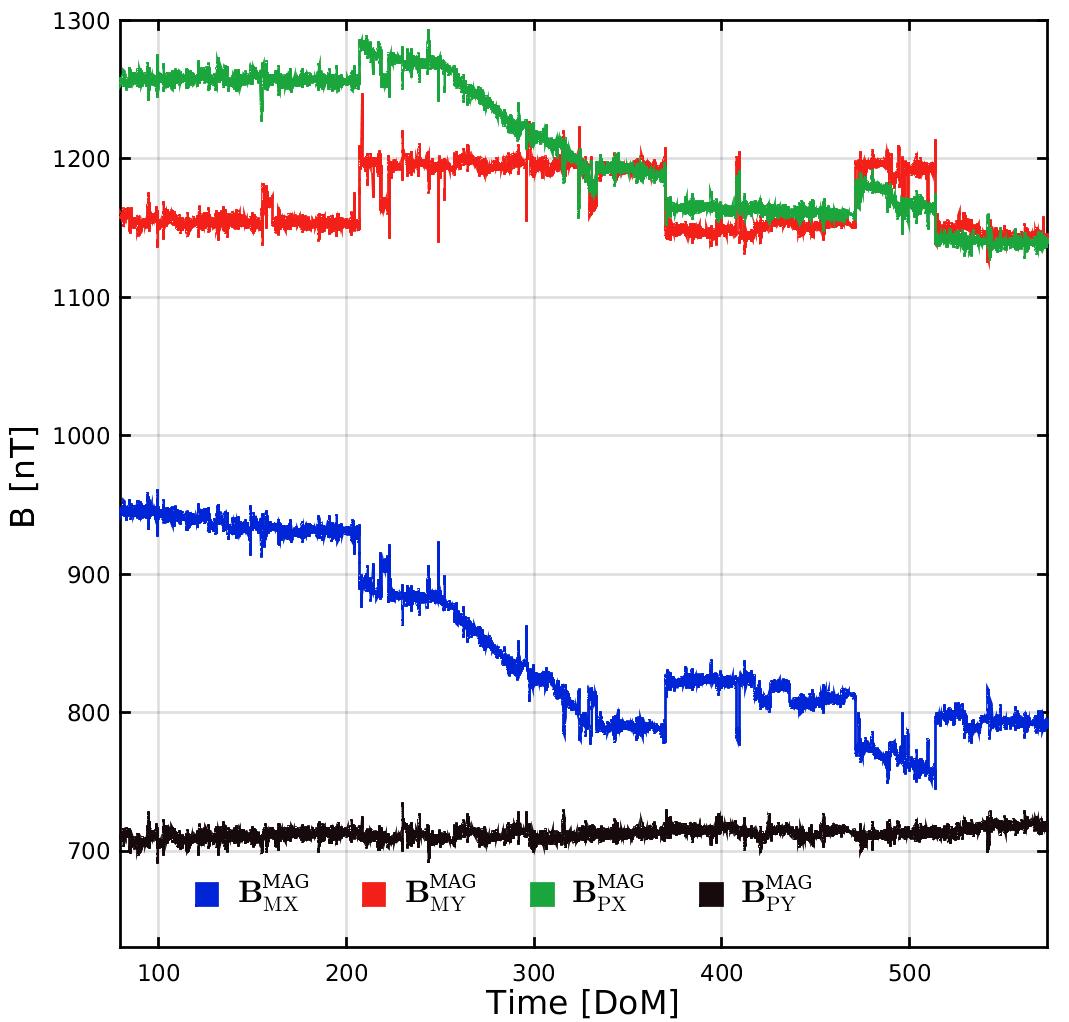}\\
	\caption{Magnetic field intensities recorded by the four LPF 3-axis fluxgate magnetometers during the time the S/C was orbiting around the Lagrange point L1 from  February 16, 2016 through June 30, 2017. Datasets were sampled at $2\times 10^{-1}\,\mathrm{Hz}$ and downsampled at $3\times 10^{-4}\,\mathrm{Hz}$.
	Time is indicated in days of mission (DoM)  after the LPF launch (4:04 UTC, December 3, 2015).}
	\label{fig_lpf_four_mags}
\end{figure}

\begin{figure}[thp]
   \centering
	%\hspace{-2cm}
	%\hspace{-0.3cm}
	\hspace{-1.3cm}
	\includegraphics[width=8.5cm,height=8cm]{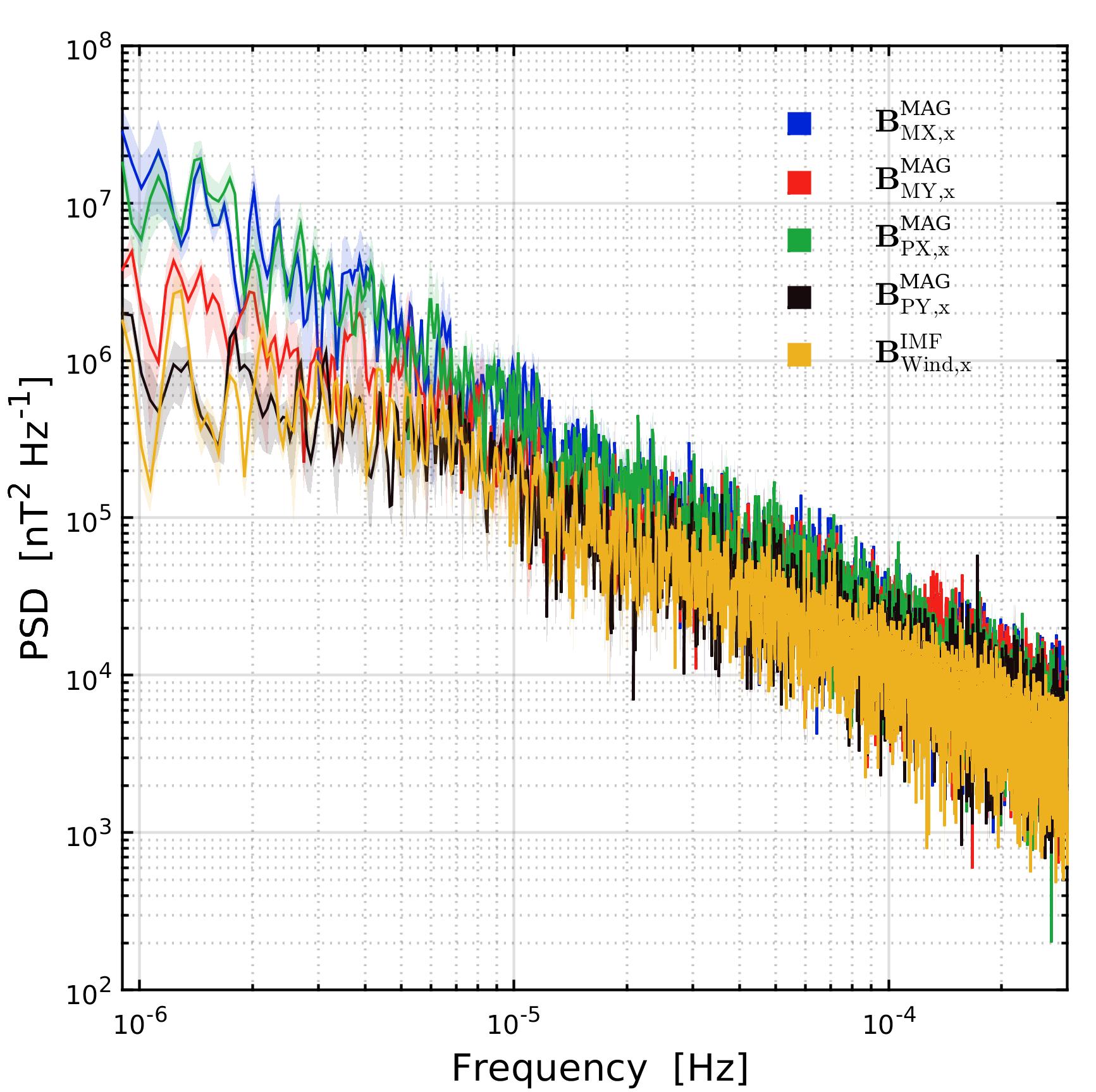}
	\includegraphics[width=8.5cm,height=8cm]{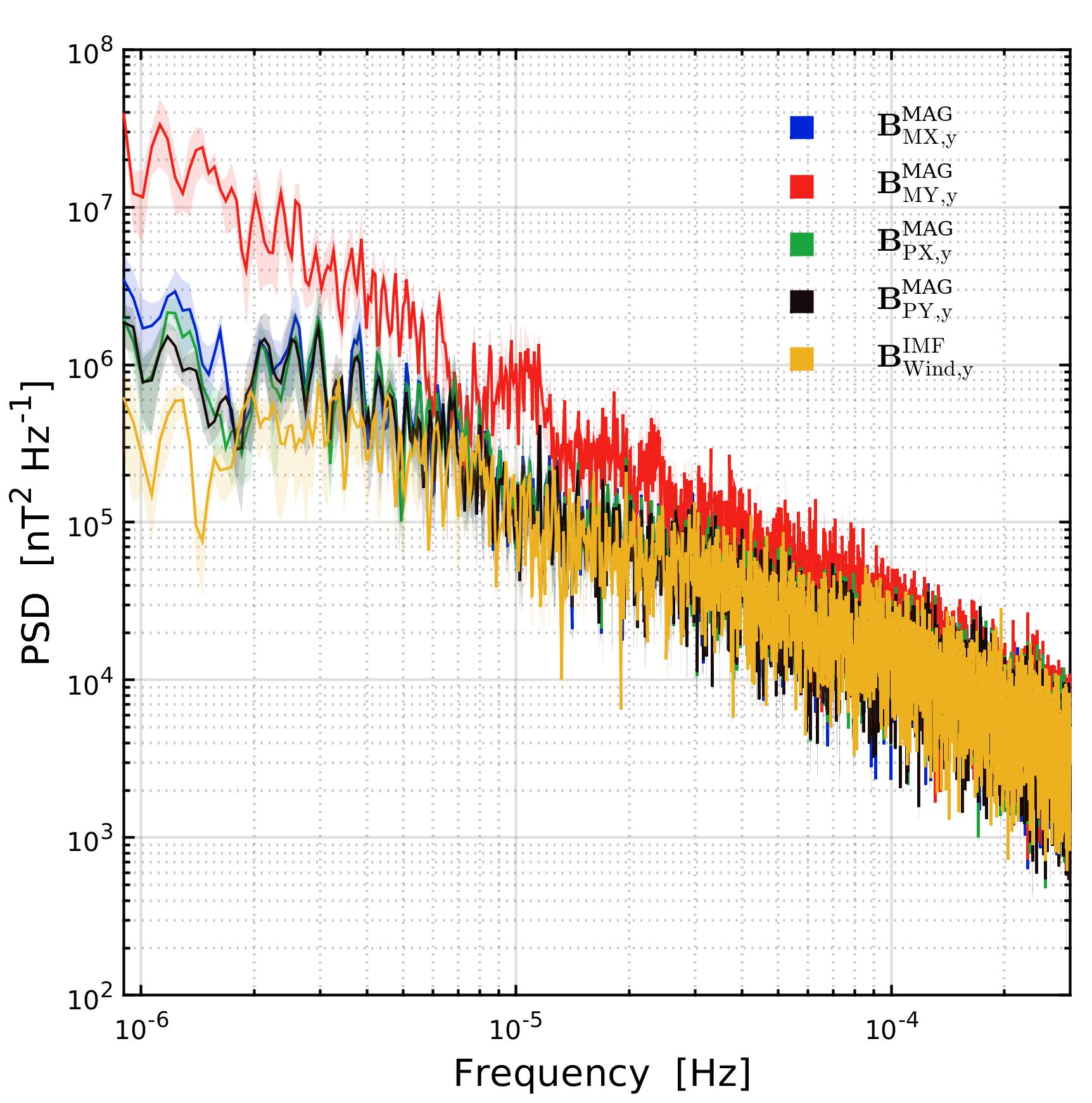}\\
	\includegraphics[width=8.5cm,height=8cm]{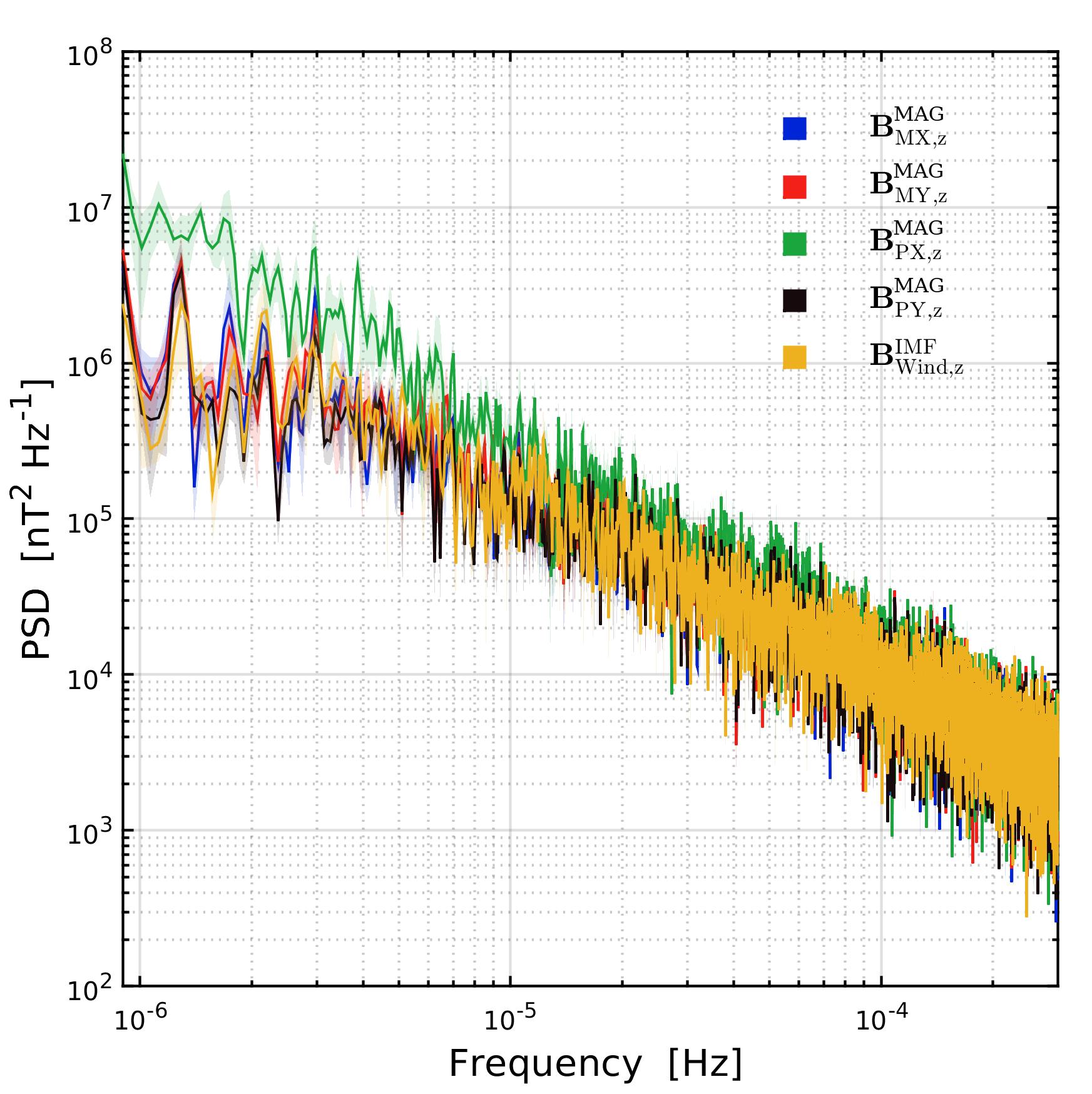}\\
	\caption{PSDs of LPF and Wind 3-axis magnetometer simultaneous measurements (from February 20, 2016 through  June 21, 2017). All measurements were reported in the $\pmb{\mathrm{R}}_\mathrm{LTP}$ reference frame for proper comparison.}
	\label{fig_lpf_four_mags_components_PSD}
\end{figure}

The LPF magnetometer calibrated datasets were sampled at 2$\times$10$^{-1}$ Hz and downsampled at $3\times 10^{-4}\,\mathrm{Hz}$. Aliasing was prevented by applying a first-order bidirectional low-pass filter. 
The intensity of the magnetic field measured by MX, MY, PX and PY magnetometers during the LPF mission are shown in Figure~\ref{fig_lpf_four_mags}. 
Observations varied by almost a factor of two between $700\,\mathrm{nT}$ and about $1300\,\mathrm{nT}$. 
The MX, MY and PX magnetometer recorded magnetic field values ($\mathbf{B}_\mathrm{MX}^\mathrm{MAG}$, $\mathbf{B}_\mathrm{MY}^\mathrm{MAG}$ and $\mathbf{B}_\mathrm{PX}^\mathrm{MAG}$, respectively) ranging between $800\,\mathrm{nT}$ and $1300\,\mathrm{nT}$  with sudden variations between different magnetic field intensities. 
Conversely, the PY magnetometer recorded stable values of the magnetic field around $700\,\mathrm{nT}$ during the whole mission lifetime ($\mathbf{B}_\mathrm{PY}^\mathrm{MAG}$). 
Magnetic field patterns and sudden changes were not ascribable to the IMF that was observed to present typical values in the $1-25\, \mathrm{nT}$ range \citep{2019ApJ...874..167A}.
In general, the magnetometers hosted on board the LPF S/C  sensed a larger noise along the three cartesian axes with respect to the Magnetic Field Investigation (MFI) instrument hosted on board the Wind mission also orbiting around L1 \citep{1995SSRv...71..207L} during the same period of LPF. Despite that, the approach presented here allows us to disentangle the IMF from the magnetic field generated on board the LPF S/C.
The power spectral densities (PSDs) of magnetic field measurements carried out with LPF during the whole mission operation period are reported in Figure~\ref{fig_lpf_four_mags_components_PSD}. 
In the same figure the PSDs of the IMF measurements gathered with MFI are reported for comparison.
It can be observed that the onboard noise overcomes that of interplanetary origin at low frequencies while it appears consistent with it above $10^{-5}\,\mathbf{Hz}$.

\section{PARAMETERIZATION OF THE MAGNETIC FIELD MEASUREMENTS GATHERED WITH LPF}
\label{section_model_parameterization}
%For the majority of the LPF in-flight operations of more than 500 days, 
%For more than 500 days after the mission launch, the   
%temperature remained nearly constant around a typical value of $297.15\pm0.75\,\mathrm{K}$ in the LISA S/C \cite{2019MNRAS.486.3368A}. 
%This temperature variation corresponded to a magnetic field  offset smaller  than $0.9\,\mathrm{nT}$ (see Table~\ref{tab0}). 
%As a result, no temperature corrections to MF measurements were necessary.
%Seldom during the S/C commissioning phase  and in particular for a period of 20 days after April 29, 2017 (DoM 513), the temperature was purposely maintained between $281\,\mathrm{K}$ and $285\,\mathrm{K}$. Magnetic field  onboard measurements gathered during these  intervals of time are shown in the following but not considered for the analysis.

The timeseries of the calibrated magnetic field  measurements gathered on-board LPF have been parameterized as follows:
\begin{eqnarray}
\label{eq_B_MAG_1}
\mathbf{B}^\mathrm{LPF}_{m}[t] =  \mathbf{B}^\mathrm{EW}_{m}[t] + \mathbf{B}^\mathrm{ED}_{m}[t] + \mathbf{B}^\mathrm{M}_{m}[t]   + \mathbf{B}^\mathrm{IMF}_{m}[t] + \mathbf{v}^c_m [t] \ .
\end{eqnarray} % result from the superposition of
In the above equation the magnetic field measured on board the LPF S/C is considered to consist of different contributions generated by the main currents flowing through wiring and electronics in the satellite ($\mathbf{B}^\mathrm{EW}_{m}[t]$ and $\mathbf{B}^\mathrm{ED}_{m}[t]$, respectively), by the magnetized materials ($\mathbf{B}^\mathrm{M}_{m}[t]$) and by the IMF ($\mathbf{B}^\mathrm{IMF}_{m}[t]$). Finally, $\mathbf{v}^c_m [t]$ represents the residue.
% or, in other words, the contribution to the LPF S/C on-board MF arising from measurement noise.
In the case the considered contributions to the LPF magnetic field observations are not overcome by other neglected components, the residue 
%$\mathbf{v}^c_m [t]$ 
is expected to be negligible and mainly associated with the measurement noise.

No measurement correction due to temperature variations were considered.
As a matter of fact, for more than 500 days after the mission launch, the temperature remained nearly constant around a typical value of $297.15\pm0.75\,\mathrm{K}$ in the LPF S/C \citep{2019MNRAS.486.3368A}.
This temperature variation corresponded to a magnetic field  offset smaller  than $0.9\,\mathrm{nT}$ (Tab.~\ref{tab0}).
%As a result, no temperature corrections to MF measurements were necessary.
Seldom during the S/C commissioning phase  and in particular for a period of 20 days after April 29, 2017 (DoM 513), the temperature was purposely maintained between $281\,\mathrm{K}$ and $285\,\mathrm{K}$. Magnetic field measurements gathered during these  intervals of time are shown in the following but excluded from the analysis.

\subsection{Magnetic field generated by wiring and electronics on board the LPF spacecraft}
\label{subs_a}

\begin{figure}[htbp]
  \centering
  \includegraphics[width=8.5cm,height=8.5cm]{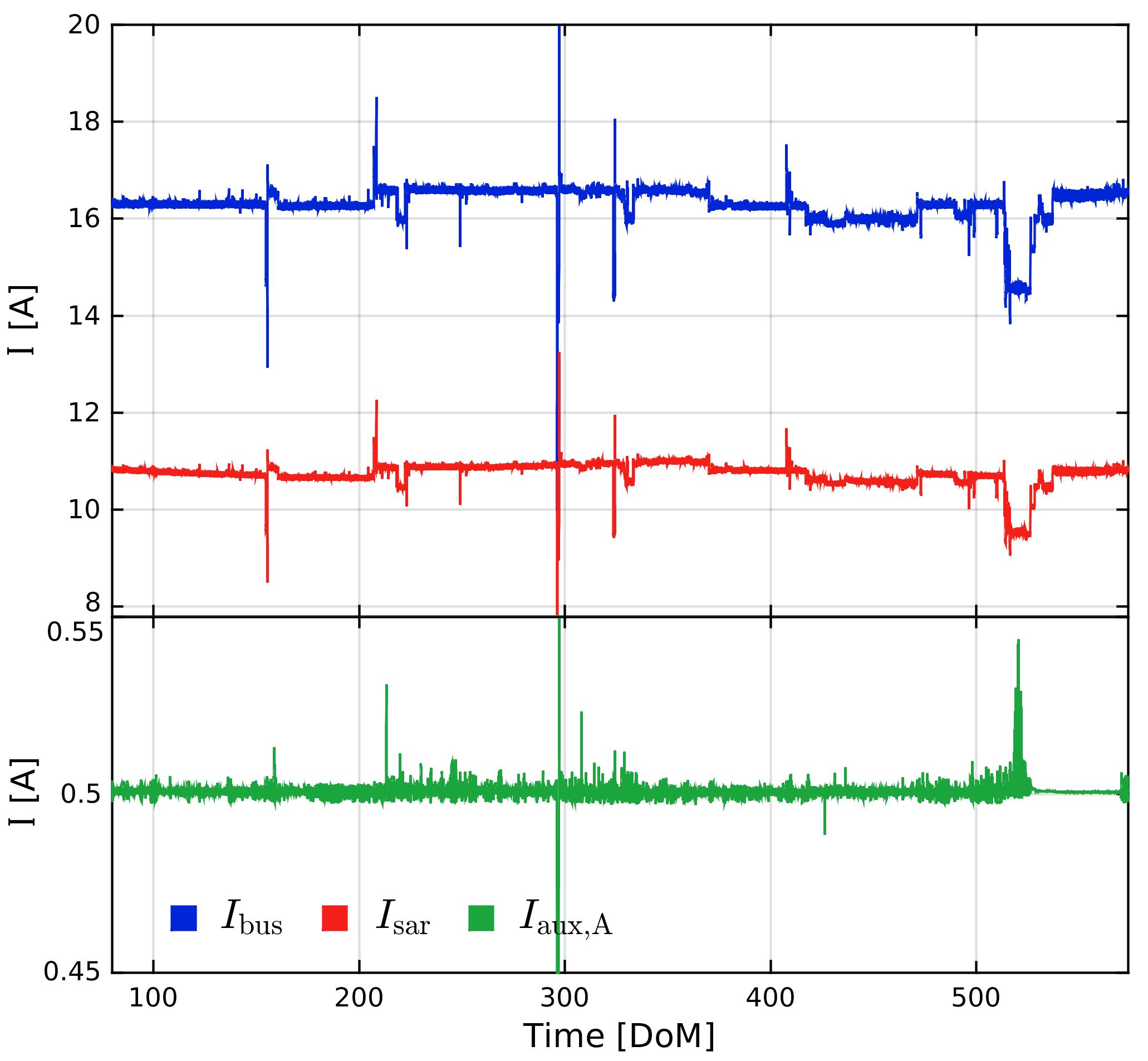}\\
  \caption{LPF main bus current ($I_{bus}$), solar array bus current ($I_{sar}$) and auxiliary power supply A current ($I_{aux,A}$) during the LPF mission operations from February 16, 2016 through June 30, 2017. At DoM near 300 an isolated overload is observed. Data in the corresponding interval of time have not been included in this analysis.}
  \label{fig_lpf_channels} 
\end{figure}

\begin{figure}[htbp]
  \centering
  \includegraphics[width=8.5cm,height=8cm]{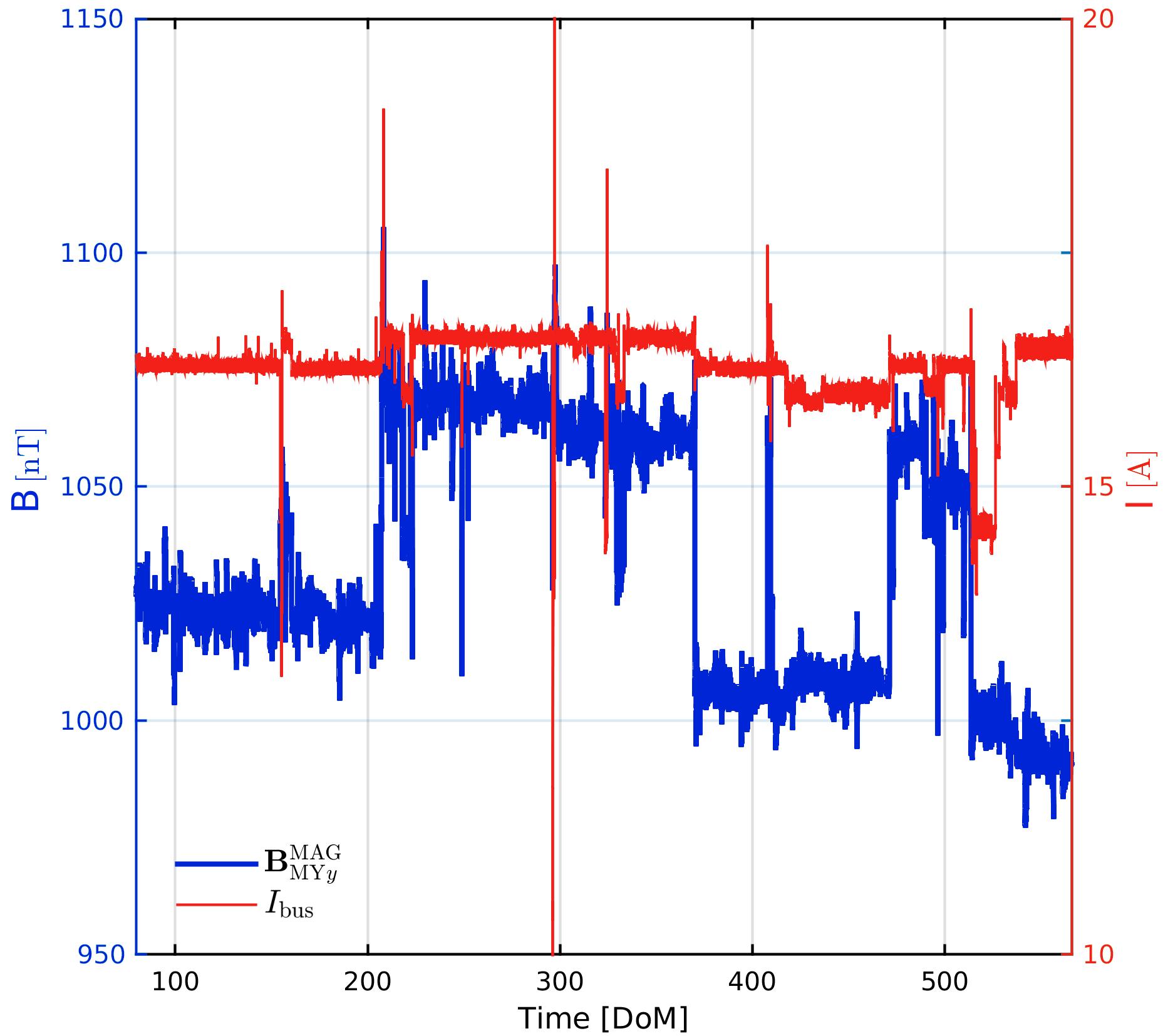}\\
  \caption{LPF  magnetic field measured by the MY magnetometer along the LTP $y$-axis (blue) and main bus current (red). The two physical quantities appear correlated and show common sudden changes.}
  \label{fig_ibus_mag} 
\end{figure}

The main contribution to the  magnetic field measured  with the  MX, MY, PX and PY magnetometers in the LPF S/C is associated with currents of $10-20\,\mathrm{A}$ flowing through the main satellite and solar array buses ($I_{bus}$ and $I_{sar}$, respectively). One secondary bus ($I_{aux,A}$) distributing a current smaller than $1\,\mathrm{A}$ was also found to play an important role. % {\blue (mainly in the case of the MX magnetometer)}. 
Figure~\ref{fig_lpf_channels} shows the timeseries of the currents flowing in the buses indicated above during the mission operations.
As an example, the correlated trend of the magnetic field measured by the MY magnetometer along the LTP $y$-axis and the main LPF current bus trend is shown in Figure~\ref{fig_ibus_mag}.
The IMF intensity shows both hourly variability and sudden jumps associated with the bus current changes.

The magnetic field components generated by the circuitry in each measurement point  are observed along the axes of the $m$-th magnetometer in the $\pmb{\mathrm{R}}_\mathrm{LTP}$ reference frame. %
At the magnetometer locations, each current bus produces a magnetic field proportional to the current flowing through the circuit that can be estimated with  the Biot-Savart equation. As a result, the sensed magnetic field change is linearly proportional to the current change in the circuitry. % per on-board relevant current bus. 

\begin{figure*}[htbp]
  \centering
  \includegraphics[width=17cm,height=6.0cm]{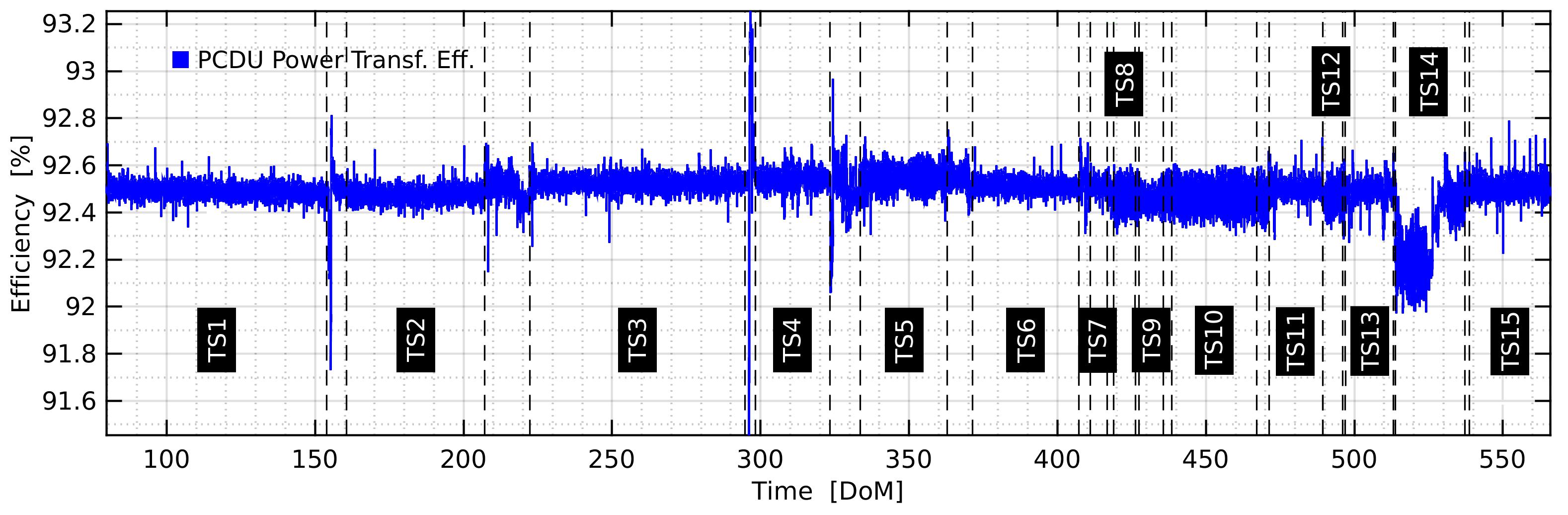}\\
  \caption{PCDU power transfer efficiency, $\eta[t]$.} 
  \label{fig_pcdu_efficiency} 
\end{figure*}

These considerations apply to the magnetic field generated by both wiring
%, $\mathbf{B}^\mathrm{EW}_{m}[t]$ 
and electronics.
%, $\mathbf{B}^\mathrm{ED}_{m}[t]$.
\iffalse
\dgreen
Orbiting around L1, the S/C was constantly illuminated by the Sun and the solar array on board LPF provided a continuous supply of power. 
This arrangement allowed to reduce the number of needed battery cells thus lowering the on-board weight and offered the possibility of using an on-board unregulated power conditioning and distribution unit (PCDU) \citep{2008ESASP.661E..45S}.
The LPF PCDU maximizes the power generation and the power storage while handling power distribution to S/C subsystems.
\fi
%The magnetic field intensity and variations on board LPF are found to be mainly associated with currents. 
With respect to the magnetic field generated by the electronics, LPF hosted an unregulated power conditioning and distribution unit (PCDU) \citep{2008ESASP.661E..45S} which showed a nearly constant power conversion efficiency, $\eta$, of 92.5\% as it is shown in Figure~\ref{fig_pcdu_efficiency}.
%\iffalse
%A nearly constant power conversion efficiency, $\eta$, of 92.5\% on LPF S/C the LPF PCDU delivers a constant power while the output voltage decreases as the output current increases (and viceversa).
%Typically, the adoption of an unregulated PCDU would force S/C subsystems to operate over a range of distinct tension levels and to compensate the current to maintain a constant power load.
%Electronics devices operating at suboptimal tension levels present different power conversion efficiencies and require larger currents which, in turn, increase the MF. 
%{\red Time-dependent changes in the main supply bus voltage, $V_{bus}[t]$, was in fact stabilized very efficiently by }
%would lead to nonlinear changes of the magnetic field generated by each electronic device with respect to the supplied current.
%Contrarily to the expectations, it was observed that 
%\fi
The main supply bus voltage was stabilized very efficiently at $V_\mathrm{bus} = 27.903\pm 0.002\, \mathrm{V}$ by a Li-Ion  battery  which, due to a constant solar illumination, was handled at a fix charge/discharge point. 
Sudden changes of the power conversion efficiency and current variations were observed during electronics configuration switches only. % sudden cahnges arising from electronics switchings.
Table~\ref{tab1} reports the complex changes of electronics configurations during the mission operations.
The S/C was driven by the drag-free attitude control system (DFACS) or by the disturbance reduction system (DRS).
Both these attitude control systems were used to command with forces and torques the S/C and the two proof TMs \citep{PhysRevLett.116.231101}.
The DFACS controlled a set of cold-gas thrusters \citep{MorrisEdwardsThrusters2013} and the electronics for the TM electrostatic actuation.
The DRS utilized a distinct control scheme implemented by a dedicated on-board computer with electronics and a second set of $8$ colloidal micro-newton thrusters (CMNTs) \citep{2018PhRvD..98j2005A}. 

%Time periods indicated by ``DFACS+'' were those during which the DRS related subsystems were  switched on, even when the actuation was assigned to the DFACS. 
During time periods indicated by “DFACS+”, the DRS related subsystems were left switched on, even when the actuation was assigned to the DFACS.
For instance, this configuration was applied during the months of July and August 2016.
Analogously, the periods indicated by ``DRS+'' were those associated with actuation assigned to the DRS with the DFACS  powered on.

%

%Table~\ref{tab1} reports the complex changes of electronics configurations during the mission operations.
The timespans were selected on the basis of the electronic switches, the current bus trends and the power transfer efficiency $\eta[t]$ in order to select proper magnetic field  pattern changes (Fig.~\ref{fig_lpf_four_mags}).
%{0.9\textheight}
\begin{table}[tbhp]
	\centering
	\resizebox{8.5cm}{!}{ 
		\begin{tabular}{|c|c|c|c|c|}\hline
			Timespan & Confs & Start Time & Stop Time & Duration\\
			 & &{[UTC]} & {[UTC]} & {[Days]}\\ \hline \hline
			1  & DFACS & 2016-02-20, 22:20 & 2016-05-04, 23:45 & 74\\
			2  & DFACS & 2016-05-11, 16:35 & 2016-06-27, 06:00 & 46\\
			3  & DRS   & 2016-07-12, 09:55 & 2016-09-23, 00:05 & 72\\
			4  & DRS   & 2016-09-26, 10:00 & 2016-10-21, 16:40 & 25\\
			5  & DFACS+& 2016-10-31, 16:35 & 2016-12-07, 03:35 & 36\\
			6  & DFACS+& 2016-12-08, 13:45 & 2017-01-13, 10:00 & 35\\
			7  & DFACS+& 2017-01-17, 06:30 & 2017-01-22, 23:45 & 5\\
			8  & DFACS+& 2017-01-25, 00:45 & 2017-02-01, 11:45 & 7\\
			9  & DFACS+& 2017-02-02, 14:15 & 2017-02-10, 19:55 & 8\\
			10 & DFACS+& 2017-02-13, 17:35 & 2017-03-14, 06:00 & 28\\
			11 & DRS+  & 2017-03-18, 10:00 & 2017-04-05, 10:50 & 18\\
			12 & DFACS+& 2017-04-05, 12:55 & 2017-04-12, 04:05 & 6\\
			13 & DRS+  & 2017-04-13, 02:15 & 2017-04-29, 07:30 & 16\\
			14 & DFACS+& 2017-04-30, 00:05 & 2017-05-23, 07:30 & 23\\
			15 & DFACS+& 2017-05-24, 20:25 & 2017-06-21, 20:00 & 27\\\hline
		\end{tabular}}
	\caption{LPF electronics configuration during different time intervals. 
	}
	\label{tab1}
\end{table}

In conclusion, according to the Biot-Savart law and the superposition principle \citep{yu2013}, the magnetic field intensity produced at each magnetometer position ($m$) by wiring and electronics is empirically parameterized as indicated in the following equation:
%\begin{eqnarray}\nonumber
%\label{eq_BH_BBS}
%\displaystyle \mathbf{B}^\mathrm{EM}_{m}[t] &=& \mathbf{B}^\mathrm{EW}_{m}[t] + \mathbf{B}^\mathrm{ED}_{m}[t] =\\\nonumber
%&=& \sum_{j=1}^{3} \frac{\mu_0}{4 \pi} \int_{\mathcal{C}_j} \frac{I_j [t]}{r_{m,j}^2} d\mathbf{l}_j \times \mathbf{u}_{m,j} = \\
%&=& \sum_{j=1}^{3}\, \mathbf{a}_{m,j}\,  I_j [t]
%\end{eqnarray}
\begin{eqnarray}\nonumber
\label{eq_BH_BBS}
\displaystyle \mathbf{B}^\mathrm{EM}_{m}[t] &=& \mathbf{B}^\mathrm{EW}_{m}[t] + \mathbf{B}^\mathrm{ED}_{m}[t] =\\\nonumber
%&=& \sum_{j=1}^{3} \frac{\mu_0}{4 \pi} \int_{\mathcal{C}_j} \frac{I_j [t]}{r_{m}^2(j)} d\mathbf{l}(j) \times \mathbf{u}_{m}(j) = \\
&=& \sum_{j=1}^{3} \frac{\mu_0}{4 \pi} \int_{\mathcal{C}_j} \frac{I_j [t]}{r_{m,j}^2} d\mathbf{l}_j \times \mathbf{u}_{m,j} = \\
&=& \sum_{j=1}^{3}\, \mathbf{a}_{m,j}\,  I_j [t]
\end{eqnarray}
where $d\mathbf{l}_j$ represents the infinitesimal current element along the $\mathcal{C}_j$ integration path associated with the $j$-th bus; $r_{m,j}$ indicates the distance of each $d\mathbf{l}_j$ current segment of each current bus from the $m$-th position of each magnetometer and $\mathbf{u}_{m,j}$ is the unit vector associated with $\mathbf{r}_{m,j}$ indicating the distance from the bus element to the point where the field is calculated. The $\mathbf{a}_{m,j}$ parameters allow to account for the magnetic field generated by the $j$-th bus current, $I_j [t]$. These parameters are estimated in Section~V.
% as $\displaystyle  \mathbf{a}_{m,j} =[\mathrm{a}_{m,j,x}\, \mathrm{a}_{m,j,y}\, \mathrm{a}_{m,j,z}]^\mathrm{T}$ which allows to establish a linear relation between the current flowing throughout each bus and the magnetic field along the selected Cartesian axis in the $\pmb{\mathrm{R}}_\mathrm{LTP}$ reference frame.

\subsection{LPF  magnetic field generated by magnetized materials}
\label{subs_b}
%\iffalse
%The overall magnetic field originating from on-board magnetized materials, $\mathbf{B}^\mathrm{M}_\mathrm{m}[t]$, is ascribable to the S/C residual demagnetization and to magnetized moving mechanical parts.
%
%On ground, an effective demagnetization of the S/C magnetized parts ($<100\,\mathrm{nT}$) cannot be achieved because of the presence of the relatively strong Earth's magnetic field (ranging between $25-65\,\mathrm{\mu T}$) with respect to the magnetic field of the space environment ($< 50\,\mathrm{nT}$). In-flight, soft magnetic materials on board LPF demagnetize depending on the intensity and variation of the on-board generated MF. 
%However, due to the limited mission elapsed time and to the subdivision of the analysis intervals (presented in Subsection~\ref{subs_a}), such a demagnetization effect becomes negligible.
%\fi

\begin{figure}[htbp]
  \centering
  \includegraphics[width=8.5cm,height=8cm]{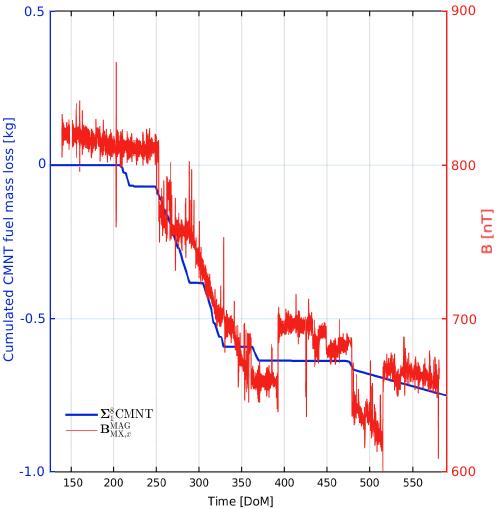}\\ %COLOR
  \caption{Total CMNT fuel mass ejected and on-board induction magnetic field timeseries $\mathbf{B}^\mathrm{MAG}_{\mathrm{MX},x}[t]$ (in blue and red, respectively) recorded by the MX magnetometer along the LTP $x$-axis.}
  \label{fig_lpf__fuelmass_mx_comparison} 
\end{figure}

Two distinct independent micro propulsion systems were hosted on board the LPF S/C: cold-gas micro propulsion thrusters, as inherited from the Gaia mission \citep{2011AcAau..69..822S}, and $8$ CMNTs developed by Busek and managed by NASA at the Jet Propulsion Laboratory \citep{ZiemerEtAl2006} which were part of the DRS. 
The DRS included four subsystems: the Integrated Avionics Unit (IAU), two clusters of four CMNTs each, the Dynamic Control Software (DCS) and the Flight Software (FSW).
CMNTs were commanded by applying an electric potential to a charged liquid in order to emit a stream of %tiny, 
droplets for the thrust. During the mission, a major contribution to the magnetic field disturbance was found to be strongly correlated with the CMNT fuel mass ejection (Fig.~\ref{fig_lpf__fuelmass_mx_comparison}).
A  magnetized assembly pushes out the propellant by moving out from the magnetometers and generating a slowly decreasing magnetic field \citep{10.1093/mnras/staa830}. This decrease is mostly sensed by the MX and PX magnetometers along the LTP $x$-axis.
On the basis of the correlated behavior between the on-board magnetic field variation and the fuel mass ejection timeseries, the contribution to the on-board magnetic field arising from the CMNT tank moving assemblies is modeled as follows:
\begin{eqnarray}
\label{eq_thrusters}
\mathbf{B}_m^\mathrm{M} [t] = \mathbf{l}_{m}\ M[t]
\end{eqnarray}	
where $M[t] = \sum_{i=1}^{8} {C\!M\!N\!T}_i [t]$ accounts for the fuel mass loss obtained by summing up  the fuel mass quantity ejected from the CMNTs (Fig.~\ref{fig_lpf__fuelmass_mx_comparison}). In Equation~\ref{eq_thrusters}, $\mathbf{l}_{m}$ is a parameter tuned via numerical fitting (see Section~\ref{section_bayesian_parameter_estimation}) and expressed in units of $\mathrm{nT\, kg^{-1}}$.

\section{Interplanetary magnetic field monitoring with LISA Pathfinder}

\subsection{Data analysis}
\label{section_bayesian_parameter_estimation}

% % % % % % % % % % % % % % % % % % % %
% % % % % % % % % % % % % % % % % % % %
% % % % % % % % % % % % % % % % % % % %
% % % % % % % % % % % % % % % % % % % %
% % % % % % % % % % % % % % % % % % % % DATA FITTING
% % % % % % % % % % % % % % % % % % % %
% % % % % % % % % % % % % % % % % % % %
% % % % % % % % % % % % % % % % % % % %
% % % % % % % % % % % % % % % % % % % %

The magnetic field measured on board the LPF S/C expressed in Equation~\ref{eq_B_MAG_1} can be now parameterized as follows: %is obtained by combining Equations~\ref{eq_BH_BBS} and~\ref{eq_thrusters} with Equation~\ref{eq_B_MAG_1},
\begin{eqnarray}
\label{eq_B_MAG_5}\nonumber
\mathbf{B}^\mathrm{LPF}_{m}[t] &=& \mathbf{B}^\mathrm{EM}_{m}[t] + \mathbf{B}^\mathrm{M}_{m}[t] + \mathbf{B}^\mathrm{IMF}_{m}[t] + \mathbf{v}^c_m [t] \\
&\cong&  \sum_{j=1}^{3} \mathbf{a}_{j,m} I_j[t] + \mathbf{l}_{m}\ M[t] + \mathbf{B}^\mathrm{IMF}_{m}[t] + \mathbf{v}^c_m [t]\ .
\end{eqnarray} 

From the above equation, it is possible to estimate the IMF component in the time domain.
% is:
%\begin{eqnarray}
%\label{eq_B_MAG_applied}\nonumber
%\mathbf{B}^\mathrm{IMF}_{m}[t] &\cong& \mathbf{B}^\mathrm{MAG}_{m}[t] - \sum_{j=1}^{3} \mathbf{a}_{j,m} I_j[t] - \mathbf{l}_{m}\ M[t] + \mathbf{v}^c_m [t]\, .
%\end{eqnarray} 
The same equation can be written in the frequency domain in order  to remove the colored noise from the data. In the frequency domain each term is indicated with the {\it tilde} symbol and the IMF can be expressed as indicated below:
\begin{eqnarray}
\label{eq_BIMFmodel_freq_applied}
\mathbf{\tilde{B}}^\mathrm{IMF}_{m}[f] &\cong& \mathbf{\tilde{B}}^\mathrm{LPF}_{m}[f] - \sum_{j=1}^{3} \mathbf{a}_{j,m} \tilde{I}_j[f] - \mathbf{l}_{m}\ \tilde{M}[f] + \mathbf{\tilde{v}}^c_m [f]\,
\end{eqnarray}
where $\mathbf{\tilde{v}}^c_m [f]$ is expected to be the overall noise amplitude in case all the main components of the magnetic field have been taken into account.
The LPF Data Analysis toolbox (LTPDA, publicly available at \url{https://www.elisascience.org/ltpda/}) is a toolbox implemented in Matlab (\url{http://www.mathworks.com}) and was specifically developed for the analysis of the LPF mission data \citep{Antonucci_2011}. 
A Markov Chain Monte Carlo (MCMC) algorithm was used to estimate the posterior distribution of the model parameters on the basis of the data timeseries provided in input \citep{PhysRevD.90.042003}.
The parameter distribution  $\mathrm{\hat{\theta}}$ was obtained with an iterative least square error estimate after the PSDs of $\tilde{I}_j[f]$ and $\tilde{M}[f]$ 
were calculated from the 
equivalent  timeseries in the time domain. %Parameters $\mathbf{a}_{j,m}$ and $\mathbf{l}_{m}$ 
%
% "Se uno sa quello che vuole dire le parole seguono..."
%
% 1- Trasforma le timeseries in frequenza ottenendo uno spettro
% 2- 
% 3- 
%
%
By indicating with $\theta_m$ the column vector of the input parameters:
{
\begin{eqnarray}
\label{eq_THETA}
\pmb{\mathrm{\theta}}_m = \left[\mathbf{a}_{m,j} \, , \, \mathbf{l}_m \right]^T
\end{eqnarray}
}
the IMF has been estimated by minimizing the residue among collected data and empirical modelization of the LPF S/C on-board magnetic field, as follows:
\begin{eqnarray}
\label{eq_THETA_ESTI}
\pmb{\mathrm{\hat{\theta}}}_m &\simeq& \argmin_{\theta_{m}} \left| \mathbf{\tilde{B}}^\mathrm{MAG}_{m}(\theta_{m})[f] - \mathbf{\tilde{B}}^\mathrm{LPF}_{m}(\theta_{m})[f] \right|\, .
\end{eqnarray}
%where the overall LPF magnetic field contribution arising from wiring, electronics and magnetized materials is indicated with $\mathbf{\tilde{B}}^\mathrm{LPF}_{m}(\theta_{m})[f]$.

Separate sets of parameters were estimated for each magnetometer and for each time segment listed in Table~\ref{tab1}. 
The fitting program returned  the IMF intensity values in the range $1-25\,\mathrm{nT}$ in line with the observations gathered by the Wind experiment during the LPF mission 
lifetime. %listed in 
Parameter estimates and corresponding uncertainties are reported in Tables~\ref{tab:evalparmag1}, \ref{tab:evalparmag2}, \ref{tab:evalparmag3} and~\ref{tab:evalparmag4}. %It 
As it was recalled in Section~\ref{section_model_parameterization}, the parameters associated with TS13 and TS14 timespans were ignored because of an abrupt fall of the temperature on board the S/C.

\begin{table*}[ht]
%\centering
\resizebox{17cm}{!}{ 
%\multicolumn{5}{c}{} & \multicolumn{8}{c}{j} \tabularnewline
%\multicolumn{5}{c}{Magnetometer 1, MX}
$
\begin{array}{|c|c|c|c|c|c|} \hline
\begin{array}{c} \\ \mathbf{MAG1}\\ \mathbf{Coefficients} \end{array} & \mathbf{Axis} & \mathbf{a_1} \mathrm{[\mathrm{nT\,nA^{-1}}]}& \mathbf{a_2} \mathrm{[\mathrm{nT\,nA^{-1}}] } & \mathbf{a_3} \mathrm{[\mathrm{nT\,nA^{-1}}]} & \mathbf{l}\, \mathrm{[\mathrm{nT\,kg^{-1}}]} \\ 
%\mathbf{Coefficients} & [\mathrm{nT\,nA^{-1}}] & [\mathrm{nT\,nA^{-1}}] & [\mathrm{nT\,nA^{-1}}] & [\mathrm{nT\,kg^{-1}}] \\ %\hline\hline

%\mathbf{Coefficients}
&
&
\begin{array}{cccc}
value & \pm & error & (\%)\\ %\hline
\end{array} 
&
\begin{array}{cccc}
value & \pm & error & (\%)\\ %\hline
\end{array} 
&
\begin{array}{cccc}
value & \pm & error & (\%)\\ %\hline
\end{array} 
&
\begin{array}{cccc}
value & \pm & error & (\%)\\ %\hline
\end{array} 
\\ \hline

\begin{array}{c} \\ \mathbf{TS1}\\ \mathrm{(74\, days)} \end{array} & 
\begin{array}{c}
  x\\ 
  y\\ 
  z\\
 \end{array} 
 & 
\begin{array}{cccc}
%  & value & \pm & error & (\%)\\ \hline
-69.638 & \pm & 5.870 &(8\%)\\ 
~75.271 & \pm & 5.682 &(7\%)\\ 
~35.670 & \pm & 7.591 &(21\%)\\
 \end{array} 
 & 
\begin{array}{cccc}
%  & value & \pm & error & (\%)\\ \hline
~108.11 & \pm & 8.06 &(7\%)\\ 
-126.72 & \pm & 8.49 &(7\%)\\ 
-36.602 & \pm & 10.360 &(28\%)\\ 
 \end{array} 
 & 
 \begin{array}{cccc}
%  & value & \pm & error & (\%)\\ \hline
~1997.0 & \pm & 86.6 &(4\%)\\ 
-812.57  & \pm & 67.51  &(8\%)\\ 
-240.66 & \pm & 97.07 &(40\%)\\ 
 \end{array} 
& 
 \\ \hline

\begin{array}{c} \\ \mathbf{TS2}\\ \mathrm{(46\, days)} \end{array} & 
\begin{array}{c}
  x\\ 
  y\\ 
  z\\
 \end{array} 
 & 
\begin{array}{cccc}
~65.043 & \pm &  24.560  &(38\%)\\ 
~       &     &       &     \\ 
~15.921 & \pm &  2.684 &(17\%)\\ 
 \end{array} 
 & 
\begin{array}{cccc}
-80.400 & \pm &  36.740 &(46\%)\\ 
~       &     &       &     \\ 
~       &     &       &     \\ 
 \end{array} 
 & 
 \begin{array}{cccc}
~1545.4 & \pm &  103.1 &(15\%)\\ 
-1134.6 & \pm &  0.7 &(0\%)\\ 
-403.53 & \pm &  110.80 &(27\%)\\ 
 \end{array} 
& \\ \hline

\begin{array}{c}\\ \mathbf{TS3}\\ \mathrm{(72\, days)} \end{array} & 
\begin{array}{c}
  x\\ 
  y\\ 
  z\\
 \end{array} 
 & 
\begin{array}{cccc}
~       &     &       &      \\ 
-53.988 & \pm & 19.020 &(35\%)\\ 
~       &     &       &      \\ 
 \end{array} 
 & 
\begin{array}{cccc}
-41.448 & \pm &  1.352 &(3\%)\\ 
~82.558 & \pm & 28.040 &(34\%)\\ 
~       &     &       &     \\ 
 \end{array} 
 & 
 \begin{array}{cccc}
~3086.1 & \pm & 36.8 &(1\%)\\ 
-1127.5 & \pm & 66.7 &(6\%)\\ 
~242.26 & \pm & 1.97 &(1\%)\\ 
 \end{array} 
 & 
\begin{array}{cccc}
~343.64 & \pm & 5.57  &(2\%)\\ 
~       &     &       &     \\ 
~73.302 & \pm & 8.083 &(11\%)\\ 
 \end{array} 
\\ \hline

\begin{array}{c}\\ \mathbf{TS4}\\ \mathrm{(25\, days)} \end{array} & 
\begin{array}{c}
  x\\ 
  y\\ 
  z\\
 \end{array} 
 & 
\begin{array}{cccc}
 -47.867 & \pm &  26.260 &(55\%)\\ 
~       &     &       &     \\ 
~       &     &       &     \\ 
 \end{array} 
 & 
\begin{array}{cccc}
~89.593 & \pm &  38.550 &(43\%)\\ 
~       &     &       &     \\ 
~       &     &       &     \\ 
 \end{array} 
 & 
\begin{array}{cccc}
~1471.5 & \pm &  91.1 &(6\%)\\ 
-1099.3 & \pm &  1.4 &(0\%)\\ 
~260.65 & \pm &  1.5 &(1\%)\\ 
\end{array} 
 & \\ \hline

\begin{array}{c} \\ \mathbf{TS5}\\ \mathrm{(36\, days)} \end{array} & 
\begin{array}{c}
  x\\ 
  y\\ 
  z\\
 \end{array} 
 & 
\begin{array}{cccc}
-25.582 & \pm &  10.990 &(43\%)\\ 
-25.882 & \pm &  10.700 &(41\%)\\ 
~       &     &       &     \\ 
\end{array} 
& 
\begin{array}{cccc}
~53.523 & \pm &  15.940 &(30\%)\\ 
~31.480 & \pm &  15.830 &(52\%)\\ 
~       &     &       &     \\ 
\end{array} 
& 
\begin{array}{cccc}
~1525.4 & \pm &  100.8 &(7\%)\\ 
-867.43 & \pm &  101.7 &(12\%)\\ 
~278.89 & \pm &  0.63 &(0\%)\\ 
\end{array} 
& \\ \hline

\begin{array}{c} \\ \mathbf{TS6}\\ \mathrm{(35\, days)} \end{array} &  
\begin{array}{c}
  x\\ 
  y\\ 
  z\\
 \end{array} 
 & 
\begin{array}{cccc}
%  & value & \pm & error & (\%)\\ \hline
~       &     &       &     \\ 
 -14.609 & \pm &  4.214 &  (29\%)\\ 
~       &     &       &     \\ 
\end{array}
&
\begin{array}{cccc}
%  & value & \pm & error & (\%)\\ \hline
~       &     &       &     \\ 
~       &     &       &     \\ 
~       &     &       &     \\ 
\end{array}
&
\begin{array}{cccc}
%  & value & \pm & error & (\%)\\ \hline
~1767.0 & \pm &  0.8 &  (0\%)\\ 
-474.09 & \pm &  173.90 &  (37\%)\\ 
~302.27 & \pm &  0.89 &  (0\%)\\ 
\end{array}
& \\ \hline

\begin{array}{c} \\ \mathbf{TS7}\\ \mathrm{(5\, days)} \end{array} & 
\begin{array}{c}
  x\\ 
  y\\ 
  z\\
 \end{array} 
 & 
\begin{array}{cccc}
~       &     &       &     \\ 
~       &     &       &     \\ 
~       &     &       &     \\ 
\end{array} 
& 
\begin{array}{cccc}
~       &     &       &     \\ 
~       &     &       &     \\ 
~       &     &       &     \\ 
\end{array} 
& 
\begin{array}{cccc}
~1766.7 & \pm &  2.5 &(1\%)\\ 
-1088.8 & \pm &  4.1 &(0\%)\\ 
~300.39 & \pm &  2.98 &(1\%)\\ 
\end{array} 
& \\ \hline

%8 LASCIATO VECCHIO
\begin{array}{c} \\ \mathbf{TS8}\\ \mathrm{(7\, days)} \end{array} & 
\begin{array}{c}
  x\\ 
  y\\ 
  z\\
 \end{array} 
 & 
\begin{array}{cccc}
~       &     &       &     \\ 
~       &     &       &     \\ 
~       &     &       &     \\ 
 \end{array} 
 & 
\begin{array}{cccc}
~       &     &       &     \\ 
~       &     &       &     \\ 
~       &     &       &     \\ 
 \end{array} 
 & 
\begin{array}{cccc}
~1730.6 & \pm &  1.5  &(0\%)\\ 
-1070.8 & \pm &  1.9  &(0\%)\\ 
~313.74 & \pm &  2.40  &(1\%)\\ 
\end{array} 
& \\ \hline

\begin{array}{c} \\ \mathbf{TS9}\\ \mathrm{(8\, days)} \end{array} & 
\begin{array}{c}
  x\\ 
  y\\ 
  z\\
 \end{array} 
 & 
\begin{array}{cccc}
~32.841 & \pm &  4.103 &(12\%)\\ 
~       &     &       &     \\ 
~       &     &       &     \\ 
 \end{array} 
 & 
\begin{array}{cccc}
~       &     &       &     \\ 
~       &     &       &     \\ 
~       &     &       &     \\ 
 \end{array} 
 & 
\begin{array}{cccc}
~723.90 & \pm &  165.50 &(23\%)\\ 
-1088.4 & \pm &  1.0 &(0\%)\\ 
~309.93 & \pm &  0.89 &(0\%)\\ 
\end{array} 
& \\ \hline

\begin{array}{c} \\ \mathbf{TS10}\\ \mathrm{(28\, days)} \end{array} & 
\begin{array}{c}
  x\\ 
  y\\ 
  z\\
 \end{array} 
 & 
\begin{array}{cccc}
~84.19 & \pm &  18.34 &(22\%)\\ 
~       &     &       &     \\ 
~       &     &       &     \\ 
 \end{array} 
 & 
\begin{array}{cccc}
-113.01 & \pm &  26.52 &(23\%)\\ 
~       &     &       &     \\ 
~       &     &       &     \\ 
 \end{array} 
 & 
\begin{array}{cccc}
~1637.6 & \pm &  155.2 &(9\%)\\ 
-1077.6 & \pm &  0.7 &(0\%)\\ 
~318.52 & \pm &  0.84 &(0\%)\\ 
\end{array} 
& \\ \hline

\begin{array}{c} \\ \mathbf{TS11}\\ \mathrm{(18\, days)} \end{array} & 
\begin{array}{c}
  x\\ 
  y\\ 
  z\\
 \end{array} 
 & 
\begin{array}{cccc}
~30.807 & \pm &  20.480 &(66\%)\\ 
~       &     &       &     \\ 
~76.224 & \pm &  27.610 &(36\%)\\ 
 \end{array} 
 & 
\begin{array}{cccc}
-59.847 & \pm &  30.920 &(52\%)\\ 
~       &     &       &     \\ 
-125.41 & \pm &  42.28 &(34\%)\\ 
 \end{array} 
 & 
\begin{array}{cccc}
~2276.3 & \pm &  107.5 &(5\%)\\ 
-1071.6 & \pm &  1.1 &(0\%)\\ 
~559.59 & \pm &  138.10 &(25\%)\\ 
\end{array} 
 & \\ \hline

\begin{array}{c} \\ \mathbf{TS12}\\ \mathrm{(6\, days)} \end{array} & 
\begin{array}{c}
  x\\ 
  y\\ 
  z\\
 \end{array} 
 & 
\begin{array}{cccc}
%  & value & \pm & error & (\%)\\ \hline
~       &     &       &     \\ 
~       &     &       &     \\ 
~       &     &       &     \\ 
\end{array}
&
\begin{array}{cccc}
%  & value & \pm & error & (\%)\\ \hline
~       &     &       &     \\ 
~       &     &       &     \\ 
~       &     &       &     \\ 
\end{array}
&
\begin{array}{cccc}
%  & value & \pm & error & (\%)\\ \hline
~1826.4 & \pm &  78.3 &  (4\%)\\ 
-1339.2 & \pm &  77.5 &  (17\%)\\ 
~       &     &       &     \\ 
\end{array}
& \\ \hline

\begin{array}{c} \\ \mathbf{TS13}\\ \mathrm{(16\, days)} \end{array} & 
\begin{array}{c}
  x\\ 
  y\\ 
  z\\
 \end{array} 
 & 
%\begin{array}{cccc}
%%  & value & \pm & error & (\%)\\ \hline
%33.175 &  \pm &  45.62 &  (\uparrow)\\ 
%-18.262 &  \pm &  57.1 &  (\uparrow)\\ 
%6.3719 &  \pm &  52.34 &  (\uparrow)\\ 
%\end{array}
&
%\begin{array}{cccc}
%%  & value & \pm & error & (\%)\\ \hline
%-87.068 &  \pm &  68.13 &  (78\%)\\ 
%20.416 &  \pm &  85.12 &  (\uparrow)\\ 
%-23.83 &  \pm &  78.73 &  (\uparrow)\\ 
%\end{array}
&
%\begin{array}{cccc}
%%  & value & \pm & error & (\%)\\ \hline
%2887.6 &  \pm &  110.6 &  (4\%)\\ 
%-852.84 &  \pm &  118.1 &  (14\%)\\ 
%674.51 &  \pm &  138.1 &  (20\%)\\ 
%\end{array}
& \\ \hline

\begin{array}{c} \\ \mathbf{TS14}\\ \mathrm{(23\, days)} \end{array} & 
\begin{array}{c}
  x\\ 
  y\\ 
  z\\
 \end{array} 
 & 
%\begin{array}{cccc}
%%  & value & \pm & error & (\%)\\ \hline
%165.32 &  \pm &  51.49 &  (31\%)\\ 
%2.0824 &  \pm &  39.81 &  (\uparrow)\\ 
%-30.758 &  \pm &  77.58 &  (\uparrow)\\ 
%\end{array}
&
%\begin{array}{cccc}
%%  & value & \pm & error & (\%)\\ \hline
%-248.25 &  \pm &  79.95 &  (32\%)\\ 
%-4.3356 &  \pm &  61.87 &  (\uparrow)\\ 
%46.789 &  \pm &  120.3 &  (\uparrow)\\ 
%\end{array}
&
%\begin{array}{cccc}
%%  & value & \pm & error & (\%)\\ \hline
%1879.4 &  \pm &  46.96 &  (2\%)\\ 
%-1026.6 &  \pm &  47.91 &  (5\%)\\ 
%310.83 &  \pm &  62.46 &  (20\%)\\ 
%\end{array}
& \\ \hline

\begin{array}{c} \\ \mathbf{TS15}\\ \mathrm{(34\, days)} \end{array} & 
\begin{array}{c}
  x\\ 
  y\\ 
  z\\
 \end{array} 
 & 
\begin{array}{cccc}
%  & value & \pm & error & (\%)\\ \hline
-84.556 & \pm &  17.160 &  (20\%)\\ 
-40.899 & \pm &  21.410 &  (52\%)\\ 
~       &     &       &     \\ 
\end{array}
&
\begin{array}{cccc}
%  & value & \pm & error & (\%)\\ \hline
~122.18 & \pm &  25.14 &  (20\%)\\ 
~47.726 & \pm &  31.070 &  (65\%)\\ 
~       &     &       &     \\ 
\end{array}
&
\begin{array}{cccc}
%  & value & \pm & error & (\%)\\ \hline
~2194.5 & \pm &  108.5 &  (5\%)\\ 
-663.54 & \pm &  97.71 &  (7\%)\\ 
~324.54 & \pm &  0.85 &  (0\%)\\ 
\end{array}
& \\ \hline

\end{array}
$
}
\caption{MCMC parameter and parameter uncertainty estimates obtained by performing a joint fit on the measurements of the MX magnetometer. The parameters appear in Equation~\ref{eq_BIMFmodel_freq_applied}.}
\label{tab:evalparmag1} %eq_BIMFmodel_freq_applied
\end{table*}

% % % % % % % % % % % % % % % % % % % % % % % % % % % % % % % % % % % % % % % % % % % % % % % % % % % % % % % % % 
\iftrue
\begin{table*}[ht]
%\centering
\resizebox{16cm}{!}{ 
%\multicolumn{5}{c}{} & \multicolumn{8}{c}{j} \tabularnewline
$
\begin{array}{|c|c|c|c|c|c|} \hline
\begin{array}{c} \\ \mathbf{MAG2}\\ \mathbf{Coefficients} \end{array} & \mathbf{Axis} & \mathbf{a_1} \mathrm{[\mathrm{nT\,nA^{-1}}]}& \mathbf{a_2} \mathrm{[\mathrm{nT\,nA^{-1}}] } & \mathbf{a_3} \mathrm{[\mathrm{nT\,nA^{-1}}]} & \mathbf{l}\, \mathrm{[\mathrm{nT\,kg^{-1}}]} \\
%\mathbf{Coefficients} & [\mathrm{nT\,nA^{-1}}] & [\mathrm{nT\,nA^{-1}}] & [\mathrm{nT\,nA^{-1}}] & [\mathrm{nT\,kg^{-1}}] \\ %\hline\hline

%\mathbf{Coefficients}
&
&
\begin{array}{cccc}
value & \pm & error & (\%)\\ %\hline
\end{array} 
&
\begin{array}{cccc}
value & \pm & error & (\%)\\ %\hline
\end{array} 
&
\begin{array}{cccc}
value & \pm & error & (\%)\\ %\hline
\end{array} 
&
\begin{array}{cccc}
value & \pm & error & (\%)\\ %\hline
\end{array} 
\\ \hline

\begin{array}{c} \\ \mathbf{TS1}\\ \mathrm{(74\, days)} \end{array} &  
\begin{array}{c}
  x\\ 
  y\\ 
  z\\
 \end{array} 
 & 
\begin{array}{cccc}
%  & value & \pm & error & (\%)\\ \hline
-26.903 & \pm & 5.222 &  (19\%)\\ 
~       &     &       &     \\ 
~       &     &       &     \\ 
\end{array}
&
\begin{array}{cccc}
%  & value & \pm & error & (\%)\\ \hline
~32.724 & \pm & 7.907 &  (24\%)\\ 
~       &     &       &     \\ 
~       &     &       &     \\ 
\end{array}
&
\begin{array}{cccc}
%  & value & \pm & error & (\%)\\ \hline
~       &     &       &     \\ 
~2593.8 & \pm & 0.6 &  (0\%)\\ 
~1339.2 & \pm & 0.7 &  (0\%)\\ 
\end{array}
& \\ \hline

\begin{array}{c} \\ \mathbf{TS2}\\ \mathrm{(46\, days)} \end{array} &  
\begin{array}{c}
  x\\ 
  y\\ 
  z\\
 \end{array} 
 & 
\begin{array}{cccc}
%  & value & \pm & error & (\%)\\ \hline
~       &     &       &     \\ 
~       &     &       &     \\ 
~       &     &       &     \\ 
\end{array}
&
\begin{array}{cccc}
%  & value & \pm & error & (\%)\\ \hline
~36.931 &  \pm &  36.090 &  (99\%)\\ 
~       &     &       &     \\ 
~       &     &       &     \\ 
\end{array}
&
\begin{array}{cccc}
%  & value & \pm & error & (\%)\\ \hline
-212.10 & \pm &  0.75 &  (1\%)\\ 
~2587.1 & \pm &  0.6 &  (0\%)\\ 
~974.16 & \pm &  94.70 &  (10\%)\\ 
\end{array}
& \\ \hline

\begin{array}{c} \\ \mathbf{TS3}\\ \mathrm{(72\, days)} \end{array} &  
\begin{array}{c}
  x\\ 
  y\\ 
  z\\
 \end{array} 
 & 
\begin{array}{cccc}
%  & value & \pm & error & (\%)\\ \hline
-16.675 &  \pm &  1.560 &  (10\%)\\ 
~238.71 &  \pm &  30.06 &  (13\%)\\ 
~       &     &       &     \\ 
\end{array}
&
\begin{array}{cccc}
%  & value & \pm & error & (\%)\\ \hline
~       &     &       &     \\ 
-296.12 &  \pm &  44.96 &  (15\%)\\ 
~       &     &       &     \\ 
\end{array}
&
\begin{array}{cccc}
%  & value & \pm & error & (\%)\\ \hline
~464.84 &  \pm &  65.09 &  (14\%)\\ 
~827.64 &  \pm &  49.48 &  (6\%)\\ 
~983.87  &  \pm & 0.84 &  (0\%)\\ 
\end{array}
&
\begin{array}{cccc}
~57.846 & \pm & 6.598 &(11\%)\\ 
-65.990 & \pm & 10.370 &(16\%)\\ 
~       &     &       &     \\ 
 \end{array} 
\\ \hline

\begin{array}{c} \\ \mathbf{TS4}\\ \mathrm{(25\, days)} \end{array} &  
\begin{array}{c}
  x\\ 
  y\\ 
  z\\
 \end{array} 
 & 
\begin{array}{cccc}
%  & value & \pm & error & (\%)\\ \hline
~       &     &       &     \\ 
~119.43 & \pm &  23.12 &  (19\%)\\ 
~112.37 & \pm &  19.57 &  (17\%)\\ 
\end{array}
&
\begin{array}{cccc}
%  & value & \pm & error & (\%)\\ \hline
-9.1909 & \pm &  0.0261 &  (0\%)\\ 
-132.65 & \pm &  34.75 &  (26\%)\\ 
-144.67 & \pm &  29.64 &  (20\%)\\ 
\end{array}
&
\begin{array}{cccc}
%  & value & \pm & error & (\%)\\ \hline
~       &     &       &     \\ 
~1309.4 & \pm &  70.77 &  (5\%)\\ 
~647.39 & \pm &  89.37 &  (14\%)\\ 
\end{array}
 & 
 \\ \hline

\begin{array}{c} \\ \mathbf{TS5}\\ \mathrm{(36\, days)} \end{array} &  
\begin{array}{c}
  x\\ 
  y\\ 
  z\\
 \end{array} 
 & 
\begin{array}{cccc}
%  & value & \pm & error & (\%)\\ \hline
-5.6256 & \pm &  0.0115 &  (0\%)\\ 
~       &     &       &     \\ 
~       &     &       &     \\ 
\end{array}
&
\begin{array}{cccc}
%  & value & \pm & error & (\%)\\ \hline
~       &     &       &     \\ 
~29.212 & \pm &  3.182 &  (11\%)\\ 
~       &     &       &     \\ 
\end{array}
&
\begin{array}{cccc}
%  & value & \pm & error & (\%)\\ \hline
~       &     &       &     \\ 
~1820.4 & \pm &  88.6 &  (5\%)\\ 
~1359.8 & \pm &  0.5 &  (0\%)\\ 
\end{array}
& \\ \hline

\begin{array}{c} \\ \mathbf{TS6}\\ \mathrm{(35\, days)} \end{array} &  
\begin{array}{c}
  x\\ 
  y\\ 
  z\\
 \end{array} 
 & 
\begin{array}{cccc}
%  & value & \pm & error & (\%)\\ \hline
-95.289 & \pm &  42.090 &  (44\%)\\ 
~19.384 & \pm &  3.460 &  (18\%)\\ 
~8.7730 & \pm &  3.271 &  (37\%)\\ 
\end{array}
&
\begin{array}{cccc}
%  & value & \pm & error & (\%)\\ \hline
~135.09 & \pm &  63.32 &  (47\%) \\ 
~       &     &       &     \\ 
~       &     &       &     \\ 
\end{array}
&
\begin{array}{cccc}
%  & value & \pm & error & (\%)\\ \hline
~1696.1 & \pm &  142.7 &  (8\%)\\ 
-474.09 & \pm &  173.90 &  (37\%)\\ 
~1011.6 & \pm &  134.9 &  (13\%)\\ 
\end{array}
& \\ \hline

\begin{array}{c} \\ \mathbf{TS7}\\ \mathrm{(5\, days)} \end{array} &  
\begin{array}{c}
  x\\ 
  y\\ 
  z\\
 \end{array} 
 & 
\begin{array}{cccc}
%  & value & \pm & error & (\%)\\ \hline
~       &     &       &     \\ 
~307.59 & \pm &  95.95  &  (31\%)\\ 
~       &     &       &     \\ 
\end{array}
&
\begin{array}{cccc}
%  & value & \pm & error & (\%)\\ \hline
~       &     &       &     \\ 
-430.86 & \pm &  133.70  &  (31\%)\\ 
~       &     &       &     \\ 
\end{array}
&
\begin{array}{cccc}
%  & value & \pm & error & (\%)\\ \hline
-232.21 & \pm &  1.52 &  (1\%)\\ 
~1601.0 & \pm &  471.4 &  (29\%)\\ 
~1368.7 & \pm &  3.0 &  (0\%)\\ 
\end{array}
& \\ \hline

%8 LASCIATO VECCHIO
\begin{array}{c} \\ \mathbf{TS8}\\ \mathrm{(7\, days)} \end{array} &  
\begin{array}{c}
  x\\ 
  y\\ 
  z\\
 \end{array} 
 & 
\begin{array}{cccc}
%  & value & \pm & error & (\%)\\ \hline
~       &     &       &     \\ 
~       &     &       &     \\ 
~62.563 & \pm &  43.130 &  (69\%)\\ 
\end{array}
&
\begin{array}{cccc}
%  & value & \pm & error & (\%)\\ \hline
~       &     &       &     \\ 
~64.123 & \pm &  11.620 &  (18\%)\\ 
-85.866 & \pm &  59.660 &  (69\%)\\ 
\end{array}
&
\begin{array}{cccc}
%  & value & \pm & error & (\%)\\ \hline
-192.31 & \pm &  1.134 &  (1\%)\\ 
~775.34 & \pm &  314.10 &  (40\%)\\ 
~1166.2 & \pm &  231.1 &  (20\%)\\ 
\end{array}
& \\ \hline

\begin{array}{c} \\ \mathbf{TS9}\\ \mathrm{(8\, days)} \end{array} &  
\begin{array}{c}
  x\\ 
  y\\ 
  z\\
 \end{array} 
 & 
\begin{array}{cccc}
%  & value & \pm & error & (\%)\\ \hline
~       &     &       &     \\ 
~40.555 & \pm &  6.551 &  (16\%)\\ 
~       &     &       &     \\ 
\end{array}
&
\begin{array}{cccc}
%  & value & \pm & error & (\%)\\ \hline
~       &     &       &     \\ 
~       &     &       &     \\ 
~       &     &       &     \\ 
\end{array}
&
\begin{array}{cccc}
%  & value & \pm & error & (\%)\\ \hline
-1548.4 & \pm &  1.3 &  (0\%)\\ 
~869.62 & \pm &  263.90 &  (30\%)\\ 
~1057.4 & \pm &  1.5 &  (0\%)\\ 
\end{array}
& \\ \hline

\begin{array}{c} \\ \mathbf{TS10}\\ \mathrm{(28\, days)} \end{array} &  
\begin{array}{c}
  x\\ 
  y\\ 
  z\\
 \end{array} 
 & 
\begin{array}{cccc}
%  & value & \pm & error & (\%)\\ \hline
~       &     &       &     \\ 
~       &     &       &     \\ 
~       &     &       &     \\ 
\end{array}
&
\begin{array}{cccc}
%  & value & \pm & error & (\%)\\ \hline
-15.513 & \pm &  6.531 &  (42\%)\\ 
~       &     &       &     \\ 
~       &     &       &     \\ 
\end{array}
&
\begin{array}{cccc}
%  & value & \pm & error & (\%)\\ \hline
~222.98 & \pm &  175.50 &  (79\%)\\ 
~2501.9 & \pm &  0.4 &  (0\%)\\ 
-15.513 & \pm &  6.531 &  (42\%)\\ 
\end{array}
& \\ \hline

\begin{array}{c} \\ \mathbf{TS11}\\ \mathrm{(18\, days)} \end{array} &  
\begin{array}{c}
  x\\ 
  y\\ 
  z\\
 \end{array} 
 & 
\begin{array}{cccc}
%  & value & \pm & error & (\%)\\ \hline
~103.43 & \pm &  28.87 &  (28\%)\\ 
~40.699 & \pm &  3.125 &  (8\%)\\ 
~51.115 & \pm &  27.690 &  (54\%)\\ 
\end{array}
&
\begin{array}{cccc}
%  & value & \pm & error & (\%)\\ \hline
-164.45 & \pm &  43.79 &  (27\%)\\ 
~       &     &       &     \\ 
-66.493 & \pm &  42.480 &  (64\%)\\ 
\end{array}
&
\begin{array}{cccc}
%  & value & \pm & error & (\%)\\ \hline
~       &     &       &     \\ 
~946.59 & \pm &  129.20 &  (14\%)\\ 
~1083.0 & \pm &  134.0 &  (12\%)\\ 
\end{array}
& \\ \hline

\begin{array}{c} \\ \mathbf{TS12}\\ \mathrm{(6\, days)} \end{array} &  
\begin{array}{c}
  x\\ 
  y\\ 
  z\\
 \end{array} 
 & 
\begin{array}{cccc}
%  & value & \pm & error & (\%)\\ \hline
-124.43 & \pm &  59.65 &  (48\%)\\ 
-125.05 & \pm &  34.99 &  (28\%)\\ 
-72.702 & \pm &  49.210 &  (68\%)\\ 
\end{array}
&
\begin{array}{cccc}
%  & value & \pm & error & (\%)\\ \hline
~169.21 & \pm &  85.61 &  (51\%)\\ 
~254.62 & \pm &  51.31 &  (20\%)\\ 
~105.93 & \pm &  73.58 &  (69\%)\\ 
\end{array}
&
\begin{array}{cccc}
%  & value & \pm & error & (\%)\\ \hline
~343.73 & \pm &  217.40 &  (63\%)\\ 
~869.91 & \pm &  163.30 &  (19\%)\\ 
~1524.4 & \pm &  186.9 &  (12\%)\\ 
\end{array}
&
\\ \hline

\begin{array}{c} \\ \mathbf{TS13}\\ \mathrm{(16\, days)} \end{array} &  
\begin{array}{c}
  x\\ 
  y\\ 
  z\\
 \end{array} 
 & 
%\begin{array}{cccc}
%%  & value & \pm & error & (\%)\\ \hline
%-8.9526 &  \pm &  55.52 &  (\uparrow)\\ 
%92.42 &  \pm &  35.72 &  (39\%)\\ 
%31.193 &  \pm &  43.44 &  (\uparrow)\\ 
%\end{array}
&
%\begin{array}{cccc}
%%  & value & \pm & error & (\%)\\ \hline
%-14.108 &  \pm &  81.55 &  (\uparrow)\\ 
%-62.659 &  \pm &  52.84 &  (84\%)\\ 
%-37.275 &  \pm &  63.87 &  (\uparrow)\\ 
%\end{array}
&
%\begin{array}{cccc}
%%  & value & \pm & error & (\%)\\ \hline
%543.1 &  \pm &  140.9 &  (26\%)\\ 
%491.84 &  \pm &  95.99 &  (20\%)\\ 
%1116.2 &  \pm &  93.28 &  (8\%)\\ 
%\end{array}
&
\\ \hline

\begin{array}{c} \\ \mathbf{TS14}\\ \mathrm{(23\, days)} \end{array} &  
\begin{array}{c}
  x\\ 
  y\\ 
  z\\
 \end{array} 
 & 
%\begin{array}{cccc}
%%  & value & \pm & error & (\%)\\ \hline
%14.303 &  \pm &  43.04 &  (\uparrow)\\ 
%146.29 &  \pm &  43.76 &  (30\%)\\ 
%-96.878 &  \pm &  23.44 &  (24\%)\\ 
%\end{array}
&
%\begin{array}{cccc}
%%  & value & \pm & error & (\%)\\ \hline
%-52.12 &  \pm &  66.43 &  (\uparrow)\\ 
%-212.46 &  \pm &  67.54 &  (32\%)\\ 
%158.43 &  \pm &  35.77 &  (23\%)\\ 
%\end{array}
&
%\begin{array}{cccc}
%%  & value & \pm & error & (\%)\\ \hline
%605.22 &  \pm &  39.42 &  (7\%)\\ 
%2190.1 &  \pm &  38.11 &  (2\%)\\ 
%1128.4 &  \pm &  32.22 &  (3\%)\\ 
%\end{array}
&
\\ \hline

\begin{array}{c} \\ \mathbf{TS15}\\ \mathrm{(34\, days)} \end{array} &  
\begin{array}{c}
  x\\ 
  y\\ 
  z\\
 \end{array} 
 & 
\begin{array}{cccc}
%  & value & \pm & error & (\%)\\ \hline
-92.406 & \pm &  20.220 &  (22\%)\\ 
-50.926 & \pm &  22.750 &  (45\%)\\ 
~       &     &       &     \\ 
\end{array}
&
\begin{array}{cccc}
%  & value & \pm & error & (\%)\\ \hline
~125.50 & \pm &  29.14 &  (23\%)\\ 
~78.335 & \pm &  33.540  &  (43\%)\\ 
~       &     &       &     \\ 
\end{array}
&
\begin{array}{cccc}
%  & value & \pm & error & (\%)\\ \hline
~157.71 & \pm &  114.00 &  (73\%)\\ 
~2443.0 & \pm &  98.3 &  (4\%)\\ 
~1406.1 & \pm &  1.0 &  (0\%)\\ 
\end{array}
&
\\ \hline

\end{array}
$
}
\caption{ %MCMC parameter and parameter uncertainty estimates obtained by performing a joint fit on the measurements of the MY magnetometer. The employed model is provided in Equation~\ref{eq_BIMFmodel_freq_applied}.
Same as Table~\ref{tab:evalparmag1} for the MY magnetometer.}
\label{tab:evalparmag2}
\end{table*}

\begin{table*}[ht]
%\centering
\resizebox{16cm}{!}{ 
%\multicolumn{5}{c}{} & \multicolumn{8}{c}{j} \tabularnewline
$
\begin{array}{|c|c|c|c|c|c|} \hline
\begin{array}{c} \\ \mathbf{MAG3}\\ \mathbf{Coefficients} \end{array} & \mathbf{Axis} & \mathbf{a_1} \mathrm{[\mathrm{nT\,nA^{-1}}]}& \mathbf{a_2} \mathrm{[\mathrm{nT\,nA^{-1}}] } & \mathbf{a_3} \mathrm{[\mathrm{nT\,nA^{-1}}]} & \mathbf{l}\, \mathrm{[\mathrm{nT\,kg^{-1}}]} \\

&
&
\begin{array}{cccc}
value & \pm & error & (\%)\\ %\hline
\end{array} 
&
\begin{array}{cccc}
value & \pm & error & (\%)\\ %\hline
\end{array} 
&
\begin{array}{cccc}
value & \pm & error & (\%)\\ %\hline
\end{array} 
&
\begin{array}{cccc}
value & \pm & error & (\%)\\ %\hline
\end{array} 
\\ \hline 

\begin{array}{c} \\ \mathbf{TS1}\\ \mathrm{(74\, days)} \end{array} &  
\begin{array}{c}
  x\\ 
  y\\ 
  z\\
 \end{array} 
 & 
\begin{array}{cccc}
%  & value & \pm & error & (\%)\\ \hline
~       &     &       &     \\ 
~       &     &       &     \\ 
~       &     &       &     \\ 
\end{array}
&
\begin{array}{cccc}
%  & value & \pm & error & (\%)\\ \hline
~ 16.300 & \pm &  3.125 &  (19\%)\\ 
~       &     &       &     \\ 
~       &     &       &     \\ 
\end{array}
&
\begin{array}{cccc}
%  & value & \pm & error & (\%)\\ \hline
~1750.8 & \pm &  85.3 &  (5\%)\\ 
-2299.8 & \pm &  0.6 &  (0\%)\\ 
~209.36 & \pm &  0.72 &  (0\%)\\ 
\end{array}
& \\ \hline

\begin{array}{c} \\ \mathbf{TS2}\\ \mathrm{(46\, days)} \end{array} &  
\begin{array}{c}
  x\\ 
  y\\ 
  z\\
 \end{array} 
 & 
\begin{array}{cccc}
%  & value & \pm & error & (\%)\\ \hline
 -428.27 & \pm &  117.7 &  (27\%)\\ 
~       &     &       &     \\ 
~       &     &       &     \\ 
\end{array}
&
\begin{array}{cccc}
%  & value & \pm & error & (\%)\\ \hline
~614.97 & \pm &  169.20 &  (27\%)\\ 
~       &     &       &     \\ 
~       &     &       &     \\ 
\end{array}
&
\begin{array}{cccc}
%  & value & \pm & error & (\%)\\ \hline
~3253.4 & \pm &  416.0 &  (8\%)\\ 
-2291.4 & \pm &  10.8 &  (7\%)\\ 
~       &     &       &     \\ 
\end{array}
& \\ \hline

\begin{array}{c} \\ \mathbf{TS3}\\ \mathrm{(72\, days)} \end{array} &  
\begin{array}{c}
  x\\ 
  y\\ 
  z\\
 \end{array} 
 & 
\begin{array}{cccc}
%  & value & \pm & error & (\%)\\ \hline
~141.27 &  \pm &  33.50 &  (24\%)\\ 
-40.107 &  \pm &  16.400 &  (41\%)\\ 
-67.865 &  \pm &  35.52 &  (52\%)\\ 
\end{array}
&
\begin{array}{cccc}
%  & value & \pm & error & (\%)\\ \hline
-166.44 &  \pm &  50.23 &  (30\%)\\ 
~60.415 &  \pm &  24.100 &  (40\%)\\ 
~78.571 &  \pm &  53.450 &  (68\%)\\ 
\end{array}
&
\begin{array}{cccc}
%  & value & \pm & error & (\%)\\ \hline
~922.87 &  \pm &  61.12 &  (7\%)\\ 
-2266.2 &  \pm &  60.2 &  (3\%)\\ 
~822.96 &  \pm &  52.78 &  (6\%)\\ 
\end{array}
&
\begin{array}{cccc}
~342.21 &  \pm &  12.11 &(3\%)\\ 
~       &     &       &     \\ 
-30.489 &  \pm &  13.360  &(44\%)\\ 
 \end{array} 

\\ \hline

\begin{array}{c} \\ \mathbf{TS4}\\ \mathrm{(25\, days)} \end{array} &  
\begin{array}{c}
  x\\ 
  y\\ 
  z\\
 \end{array} 
 & 
\begin{array}{cccc}
%  & value & \pm & error & (\%)\\ \hline
~       &     &       &     \\ 
~       &     &       &     \\ 
~       &     &       &     \\ 
\end{array}
&
\begin{array}{cccc}
%  & value & \pm & error & (\%)\\ \hline
~       &     &       &     \\ 
~       &     &       &     \\ 
~       &     &       &     \\ 
\end{array}
&
\begin{array}{cccc}
%  & value & \pm & error & (\%)\\ \hline
~2242.5 & \pm &  2.6 &  (0\%)\\ 
-2260.6 & \pm &  2.9 &  (1\%)\\ 
~136.70 & \pm &  3.67 &  (3\%)\\ 
\end{array}
& 
 \\ \hline

\begin{array}{c} \\ \mathbf{TS5}\\ \mathrm{(36\, days)} \end{array} &  
\begin{array}{c}
  x\\ 
  y\\ 
  z\\
 \end{array} 
 & 
\begin{array}{cccc}
%  & value & \pm & error & (\%)\\ \hline
~21.716 & \pm &  2.853 &  (13\%)\\ 
-62.857 & \pm &  11.680 &  (19\%)\\ 
~       &     &       &     \\ 
\end{array}
&
\begin{array}{cccc}
%  & value & \pm & error & (\%)\\ \hline
~       &     &       &     \\ 
~72.218 & \pm &  18.020 &  (25\%)\\ 
~       &     &       &     \\ 
\end{array}
&
\begin{array}{cccc}
%  & value & \pm & error & (\%)\\ \hline
~1339.7 & \pm &  120.0 &  (9\%)\\ 
-1642.6 & \pm &  102.6 &  (6\%)\\ 
~144.37 & \pm &  0.65 &  (0\%)\\ 
\end{array}
& \\ \hline

\begin{array}{c} \\ \mathbf{TS6}\\ \mathrm{(35\, days)} \end{array} &  
\begin{array}{c}
  x\\ 
  y\\ 
  z\\
 \end{array} 
 & 
\begin{array}{cccc}
%  & value & \pm & error & (\%)\\ \hline
~133.27 & \pm &  38.42 &  (29\%)\\ 
-12.112 & \pm &  4.395 &  (36\%)\\ 
~13.173 & \pm &  4.071 &  (31\%)\\ 
\end{array}
&
\begin{array}{cccc}
%  & value & \pm & error & (\%)\\ \hline
-121.82 & \pm & 57.82 & (5\%)\\ 
~       &     &       &     \\ 
~       &     &       &     \\ 
\end{array}
&
\begin{array}{cccc}
%  & value & \pm & error & (\%)\\ \hline
~       &     &       &     \\ 
-1773.4 & \pm &  181.4 &  (10\%)\\ 
-316.73 & \pm &  168.30 &  (53\%)\\ 
\end{array}
& \\ \hline

\begin{array}{c} \\ \mathbf{TS7}\\ \mathrm{(5\, days)} \end{array} &  
\begin{array}{c}
  x\\ 
  y\\ 
  z\\
 \end{array} 
 & 
\begin{array}{cccc}
%  & value & \pm & error & (\%)\\ \hline
~       &     &       &     \\ 
~       &     &       &     \\ 
~       &     &       &     \\ 
\end{array}
&
\begin{array}{cccc}
%  & value & \pm & error & (\%)\\ \hline
~       &     &       &     \\ 
~       &     &       &     \\ 
~       &     &       &     \\ 
\end{array}
&
\begin{array}{cccc}
%  & value & \pm & error & (\%)\\ \hline
~2158.6 & \pm &  1.9  &  (0\%)\\ 
-2278.6 & \pm &  3.8  &  (0\%)\\ 
~227.26 & \pm &  3.14 &  (1\%)\\ 
\end{array}
& \\ \hline

%8 LASCIATO VECCHIO
\begin{array}{c} \\ \mathbf{TS8}\\ \mathrm{(7\, days)} \end{array} &  
\begin{array}{c}
  x\\ 
  y\\ 
  z\\
 \end{array} 
 & 
\begin{array}{cccc}
%  & value & \pm & error & (\%)\\ \hline
~       &     &       &     \\ 
~       &     &       &     \\ 
~       &     &       &     \\ 
\end{array}
&
\begin{array}{cccc}
%  & value & \pm & error & (\%)\\ \hline
~19.882 & \pm &  14.260  &  (71\%)\\ 
~       &     &       &     \\ 
~       &     &       &     \\ 
\end{array}
&
\begin{array}{cccc}
%  & value & \pm & error & (\%)\\ \hline
~1619.9 & \pm &  384.6 &  (24\%)\\ 
-2270.7 & \pm &  1.8 &  (0\%)\\ 
~859.35 & \pm &  1.98 &  (0\%)\\ 
\end{array}
& \\ \hline

\begin{array}{c} \mathbf{TS9}\\ \mathrm{(8\, days)} \end{array} &  
\begin{array}{c}
  x\\ 
  y\\ 
  z\\
 \end{array} 
 & 
\begin{array}{cccc}
%  & value & \pm & error & (\%)\\ \hline
~       &     &       &     \\ 
-46.279 & \pm &  6.044 &  (13\%)\\ 
~       &     &       &     \\ 
\end{array}
&
\begin{array}{cccc}
%  & value & \pm & error & (\%)\\ \hline
~       &     &       &     \\ 
~       &     &       &     \\ 
~       &     &       &     \\ 
\end{array}
&
\begin{array}{cccc}
%  & value & \pm & error & (\%)\\ \hline
~2161.7 & \pm &  2.1 &  (0\%)\\ 
-401.76 & \pm &  243.80 &  (61\%)\\ 
~229.45 & \pm &  1.66 &  (1\%)\\ 
\end{array}
& \\ \hline

\begin{array}{c} \mathbf{TS10}\\ \mathrm{(28\, days)} \end{array} &  
\begin{array}{c}
  x\\ 
  y\\ 
  z\\
 \end{array} 
 & 
\begin{array}{cccc}
%  & value & \pm & error & (\%)\\ \hline
~124.31 & \pm &  6.09 &  (5\%)\\ 
~       &     &       &     \\ 
~       &     &       &     \\ 
\end{array}
&
\begin{array}{cccc}
%  & value & \pm & error & (\%)\\ \hline
~       &     &       &     \\ 
~9.6179 & \pm &  6.1720 &  (64\%)\\ 
~       &     &       &     \\ 
\end{array}
&
\begin{array}{cccc}
%  & value & \pm & error & (\%)\\ \hline
~1576.0 & \pm & 247.7 &  (16\%)\\ 
-2529.0 & \pm & 165.9 &  (6\%)\\ 
~877.93 & \pm & 0.79 &  (0\%)\\ 
\end{array}
& \\ \hline

\begin{array}{c} \mathbf{TS11}\\ \mathrm{(18\, days)} \end{array} &  
\begin{array}{c}
  x\\ 
  y\\ 
  z\\
 \end{array} 
 & 
\begin{array}{cccc}
%  & value & \pm & error & (\%)\\ \hline
~79.906 & \pm &  17.700 &  (22\%)\\ 
~       &     &       &     \\ 
~62.376 & \pm &  23.530 &  (38\%)\\ 
\end{array}
&
\begin{array}{cccc}
%  & value & \pm & error & (\%)\\ \hline
-81.816 & \pm &  25.600 &  (31\%)\\ 
~       &     &       &     \\ 
-114.23 & \pm &  35.26 &  (31\%)\\ 
\end{array}
&
\begin{array}{cccc}
%  & value & \pm & error & (\%)\\ \hline
~1177.6 & \pm &  127.5 &  (11\%)\\ 
-2256.2 & \pm &  1.0 &  (0\%)\\ 
~688.78 & \pm &  117.50 &  (17\%)\\ 
\end{array}
& \\ \hline

\begin{array}{c} \mathbf{TS12}\\ \mathrm{(6\, days)} \end{array} &  
\begin{array}{c}
  x\\ 
  y\\ 
  z\\
 \end{array} 
 & 
\begin{array}{cccc}
%  & value & \pm & error & (\%)\\ \hline
~       &     &       &     \\ 
-53.046 & \pm &  5.581 &  (10\%)\\ 
~       &     &       &     \\ 
\end{array}
&
\begin{array}{cccc}
%  & value & \pm & error & (\%)\\ \hline
~15.884 & \pm &  9.977 &  (63\%)\\ 
~       &     &       &     \\ 
~       &     &       &     \\ 
\end{array}
&
\begin{array}{cccc}
%  & value & \pm & error & (\%)\\ \hline
~1802.6 & \pm &  269.3 &  (15\%)\\ 
-1691.7 & \pm &  228.3 &  (13\%)\\ 
-850.34 & \pm &  2.03 &  (0\%)\\ 
\end{array}
&
\\ \hline

\begin{array}{c} \mathbf{TS13}\\ \mathrm{(16\, days)} \end{array} &  
\begin{array}{c}
  x\\ 
  y\\ 
  z\\
 \end{array} 
 & 
%\begin{array}{cccc}
%%  & value & \pm & error & (\%)\\ \hline
%38.061 &  \pm &  49.23 &  (\uparrow)\\ 
%-33.503 &  \pm &  46.57 &  (\uparrow)\\ 
%14.814 &  \pm &  43.6 &  (\uparrow)\\ 
%\end{array}
&
%\begin{array}{cccc}
%%  & value & \pm & error & (\%)\\ \hline
%7.7557 &  \pm &  73.23 &  (\uparrow)\\ 
%42.162 &  \pm &  69.11 &  (\uparrow)\\ 
%-55.36 &  \pm &  64.88 &  (\uparrow)\\ 
%\end{array}
&
%\begin{array}{cccc}
%%  & value & \pm & error & (\%)\\ \hline
%446.64 &  \pm &  124.8 &  (28\%)\\ 
%-1991.1 &  \pm &  118.9 &  (6\%)\\ 
%1066.6 &  \pm &  139.2 &  (13\%)\\ 
%\end{array}
&
\\ \hline

\begin{array}{c} \mathbf{TS14}\\ \mathrm{(23\, days)} \end{array} &  
\begin{array}{c}
  x\\ 
  y\\ 
  z\\
 \end{array} 
 & 
%\begin{array}{cccc}
%%  & value & \pm & error & (\%)\\ \hline
%116.79 &  \pm &  39.22 &  (34\%)\\ 
%-91.598 &  \pm &  31.64 &  (35\%)\\ 
%-169.95 &  \pm &  27.47 &  (16\%)\\ 
%\end{array}
&
%\begin{array}{cccc}
%%  & value & \pm & error & (\%)\\ \hline
%-167.54 &  \pm &  60.58 &  (36\%)\\ 
%130.56 &  \pm &  48.77 &  (37\%)\\ 
%257.39 &  \pm &  42.32 &  (16\%)\\ 
%\end{array}
&
%\begin{array}{cccc}
%%  & value & \pm & error & (\%)\\ \hline
%1837.8 &  \pm &  35.58 &  (2\%)\\ 
%-1983.6 &  \pm &  36.67 &  (2\%)\\ 
%308.94 &  \pm &  37.18 &  (12\%)\\ 
%\end{array}
&
\\ \hline

\begin{array}{c} \mathbf{TS15}\\ \mathrm{(34\, days)} \end{array} &  
\begin{array}{c}
  x\\ 
  y\\ 
  z\\
 \end{array} 
 & 
\begin{array}{cccc}
%  & value & \pm & error & (\%)\\ \hline
-40.822 & \pm &  26.380 &  (65\%)\\ 
~       &     &       &     \\ 
~       &     &       &     \\ 
\end{array}
&
\begin{array}{cccc}
%  & value & \pm & error & (\%)\\ \hline
~70.111 & \pm &  38.490 &  (55\%)\\ 
~       &     &       &     \\ 
~       &     &       &     \\ 
\end{array}
&
\begin{array}{cccc}
%  & value & \pm & error & (\%)\\ \hline
1906.7 & \pm &  138.9 &  (7\%)\\ 
-557.02 & \pm &  0.77 &  (0\%)\\ 
942.06 & \pm &  1.27 &  (0\%)\\ 
\end{array}
&
\\ \hline

\end{array}
$
}
\caption{ %MCMC parameter and parameter uncertainty estimates obtained by performing a joint fit on the measurements of the PX magnetometer. The employed model is provided in Equation~\ref{eq_BIMFmodel_freq_applied}.
Same as Table~\ref{tab:evalparmag1} for the PX magnetometer.}
\label{tab:evalparmag3}
\end{table*}

% % % % % % % % % % % % % % % % % % % % % % % % % % % % % % % % % % % % % % % % % % % % % % % % % % % % % % % % %

\begin{table*}[ht]
%\centering
\resizebox{16cm}{!}{ 
%\multicolumn{5}{c}{} & \multicolumn{8}{c}{j} \tabularnewline
$
\begin{array}{|c|c|c|c|c|c|} \hline
\begin{array}{c} \\ \mathbf{MAG4}\\ \mathbf{Coefficients} \end{array}  & \mathbf{Axis} & \mathbf{a_1} \mathrm{[\mathrm{nT\,nA^{-1}}]}& \mathbf{a_2} \mathrm{[\mathrm{nT\,nA^{-1}}] } & \mathbf{a_3} \mathrm{[\mathrm{nT\,nA^{-1}}]} & \mathbf{l}\, \mathrm{[\mathrm{nT\,kg^{-1}}]} \\

&
&
\begin{array}{cccc}
value & \pm & error & (\%)\\ %\hline
\end{array} 
&
\begin{array}{cccc}
value & \pm & error & (\%)\\ %\hline
\end{array} 
&
\begin{array}{cccc}
value & \pm & error & (\%)\\ %\hline
\end{array} 
&
\begin{array}{cccc}
value & \pm & error & (\%)\\ %\hline
\end{array} 
\\ \hline 

%\begin{array}{c} \mathrm 1 \\ \mathbf{DFACS1A}\\ \mathrm{(74\, days)} \end{array} & 
\begin{array}{c} \mathbf{TS1}\\ \mathrm{(74\, days)} \end{array} &  
\begin{array}{c}
  x\\ 
  y\\ 
  z\\
 \end{array} 
 & 
\begin{array}{cccc}
%  & value & \pm & error & (\%)\\ \hline
-53.893 & \pm &  4.589 &  (8\%)\\ 
~       &     &       &     \\ 
~19.747 & \pm &  6.894 &  (35\%)\\ 
\end{array}
&
\begin{array}{cccc}
%  & value & \pm & error & (\%)\\ \hline
~71.265 & \pm &  6.947 &  (10\%)\\ 
~       &     &       &     \\ 
-22.969 & \pm &  9.766 &  (42\%)\\ 
\end{array}
&
\begin{array}{cccc}
%  & value & \pm & error & (\%)\\ \hline
~       &     &       &     \\ 
~1485.5 & \pm &  0.6 &  (0\%)\\ 
~785.91 & \pm &  78.05 &  (10\%)\\ 
\end{array}
& \\ \hline

%\begin{array}{c} \mathrm 2 \\ \mathbf{DFACS1B}\\ \mathrm{(46\, days)} \end{array} & 
\begin{array}{c} \mathbf{TS2}\\ \mathrm{(46\, days)} \end{array} &  
\begin{array}{c}
  x\\ 
  y\\ 
  z\\
 \end{array} 
 & 
\begin{array}{cccc}
%  & value & \pm & error & (\%)\\ \hline
~       &     &       &     \\ 
~62.294 & \pm &  17.050 &  (27\%)\\ 
~57.968 & \pm &  16.810 &  (29\%)\\ 
\end{array}
&
\begin{array}{cccc}
%  & value & \pm & error & (\%)\\ \hline
~       &     &       &     \\ 
-75.134 & \pm & 25.600 &  (34\%)\\ 
-72.503 & \pm & 25.260 &  (35\%)\\ 
\end{array}
&
\begin{array}{cccc}
%  & value & \pm & error & (\%)\\ \hline
-293.80 & \pm &  0.52 &  (0\%)\\ 
~947.99 & \pm & 81.79 &  (9\%)\\ 
~553.11 & \pm & 79.49 &  (15\%)\\ 
\end{array}
& \\ \hline

%\begin{array}{c} \mathrm 3 \\ \mathbf{DRS1A}\\ \mathrm{(72\, days)} \end{array} & 
\begin{array}{c} \mathbf{TS3}\\ \mathrm{(72\, days)} \end{array} &  
\begin{array}{c}
  x\\ 
  y\\ 
  z\\
 \end{array} 
 &  
\begin{array}{cccc}
%  & value & \pm & error & (\%)\\ \hline
-96.087 &  \pm &  12.37 &  (13\%)\\ 
~75.328 &  \pm &  10.54 &  (14\%)\\ 
~64.095 &  \pm &  12.42 &  (19\%)\\ 
\end{array}
&
\begin{array}{cccc}
%  & value & \pm & error & (\%)\\ \hline
~142.9 &  \pm &  18.11 &  (13\%)\\ 
-105.32 &  \pm &  15.50 &  (15\%)\\ 
-83.250 &  \pm &  18.350 &  (22\%)\\ 
\end{array}
&
\begin{array}{cccc}
%  & value & \pm & error & (\%)\\ \hline
-185.71 &  \pm &  58.84 &  (32\%)\\ 
~1212.6 &  \pm &  44.5 &  (4\%)\\ 
~586.55 &  \pm &  53.54 &  (9\%)\\ 
\end{array}
&
\begin{array}{cccc}
~81.605 & \pm & 9.199 &(11\%)\\ 
-15.273 & \pm & 7.860 &(51\%)\\ 
-20.082 & \pm & 9.369 &(47\%)\\ 
 \end{array} 
\\ \hline

%\begin{array}{c} \mathrm 4 \\ \mathbf{DRS1B}\\ \mathrm{(25\, days)} \end{array} & 
\begin{array}{c} \mathbf{TS4}\\ \mathrm{(25\, days)} \end{array} &  
\begin{array}{c}
  x\\ 
  y\\ 
  z\\
 \end{array} 
 &  
\begin{array}{cccc}
%  & value & \pm & error & (\%)\\ \hline
-78.156 & \pm &  11.970 &  (15\%)\\ 
~10.226 & \pm &  1.108 &  (11\%)\\ 
~80.415 & \pm &  13.380 &  (17\%)\\ 
\end{array}
&
\begin{array}{cccc}
%  & value & \pm & error & (\%)\\ \hline
~117.34 & \pm &  17.90 &  (15\%)\\ 
~       &     &       &     \\ 
-106.73 & \pm &  20.20 &  (19\%)\\ 
\end{array}
&
\begin{array}{cccc}
%  & value & \pm & error & (\%)\\ \hline
-251.14 & \pm &  60.61 &  (24\%)\\ 
~1023.0 & \pm &  46.4 &  (5\%)\\ 
~557.94 & \pm &  53.12 &  (10\%)\\ 
\end{array}
&
 \\ \hline

%\begin{array}{c} \mathrm 5 \\ \mathbf{DFACS+1A}\\ \mathrm{(36\, days)} \end{array} & 
\begin{array}{c} \mathbf{TS5}\\ \mathrm{(36\, days)} \end{array} &  
\begin{array}{c}
  x\\ 
  y\\ 
  z\\
 \end{array} 
 &  
\begin{array}{cccc}
%  & value & \pm & error & (\%)\\ \hline
-67.832 & \pm &  10.660 &  (16\%)\\ 
~       &     &       &     \\ 
~       &     &       &     \\ 
\end{array}
&
\begin{array}{cccc}
%  & value & \pm & error & (\%)\\ \hline
~97.140 & \pm &  15.950 &  (16\%)\\ 
~9.2933 & \pm & 2.4640 &  (26\%)\\ 
~       &     &       &     \\ 
\end{array}
&
\begin{array}{cccc}
%  & value & \pm & error & (\%)\\ \hline
-125.72 & \pm &  88.43 &  (70\%)\\ 
~1200.1 & \pm &  68.7 &  (6\%)\\ 
~979.84 & \pm &  0.49 &  (0\%)\\ 
\end{array}
& \\ \hline

%\begin{array}{c} \mathrm 6 \\ \mathbf{DFACS+1B}\\ \mathrm{(35\, days)} \end{array} &
\begin{array}{c} \mathbf{TS6}\\ \mathrm{(35\, days)} \end{array} &    
\begin{array}{c}
  x\\ 
  y\\ 
  z\\
 \end{array} 
 &  
\begin{array}{cccc}
%  & value & \pm & error & (\%)\\ \hline
~       &     &       &     \\ 
~       &     &       &     \\ 
~       &     &       &     \\ 
\end{array}
&
\begin{array}{cccc}
%  & value & \pm & error & (\%)\\ \hline
~       &     &       &     \\ 
~9.9895 & \pm &  4.3330 &  (43\%)\\ 
~       &     &       &     \\ 
\end{array}
&
\begin{array}{cccc}
%  & value & \pm & error & (\%)\\ \hline
-263.19 & \pm &  0.70 &  (0\%)\\ 
~1192.1 & \pm &  118.9 &  (10\%)\\ 
~981.75 & \pm &  0.79 &  (0\%)\\ 
\end{array}
& \\ \hline

%\begin{array}{c} \mathrm 7 \\ \mathbf{DFACS+1C}\\ \mathrm{(5\, days)} \end{array} &  
\begin{array}{c} \mathbf{TS7}\\ \mathrm{(5\, days)} \end{array} &    
\begin{array}{c}
  x\\ 
  y\\ 
  z\\
 \end{array} 
 &  
\begin{array}{cccc}
%  & value & \pm & error & (\%)\\ \hline
~       &     &       &     \\ 
~259.06 & \pm &  131.70 &  (51\%)\\ 
~       &     &       &     \\ 
\end{array}
&
\begin{array}{cccc}
%  & value & \pm & error & (\%)\\ \hline
~       &     &       &     \\ 
-374.42 & \pm &  188.10  &  (50\%)\\ 
~       &     &       &     \\ 
\end{array}
&
\begin{array}{cccc}
%  & value & \pm & error & (\%)\\ \hline
-260.85 & \pm &  1.60 &  (1\%)\\ 
~1030.9 & \pm &  445.1 &  (43\%)\\ 
~980.19 & \pm &  3.01 &  (0\%)\\ 
\end{array}
& \\ \hline

%8 LASCIATO VECCHIO
%\begin{array}{c} \mathrm 8 \\ \mathbf{DFACS+2A}\\ \mathrm{(7\, days)} \end{array} & 
\begin{array}{c} \mathbf{TS8}\\ \mathrm{(7\, days)} \end{array} &     
\begin{array}{c}
  x\\ 
  y\\ 
  z\\
 \end{array} 
 &  
\begin{array}{cccc}
%  & value & \pm & error & (\%)\\ \hline
-101.00 & \pm &  0.03 &  (0\%)\\ 
~       &     &       &     \\ 
~71.706 & \pm &  46.680  &  (65\%)\\ 
\end{array}
&
\begin{array}{cccc}
%  & value & \pm & error & (\%)\\ \hline
~       &     &       &     \\ 
~       &     &       &     \\ 
-100.47 & \pm &  66.11 &  (66\%)\\ 
\end{array}
&
\begin{array}{cccc}
%  & value & \pm & error & (\%)\\ \hline
~       &     &       &     \\ 
~1254.2 & \pm &  1.4 &  (0\%)\\ 
~753.48 & \pm &  220.50 &  (29\%)\\ 
\end{array}
& \\ \hline

%\begin{array}{c} \mathrm 9 \\ \mathbf{DFACS+3}\\ \mathrm{(8\, days)} \end{array} & 
\begin{array}{c} \mathbf{TS9}\\ \mathrm{(8\, days)} \end{array} &      
\begin{array}{c}
  x\\ 
  y\\ 
  z\\
 \end{array} 
 &  
\begin{array}{cccc}
%  & value & \pm & error & (\%)\\ \hline
~       &     &       &     \\ 
~       &     &       &     \\ 
~       &     &       &     \\ 
\end{array}
&
\begin{array}{cccc}
%  & value & \pm & error & (\%)\\ \hline
~       &     &       &     \\ 
~       &     &       &     \\ 
~       &     &       &     \\ 
\end{array}
&
\begin{array}{cccc}
%  & value & \pm & error & (\%)\\ \hline
~       &     &       &     \\ 
~1466.8 & \pm &  32.4 &  (2\%)\\ 
~892.28 & \pm &  25.29 &  (3\%)\\ 
\end{array}
& \\ \hline

%\begin{array}{c} \mathrm 10 \\ \mathbf{DFACS+2B}\\ \mathrm{(28\, days)} \end{array} &  
\begin{array}{c} \mathbf{TS10}\\ \mathrm{(28\, days)} \end{array} &      
\begin{array}{c}
  x\\ 
  y\\ 
  z\\
 \end{array} 
 &  
\begin{array}{cccc}
%  & value & \pm & error & (\%)\\ \hline
~29.737 & \pm &  20.060  &  (67\%)\\ 
~23.146 & \pm &  16.000  &  (69\%)\\ 
~35.643 & \pm &  16.460  &  (46\%)\\ 
\end{array}
&
\begin{array}{cccc}
%  & value & \pm & error & (\%)\\ \hline
-54.005 & \pm &  30.330  &  (56\%)\\ 
-25.140 & \pm &  22.850  &  (91\%)\\ 
-45.578 & \pm &  21.840  &  (48\%)\\ 
\end{array}
&
\begin{array}{cccc}
%  & value & \pm & error & (\%)\\ \hline
~1210.1 & \pm &  114.6  &  (10\%)\\ 
~740.57 & \pm &  126.90  &  (17\%)\\ 
~       &     &       &     \\ 
\end{array}
& \\ \hline

%\begin{array}{c} \mathrm 11 \\ \mathbf{DRS2A}\\ \mathrm{(18\, days)} \end{array} & 
\begin{array}{c} \mathbf{TS11}\\ \mathrm{(18\, days)} \end{array} &       
\begin{array}{c}
  x\\ 
  y\\ 
  z\\
 \end{array} 
 &  
\begin{array}{cccc}
%  & value & \pm & error & (\%)\\ \hline
~96.512 & \pm &  18.980 &  (20\%)\\ 
-38.495 & \pm &  21.350 &  (55\%)\\ 
~55.059 & \pm &  22.750 &  (41\%)\\ 
\end{array}
&
\begin{array}{cccc}
%  & value & \pm & error & (\%)\\ \hline
-142.08 & \pm &  29.73 &  (21\%)\\ 
~83.607 & \pm &  31.370 &  (37\%)\\ 
-85.183 & \pm &  34.260 &  (40\%)\\ 
\end{array}
&
\begin{array}{cccc}
%  & value & \pm & error & (\%)\\ \hline
-356.92 & \pm &  106.50 &  (30\%)\\ 
~781.49 & \pm &  111.20 &  (14\%)\\ 
~1007.4 & \pm &  121.3 &  (12\%)\\ 
\end{array}
& \\ \hline

%\begin{array}{c} \mathrm 12 \\ \mathbf{DRS3}\\ \mathrm{(6\, days)} \end{array} & 
\begin{array}{c} \mathbf{TS12}\\ \mathrm{(6\, days)} \end{array} &        
\begin{array}{c}
  x\\ 
  y\\ 
  z\\
 \end{array} 
 &  
\begin{array}{cccc}
%  & value & \pm & error & (\%)\\ \hline
~       &     &       &     \\ 
-58.087 & \pm &  50.290 &  (87\%)\\ 
~       &     &       &     \\ 
\end{array}
&
\begin{array}{cccc}
%  & value & \pm & error & (\%)\\ \hline
~       &     &       &     \\ 
~93.223 & \pm &  70.120 &  (75\%)\\ 
~       &     &       &     \\ 
\end{array}
&
\begin{array}{cccc}
%  & value & \pm & error & (\%)\\ \hline
-219.67 & \pm &  1.17 &  (0\%)\\ 
~1338.9 & \pm &  315.6 &  (24\%)\\ 
~976.55 & \pm &  1.05 &  (0\%)\\ 
\end{array}
&
\\ \hline

%\begin{array}{c} \mathrm 13 \\ \mathbf{DRS2B}\\ \mathrm{(16\, days)} \end{array} &  
\begin{array}{c} \mathbf{TS13}\\ \mathrm{(16\, days)} \end{array} &        
\begin{array}{c}
  x\\ 
  y\\ 
  z\\
 \end{array} 
 &  
%\begin{array}{cccc}
%%  & value & \pm & error & (\%)\\ \hline
%-3.6935 &  \pm &  34.61 &  (\uparrow)\\ 
%72.988 &  \pm &  28.12 &  (39\%)\\ 
%19.429 &  \pm &  28.62 &  (\uparrow)\\ 
%\end{array}
&
%\begin{array}{cccc}
%%  & value & \pm & error & (\%)\\ \hline
%3.3834 &  \pm &  50.9 &  (\uparrow)\\ 
%-105.81 &  \pm &  41.54 &  (39\%)\\ 
%-22.777 &  \pm &  41.95 &  (\uparrow)\\ 
%\end{array}
&
%\begin{array}{cccc}
%%  & value & \pm & error & (\%)\\ \hline
%-166.51 &  \pm &  100.3 &  (60\%)\\ 
%1324.9 &  \pm &  86.8 &  (7\%)\\ 
%792.06 &  \pm &  82.73 &  (10\%)\\ 
%\end{array}
&
\\ \hline

%\begin{array}{c} \mathrm 14 \\ \mathbf{DFACSA+4A}\\ \mathrm{(23\, days)} \end{array} &  
\begin{array}{c} \mathbf{TS14}\\ \mathrm{(23\, days)} \end{array} &        
\begin{array}{c}
  x\\ 
  y\\ 
  z\\
 \end{array} 
 &  
%\begin{array}{cccc}
%%  & value & \pm & error & (\%)\\ \hline
%76.536 &  \pm &  16.99 &  (22\%)\\ 
%48.54 &  \pm &  16.82 &  (35\%)\\ 
%74.005 &  \pm &  31.24 &  (42\%)\\ 
%\end{array}
&
%\begin{array}{cccc}
%%  & value & \pm & error & (\%)\\ \hline
%-118.98 &  \pm &  26.05 &  (22\%)\\ 
%-64.013 &  \pm &  25.79 &  (40\%)\\ 
%-107.86 &  \pm &  48.03 &  (45\%)\\ 
%\end{array}
&
%\begin{array}{cccc}
%%  & value & \pm & error & (\%)\\ \hline
%-178.76 &  \pm &  20.41 &  (11\%)\\ 
%1206.7 &  \pm &  18 &  (1\%)\\ 
%845.88 &  \pm &  23.66 &  (3\%)\\ 
%\end{array}
&
\\ \hline

%\begin{array}{c} \mathrm 15 \\ \mathbf{DFACS+4B}\\ \mathrm{(34\, days)} \end{array} &  
\begin{array}{c} \mathbf{TS15}\\ \mathrm{(34\, days)} \end{array} &        
\begin{array}{c}
  x\\ 
  y\\ 
  z\\
 \end{array} 
 &  
\begin{array}{cccc}
%  & value & \pm & error & (\%)\\ \hline
-36.496 & \pm &  22.28 &  (61\%)\\ 
-45.033 & \pm &  21.58 &  (48\%)\\ 
~       &     &       &     \\ 
\end{array}
&
\begin{array}{cccc}
%  & value & \pm & error & (\%)\\ \hline
~56.891 & \pm &  32.450 &  (57\%)\\ 
~65.286 & \pm &  31.820 &  (49\%)\\ 
~       &     &       &     \\ 
\end{array}
&
\begin{array}{cccc}
%  & value & \pm & error & (\%)\\ \hline
-268.53 & \pm &  110.00 &  (41\%)\\ 
~1557.9 & \pm &  94.410  &  (6\%)\\ 
~1005.0 & \pm &  0.8 &  (0\%)\\ 
\end{array}
&
\\ \hline
%\midrule[2pt]
%\cmidrule{4-5}

\end{array}
$
}
\caption{%MCMC parameter and parameter uncertainty estimates obtained by performing a joint fit on the measurements of the PY magnetometer. The employed model is provided in Equation~\ref{eq_BIMFmodel_freq_applied}.
Same as Table~\ref{tab:evalparmag1} for the PY magnetometer.}
\label{tab:evalparmag4}
\end{table*}

\fi

\subsection{Comparison of LISA Pathfinder and Wind interplanetary magnetic field data}
\begin{figure}[htbp]
%\vspace{-2cm}
\centering 
	\includegraphics[width=7cm,height=6.5cm]{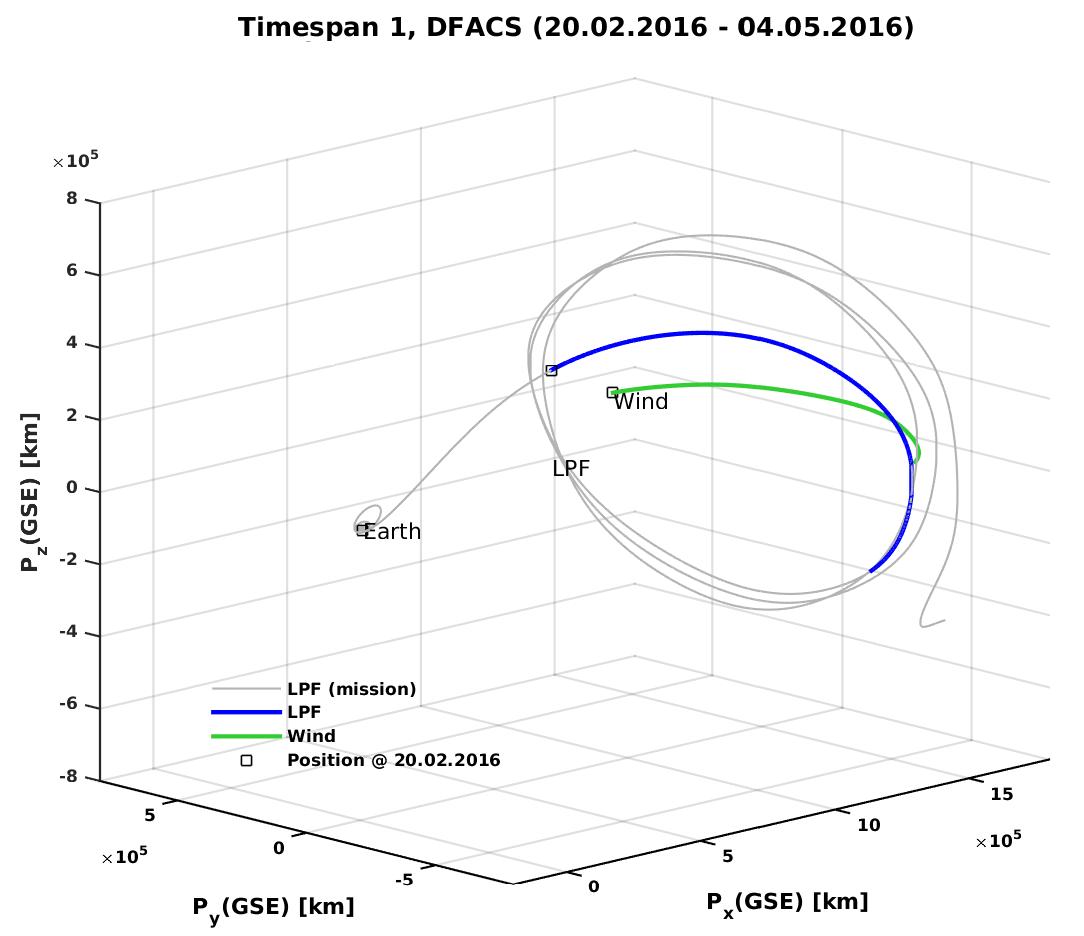}\\
	\includegraphics[width=7cm,height=6.5cm]{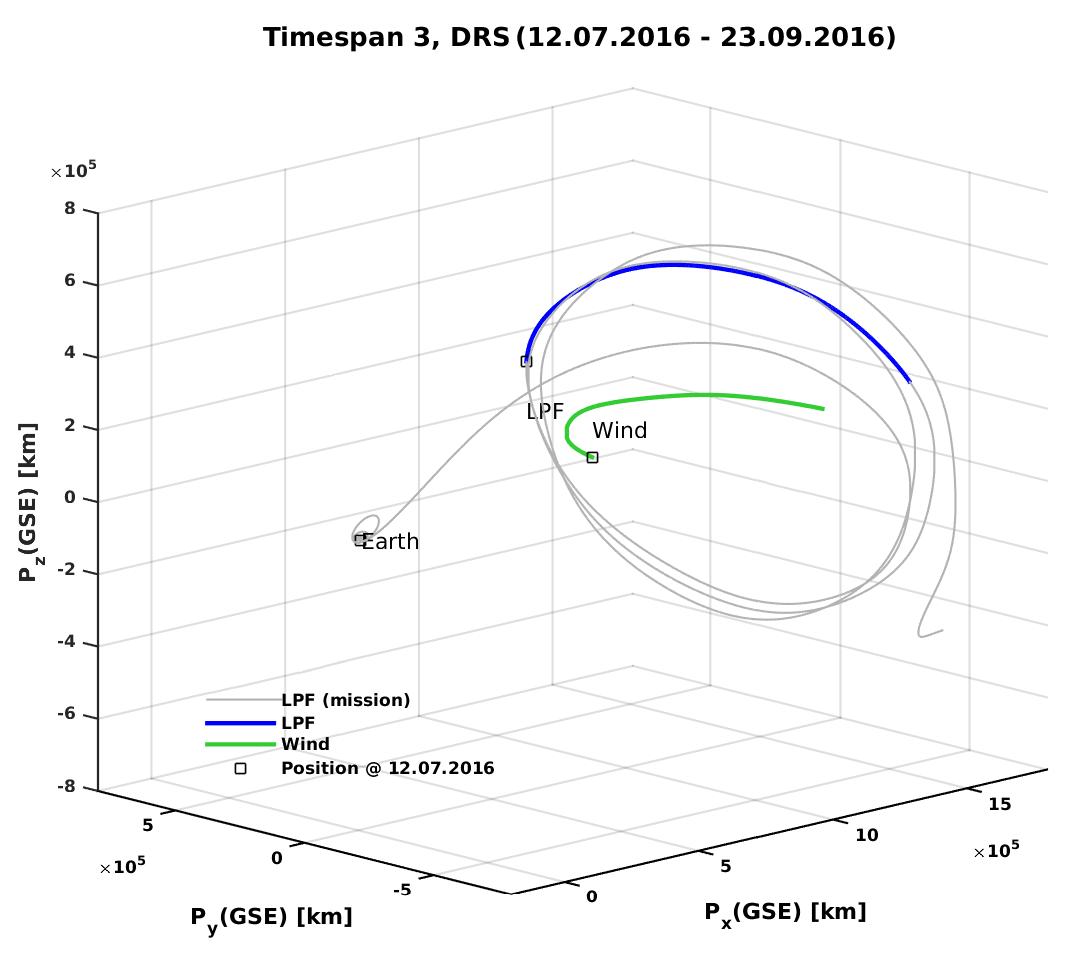}
	\caption{LPF and Wind orbits during the timespans indicated with TS1 and TS3.}
	\label{fig_lpf_ace_wind_orbits} 
\end{figure}

\begin{figure*}[htbp]
\vspace{-2cm}
  \centering
	\includegraphics[width=16cm,height=24cm]{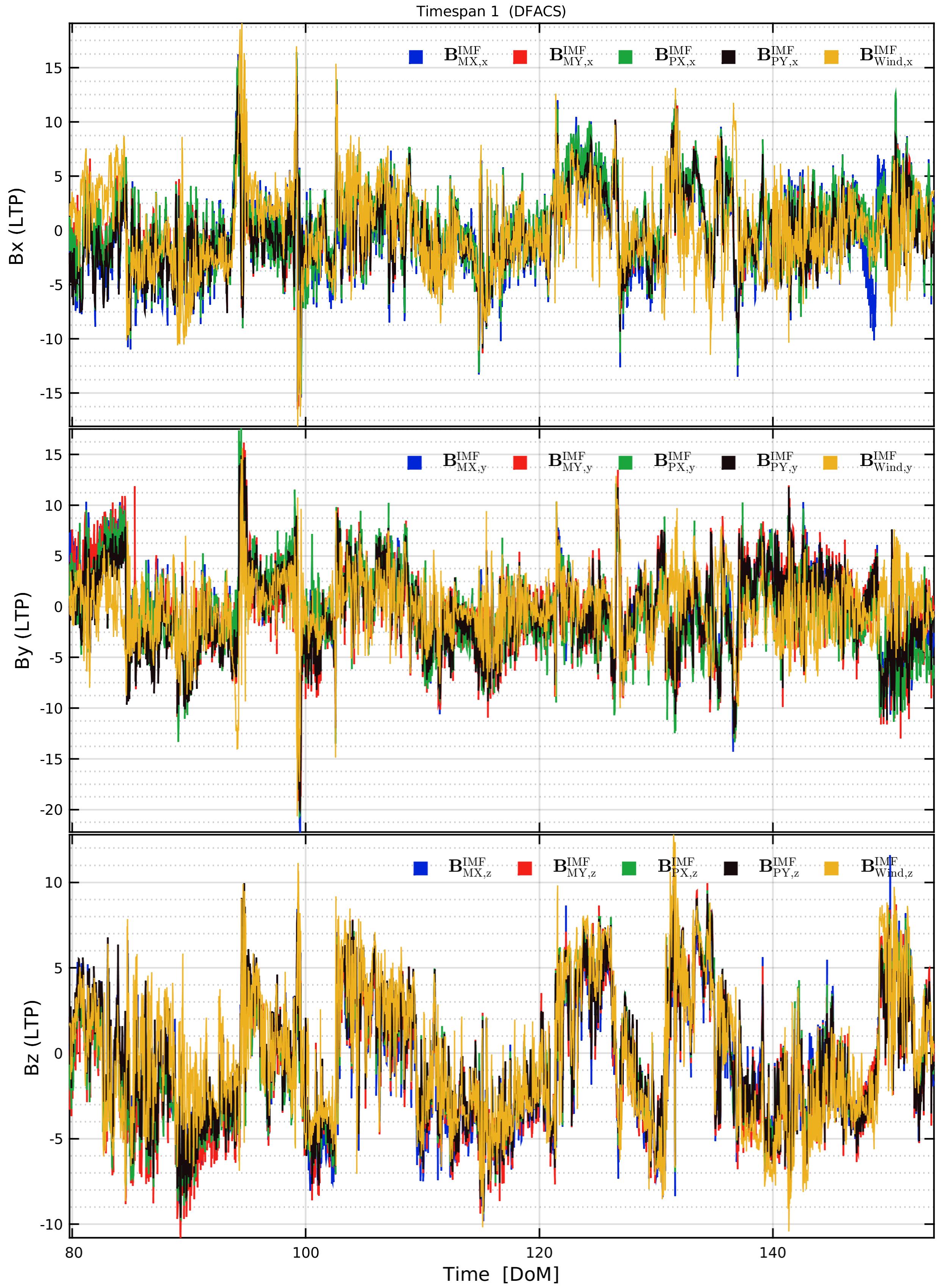}
	\caption{Estimated LPF IMF timeseries ($\mathbf{B}^\mathrm{IMF}_\mathrm{MX}$, $\mathbf{B}^\mathrm{IMF}_\mathrm{MY}$, $\mathbf{B}^\mathrm{IMF}_\mathrm{PX}$ and $\mathbf{B}^\mathrm{IMF}_\mathrm{PY}$) and Wind MFI measurements in the $\pmb{\mathrm{R}}_\mathrm{LTP}$ reference frame for the TS1 time interval (see Table~\ref{tab1}).}
	\label{fig_lpf__TD_TS1_components} 
\end{figure*}

\begin{figure*}[htbp]
\vspace{-2cm}
  \centering
	\includegraphics[width=16cm,height=24cm]{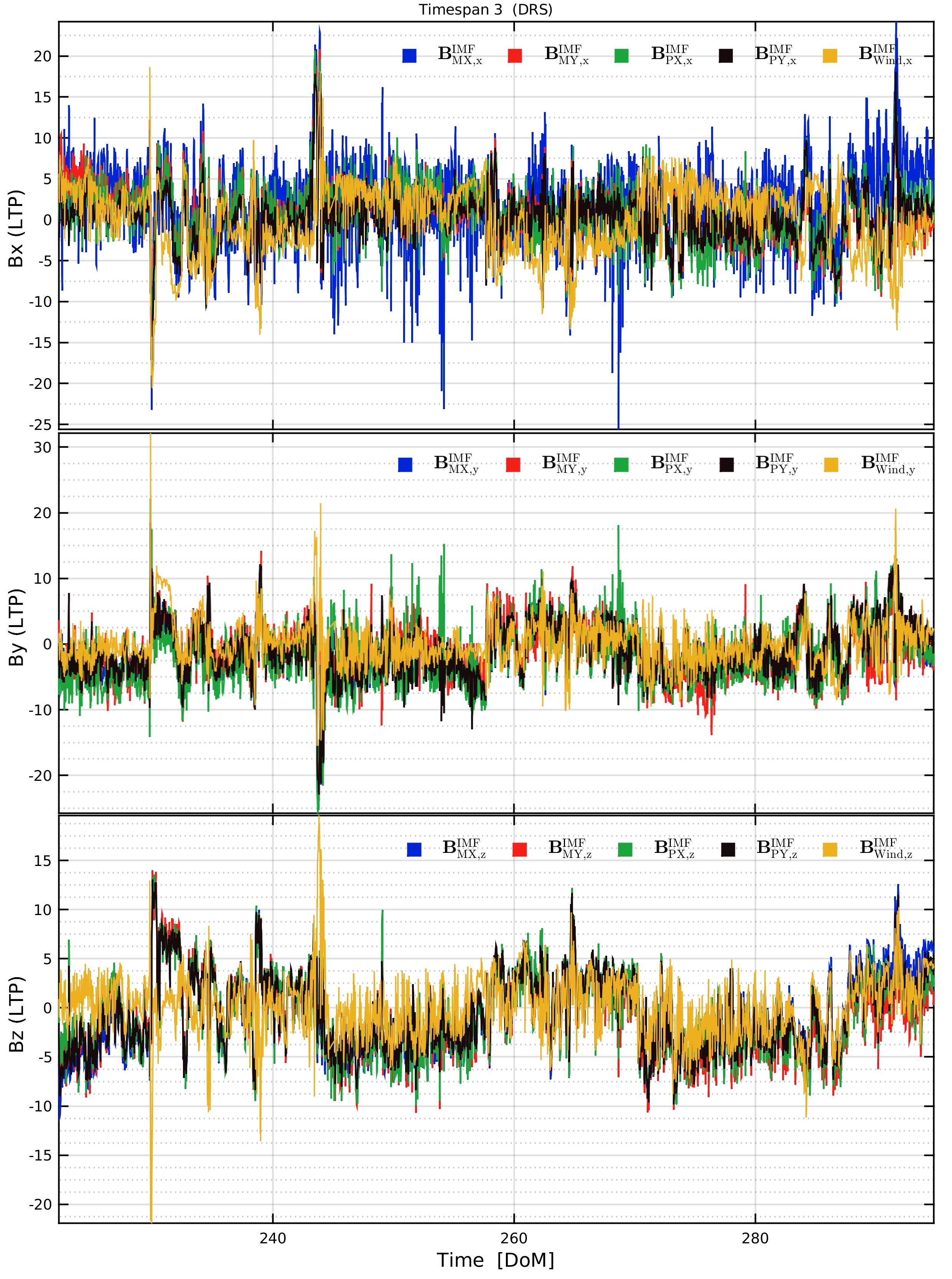}
	\caption{Same as Figure~\ref{fig_lpf__TD_TS1_components} for the TS3 time interval. 
	}
	\label{fig_lpf__TD_TS3_components} 
\end{figure*}

In order to test the reliability of our work, the IMF inferred from LPF magnetometer measurements were compared with contemporaneous observations collected by the MFI detector on board the Wind S/C. The orbits of the two S/C are shown in Figure~\ref{fig_lpf_ace_wind_orbits} during the timespans TS1 (from February 20 to May 4, 2016) and TS3 (from July 12 to September 23, 2016), as an example. In particular, we focused on these two timespans  since they were characterized by the longest data taking periods during which the DFACS and DRS controls were actuating the S/C, respectively. The average distance between LPF and Wind S/C was of $1.12 \times 10^6\, \mathrm{km}$ and the IMF is not expected to vary significantly over these length scales \citep{1990pihl.book..183M}.
In Figures~\ref{fig_lpf__TD_TS1_components} and~\ref{fig_lpf__TD_TS3_components} we have compared the contemporaneous LPF IMF estimates and Wind observations gathered during the mentioned intervals of time. 
The LPF estimates are in fairly good agreement with the Wind data along each one of the three Cartesian components. It is pointed out that the Wind MFI measurements were rotated in the LTP reference frame, $\pmb{\mathrm{R}}_\mathrm{LTP}$.
% (Fig.~\ref{fig_lpf_ace_wind_orbits}). 
Some residual flickering is observed in Figure~\ref{fig_lpf__TD_TS3_components} during the TS3 timespan for the MX magnetometer along the LTP $x$-axis. This  noise is believed to be associated with  a residual contribution of the magnetized moving assemblies (see Section~\ref{subs_b}, \citeauthor{10.1093/mnras/staa830}, \citeyear{10.1093/mnras/staa830}).

\begin{figure*}[htbp]
  \centering
\vspace{-3.0cm}
	\includegraphics[width=13cm,height=25cm]{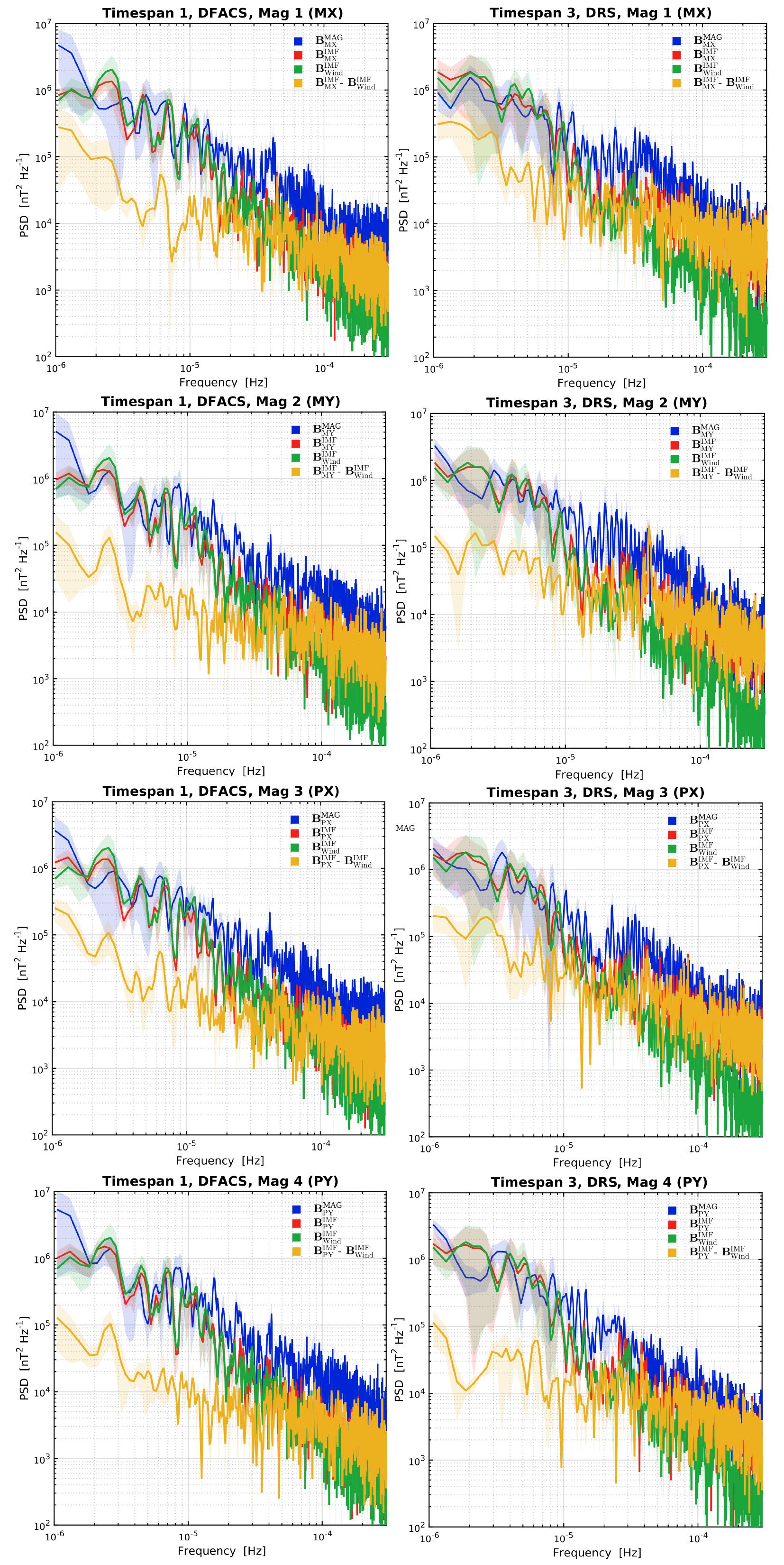}
    %\captionsetup{margin=3cm}
	\caption{PSDs of the LPF magnetometer measurements ($\mathbf{B}^\mathrm{MAG}_\mathrm{MX}$, $\mathbf{B}^\mathrm{MAG}_\mathrm{MY}$, $\mathbf{B}^\mathrm{MAG}_\mathrm{PX}$, $\mathbf{B}^\mathrm{MAG}_\mathrm{PY}$) and of the estimated IMF. Datasets refer to TS1 (left panel) and to TS3 (right panel) time intervals (Tab.~\ref{tab1}). The PSDs obtained by subtracting the Wind data from each LPF IMF measurement timeseries are also provided.
	}
	\label{fig_lpf__PSD} 
\end{figure*}

In Figure~\ref{fig_lpf__PSD} are reported the PSDs obtained by processing the LPF IMF timeseries presented in Figures~\ref{fig_lpf__TD_TS1_components} and~\ref{fig_lpf__TD_TS3_components}.
The $\mathrm{PSD}$ of $\mathbf{B}^\mathrm{IMF}_\mathrm{m}-\mathbf{B}^\mathrm{IMF}_\mathrm{Wind}$ is shown in yellow in the same figure obtained by subtracting the IMF measurements carried out by MFI (in green) from the LPF IMF data (in red).
By comparing the PSDs of the calibrated LPF magnetic field measurements (in blue) with the PSDs of the IMF measured on board Wind (in green), it is possible to notice that the measurements gathered on board LPF are affected by a larger noise at all frequencies. %This evidence is obviously ascribable to both magnetometers design and locations on the Wind S/C optimized for IMF monitoring.
The noise difference between LPF estimated IMF and the IMF measured by the Wind MFI goes down 10\% below $4\times10^{-5}\,\mathrm{Hz}$, where the noise of the IMF becomes increasingly dominant. 

\begin{figure*}[hbtp]
  \centering
	\includegraphics[width=16cm]{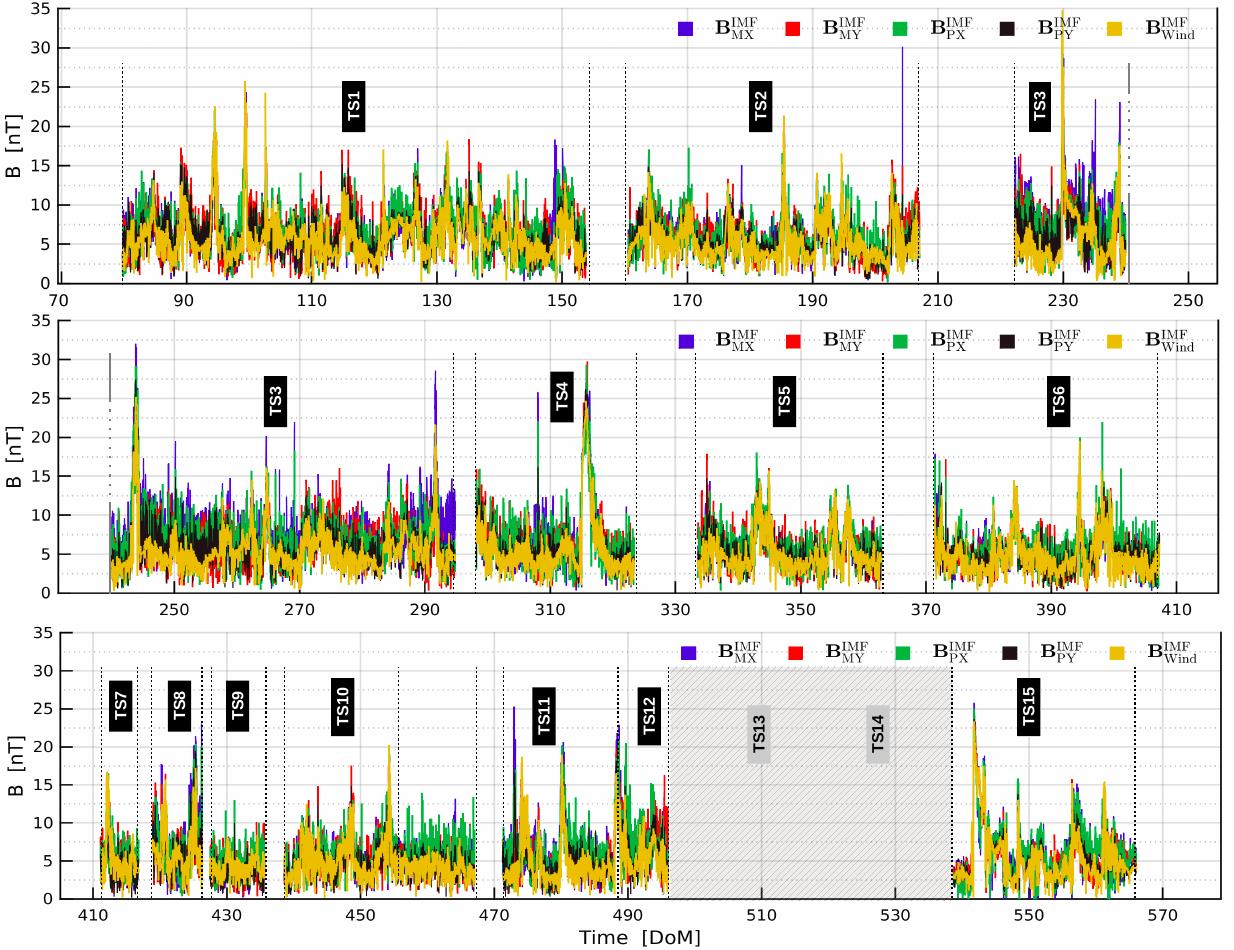}
	\caption{IMF measurement timeseries obtained with the four LPF magnetometers ($\mathbf{B}^\mathrm{IMF}_\mathrm{MX}$, $\mathbf{B}^\mathrm{IMF}_\mathrm{MY}$, $\mathbf{B}^\mathrm{IMF}_\mathrm{PX}$ and $\mathbf{B}^\mathrm{IMF}_\mathrm{PY}$ data appear in blue, red, green, and black, respectively). 
	Simultaneous Wind magnetic measurements, $\mathbf{B}^\mathrm{IMF}_\mathrm{Wind}$, are also shown in yellow for comparison. Timespans TS13 and TS14 (Tab.~\ref{tab1}) are ignored as discussed in Section~\ref{section_magnetic_monitoring} (hatched area).
	The $x$-axis indicates the mission lifetime in DOM.}
	\label{fig_lpf_IMF_comparison}
\end{figure*}

The empirical approach adopted in the present work to separate the IMF component from on-board platform magnetic field  measurements  has been applied to the entire LPF datasets.
The IMF timeseries $\mathbf{B}^\mathrm{IMF}_\mathrm{MX}[t]$, $\mathbf{B}^\mathrm{IMF}_\mathrm{MY}[t]$, $\mathbf{B}^\mathrm{IMF}_\mathrm{PX}[t]$ and $\mathbf{B}^\mathrm{IMF}_\mathrm{PY}[t]$ obtained by processing the four magnetometer datasets are reported in Figure~\ref{fig_lpf_IMF_comparison} in blue, red, green and black colors, respectively. In the same figure, the LPF IMF estimates are compared to contemporaneous measurements carried out with the MFI detector (in yellow) hosted on board the Wind S/C (NASA CDAWeb, \url{https://directory.eoportal.org/web/eoportal/satellite-missions/}). 
Data gaps are associated with magnetic experiments, electronics functional tests, faults and anomalies.
A good agreement is found between Wind MFI and LPF IMF estimates during the whole mission duration. % resulting from the LTPDA frequency-domain minimization procedure.

\begin{figure}[htbp]
  \centering
	\includegraphics[width=8.5cm,height=8.5cm]{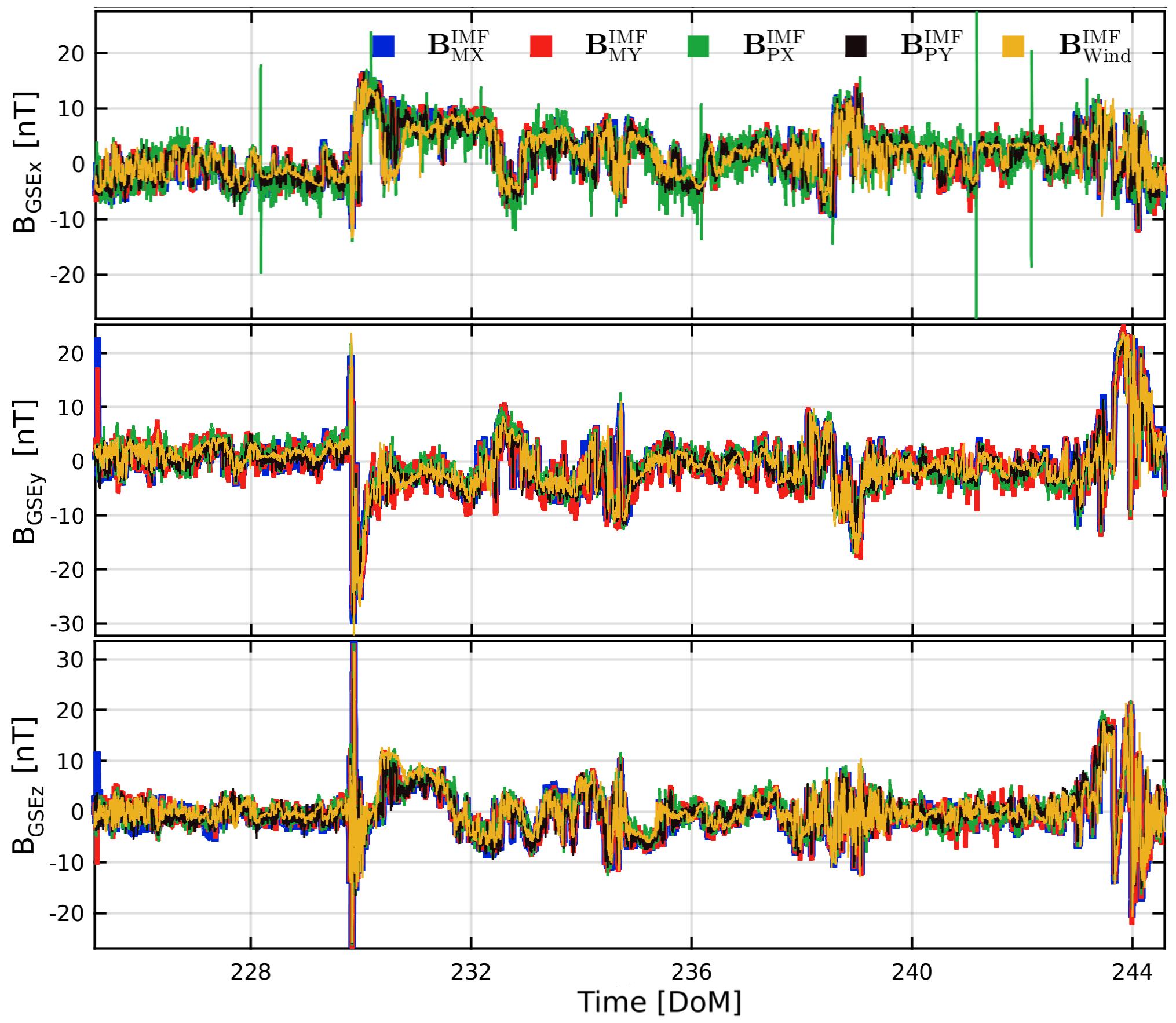}\\ %COLOR
	\caption{Comparison between LPF and Wind IMF measurements in the $\pmb{\mathrm{R}}_\mathrm{GSE}$ between July 15 and August 4, 2016.
	}
	\label{fig_lpf__wind2lpf} 
\end{figure}

\begin{figure*}[htbp]
  \centering
    \includegraphics[width=9.0cm,height=8.5cm]{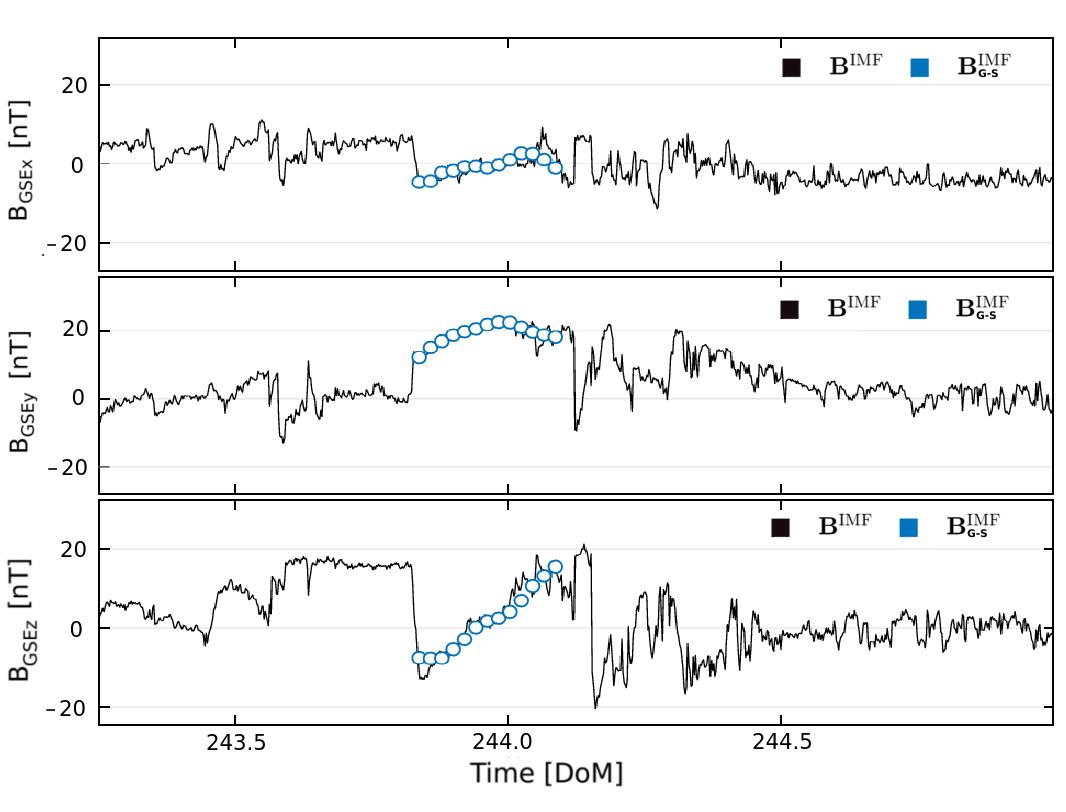}
	\caption{
	LPF IMF (black line) and IMF along the LPF orbit obtained with a Grad-Shafranov reconstruction based on Wind data of the magnetic cloud crossed by the S/C on August 2-3, 2016 (open blue circles). Each LPF IMF component was obtained by averaging the measurements of the four on-board magnetometers.}
	\label{fig_gargiulo}
\end{figure*}

In Figure~\ref{fig_lpf__wind2lpf} the time interval ranging from July 15 through August 4, 2016 is considered in particular. 
In this figure, the IMF components inferred from the LPF on-board measurements are shown shown in the Global Solar Ecliptic reference frame ($\pmb{\mathrm{R}}_\mathrm{GSE}$) in order to compare our results with the Wind IMF contemporaneous measurements. The %$\pmb{\mathrm{R}_\mathrm{LTP}$ 
LTP $z$-axis of the LPF magnetometer reference frame was aligned with the GSE $x$-axis during the mission lifetime, while the $\pmb{\mathrm{R}}_\mathrm{LTP}$ $xy$-plane rotated around the LTP $z$-axis with a six month period.
%The average distance between LPF and Wind S/C was of $1.12 \times 10^6\, \mathrm{km}$ and the IMF is not expected to vary significantly over these length scales \cite{1990pihl.book.....S}. 
Between August 2 and August 3, 2016 the LPF S/C crossed a magnetic cloud. %associated with the passage of an interplanetary counterpart of a coronal mass ejection (ICME).
It is worthwhile to recall that a magnetic cloud is a coherent plasma structure characterized by a smooth rotation of the IMF and an enhanced magnetic field intensity with respect to the average value observed in the solar wind while temperature and plasma beta parameters ({\it i.e.} the ratio of the plasma pressure to the magnetic pressure) present lower values with respect to an undisturbed interplanetary medium \citep{1990GMS....58..373B}. The LPF IMF data are compared to the magnetic field estimated by applying the Grad-Shafranov reconstruction technique based on the Wind S/C observations  \citep{Hu_Sonnerup_2002,Hu2017,2020ApJ...901...21B}. %2019NCimC..42...44B,
The comparison between the IMF components inferred from the GS reconstruction and the LPF observations is shown in Figure~\ref{fig_gargiulo}. The LPF IMF estimates are computed by averaging the data of the four magnetometers and are reported by considering a sampling time of one minute. Again, a good agreement is observed and this comparison allows for an independent validation of the results of our empirical approach  within a few percent.

A similar work could be carried out for LISA that will also host platform magnetometers in order to monitor the transit of interplanetary magnetic structures. 
It is worthwhile to point out that contrarily to LPF, LISA  will not benefit of other experiments gathering interplanetary magnetic field, plasma data and particle monitoring along its orbit.  

\section{Galactic and solar energetic particle events during the LISA mission operations}

\subsection{Galactic cosmic-ray energy spectrum during the solar cycle 26}

\begin{figure}[hbt!]
  \centering
%\hspace{-2cm}   
\hspace{-0.3cm}
\centering
\includegraphics[width=9.cm,height=9.cm]{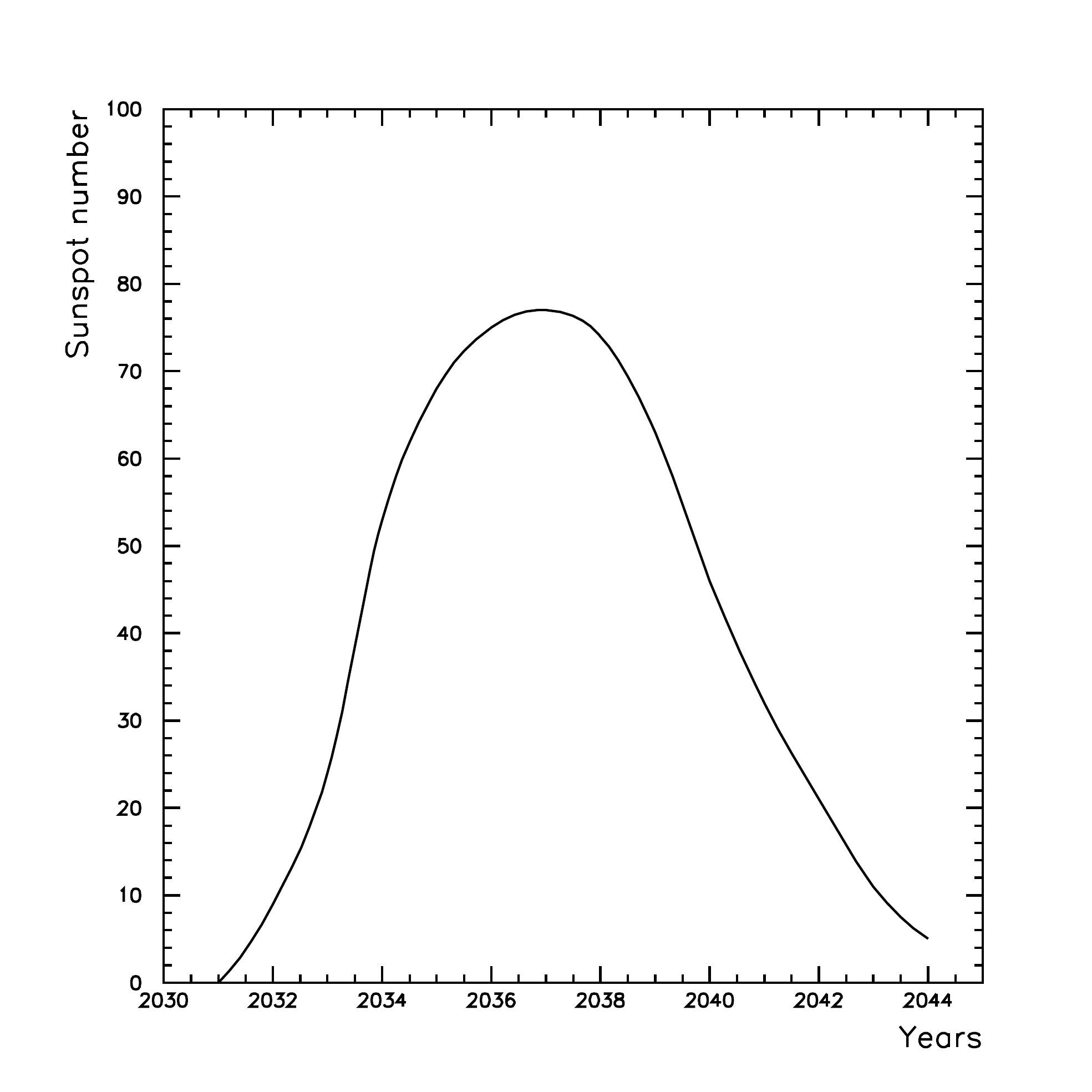}
\caption{Sunspot number prediction for the solar cycle 26 \citep{sing19}.}
\label{fig_sunspot_number_prediction}
\end{figure}

The galactic cosmic-ray (GCR) flux in the inner heliosphere appears energy, time, space, and charge dependent.
The overall GCR flux was observed to vary by a factor of four near Earth during the last three solar cycles when the majority of data to study the long-term variations of particles of galactic origin were gathered in space \citep{2021A&A...656A..15G}.
Since LISA will orbit the Sun at 1 AU,  both near-Earth GCR and SEP observations  can be used to carry out long-term predictions of the overall particle flux  during the mission operations by properly taking into account  the solar activity expected  during the solar cycle 26 \citep{sing19}. The predicted sunspot number, the most widely used proxy of the solar activity during the solar cycle 26, is reported in Figure~\ref{fig_sunspot_number_prediction} and Table~\ref{tab2}.
%As a result, we will consider the time/energy depenence of the GCR flux variations of particle observations gathered at 1 AU.                                                
%Both predictions are associated with the long-term solar activity ($>$ 1 year).

\begin{table}[tbh]
        \centering
        \resizebox{3.2cm}{!}{
                \begin{tabular}{lcc}\hline\hline
                        Year & SSN & SEP \\ \hline
                        2035 & 68 & 4.7 \\
                        2036 & 76 & 5.3 \\
                        2037 & 77 & 5.3 \\
                        2038 & 74 & 5.1 \\
                        2039 & 63 & 4.4 \\
                        2040 & 42 & 2.9 \\
                        2041 & 32 & 2.2 \\
                        2042 & 21 & 1.5 \\
                        2043 & 11 & 0.8 \\
                        2044 & 1  & 0.07 \\\hline\hline
                \end{tabular}
        }
        \caption{Expected sunspot number (SSN) and SEP events during the LISA mission with a fluence larger than the estimated GCR background.}
        \label{tab2}
\end{table}
                                                      
The past solar cycle 24 has been the weakest of the last hundred years. According to Singh and Barghawa \citep{sing19}, the next two solar cycles are expected to be even weaker. However, it is worthwhile to point out that the initial phase of the solar cycle 25 appears more intense than these predictions and other authors indicate an intensity similar to that of the solar cycle 23 \citep{Diego_Storini_Laurenza_2010,McIntoshEtAl2020,Diego_and_Laurenza_2021}. Quasi-eleven and quasi twenty-two year periodicities are observed in the GCR intensity associated with the solar activity and the GSMF polarity reversal, respectively.
The force-field model by Gleeson and Axford \citep{glax68} is adopted here for the prediction of GCR long-term modulation while the Nymmik model \citep{{Nymmik1999a},{Nymmik1999b}} is considered to estimate the occurrence of SEP events  with fluence larger than  10$^6$ protons cm$^{-2}$ above 30 MeV during the LISA mission. The Gleeson and Axford model allows us to estimate the GCR flux in the inner heliosphere by assuming a time independent interstellar spectrum and a solar modulation parameter $\phi$ describing the effect of the solar activity modulation (see \citeauthor{apj1}, \citeyear{apj1}, where the details of our approach are illustrated).  % the effect of the solar activity in modulating the GCR spectrum (see \citet{apj1} where the details of our approach are illustrated). 
The sunspot number is correlated with the solar modulation parameter.
After the LISA launch in 2035, the maximum expected number of  sunspots is about 80 and, consequently, the solar modulation parameter may possibly vary between 600 MV/c and 700 MV/c, as in September 2015 for instance (see \url{https://cosmicrays.oulu.fi/phi/Phi_mon.txt} handled by the University of Oulu, Finland). This estimate appears consistent with the solar modulation values inferred from \citet{refsitof}.
%The expected flux is interpolated according to the following equation (ref):                                                                                                 

\noindent The proton energy spectrum above 70 MeV corresponding to the above solar modulation was obtained with the Gleeson and Axford model (bottom dot-dashed line in Fig.~\ref{fig_min_and_max_proton_energy_spectra}) parameterized according to \citet{papini96} and \citet{apj1} as follows:
\begin{equation}
F(E)= 18000 \times (E+1.5)^{-3.66}\ E^{0.87}  \ \ \ {\rm particles/(m^2\ sr\ s\ GeV),}
\label{equation2}
\end{equation}
where E is the proton kinetic energy in GeV. The proton flux is considered in the figure since protons constitute 90\% of the overall cosmic-ray bulk.
%The number of sunspots during the solar cycle 26 is reported in Fig. 1.          
We verified the reliability of this approach for the LISA Pathfinder mission in 2016-2017, by comparing our cosmic-ray proton flux predictions to observations carried out above 450 MeV n$^{-1}$ by the AMS-02 experiment during the same period on the Space Station (see \citeauthor{aguilar18}, \citeyear{aguilar18}).
%, while the Nymmik model outcomes were compared to SEP occurrence in the years 1986-2004  \cite{Grimani:2012ce}.                                                                      
Our predicted GCR integral proton flux at the LPF launch in December 2015 was found only 10\% higher than the AMS-02 data \citep{grim19}.
%while the SEP occurrence was found to differ by a factor of two at most with observations \cite{Grimani:2012ce}.                                                                      

%The overall GCR flux was observed to vary by a factor of four  near Earth during the last three solar cycles when the majority of data to study the GCR flux variations were gathered in space \cite{2021A&A...656A..15G}. %resently, the LISA  test-mass  charging is estimated  with Monte Carlo simulations according to long and short-term GCR flux  variation predictions. Our work  will be refined just before the mission launch when the actual solar modulation will be  known. It must be pointed out that
%the noise associated with the cosmic-ray test-mass charging process is due to both poissonian fluctuations of the secondary particle production in the spacecraft and to stochastic short-term flux variations, observed at the transit of interplanetary counterparts of coronal mass ejections (ICMEs) and corotating interaction regions formed at the leading edge of high-speed solar wind streams  overtaking the  slow solar wind \cite{richa}. 

%\begin{figure}
%\begin{center}
%\centering \epsfig{figure=sc26.eps,width=0.6\linewidth}
%\end{center}
%\caption{\label{avg} Sunspot number prediction for the solar cycle 26 \cite{sing19}. }
%\end{figure}

\subsection{Estimates of solar energetic particle event occurrence during the LISA mission}
Nymmik \citep{{Nymmik1999a},{Nymmik1999b}} has found that the SEP fluence  distribution  follows a power-law trend with an exponential cutoff for increasing values of the fluence. This model applies to proton fluences ranging between 10$^6$ and 10$^{11}$ protons cm$^{-2}$ above 30 MeV, where 10$^6$ protons cm$^{-2}$ represents the fluence that approximately corresponds to the background of GCRs near solar minimum. The Nymmik model was developed on the basis of SEP event measurements carried out with the IMP-7 and IMP-8 S/C during the solar cycles 20-22 and from proton fluxes estimated with radionuclide observations in lunar rocks generated in the last few million years.
The method is illustrated in the following \citep{StoriniEtAl2008}. The yearly number of SEP events, $<n>$ is assessed on the basis of sunspot number predictions, $<N>$:
%during the solar cycle 24                                                                                                                                                    
%(see also  Wess, Ara\'ujo and Sumner, 2003) while for the solar cycle 25                                                                                                     
%we  used the average number of  predicted solar spots (ref)                                                                                                                  

\begin{equation}
 <n>=0.0694<N> \label{ee1}.
%\Phi=\frac{Z e R}{k_{2}(R)} \phi  \label{ee1}                                                                                                                                  
\end{equation}
\noindent In order to estimate the number of events per interval of fluence, $f$, a  normalization constant C must be then set by integrating the
equation:
\begin{equation}
%  <n>=0.0694<w>                                                                                                                                                              
dn=C f^{-1.41} e^{-f/f_c} df  \label{ee2},
\end{equation}
%\hspace{2.cm} dN=C $f^{-1.41}$ e$^{-f/f_c}$ df                                                                                                                               
\noindent and by equaling it to $<n>$. In the above equation
$f_c$ is assumed equal to 4$\times$10$^{9}$.
%Finally, after the determination of the C constant,  the                                                                                                                     
%number of  events per interval of fluence can be estimated.                                                                                                                  
It is pointed out that 65\% of the events are expected to belong in the fluence range 10$^6$-10$^7$ protons cm$^{-2}$.

In \citet{Grimani:2012ce} the expected number of SEP events satisfying the Nymmik model was compared to observations carried out between 1986 and 2004. Model and data were found in agreement  within a factor of two.
The expected  number of  SEP events  during the time the LISA mission will remain in space, possibly between 2035 and 2044, is reported in Table~\ref{tab2}.
%It is pointed out that 65\% of the events are expected to lie in the fluence range 10$^6$-10$^7$ protons cm$^{-2}$. 
                                                         
As a worst case scenario, the occurrence of a SEP event with  10$^9$ protons cm$^{-2}$ fluence in the next years should be also considered to estimate the LISA TM charging. The typical rate of occurrence of these events is one every sixty years. The last one, characterized by particle acceleration above 100 MeV n$^{-1}$, was observed on February 23, 1956 \citep{2008AdSpR..41..926V}.
%As a result, it is plausible to expect one of these events during the LISA mission.      
The proton fluxes observed during the evolution of SEP events of different intensities are shown in Figure~\ref{fig_min_and_max_proton_energy_spectra} in comparison to the minimum expected galactic flux over the LISA mission duration.

%\begin{table}
%\caption{\label{table1} Expected sunspot number (SSN) and SEP events during the LISA mission with a fluence larger than the estimated GCR background.}
%\begin{indented}
%\lineup
%\item[]\begin{tabular}{@{}*{3}{l}}
%\br
%\m  Year & \m  SSN  & \m  SEP events \cr
%\mr
%2035& 68& 5\cr
%2036&76& 5\cr
%2037&77& 5\cr
%2038&74& 5\cr
%2039&63& 4\cr
%2040&42& 3\cr
%2041&32& 2\cr
%2042&21& 1\cr
%2043&11& 1\cr
%2044&5& $<$1 \cr
%2045&& \cr                                                                                                                                     
%\br
%\end{tabular}
%\end{indented}
%\end{table}

\section{Next Generation Radiation Monitor}
\begin{figure}[t!]
%\hspace{-2cm}
\hspace{-0.3cm}
\centering
\includegraphics[width=9.cm,height=9.cm]{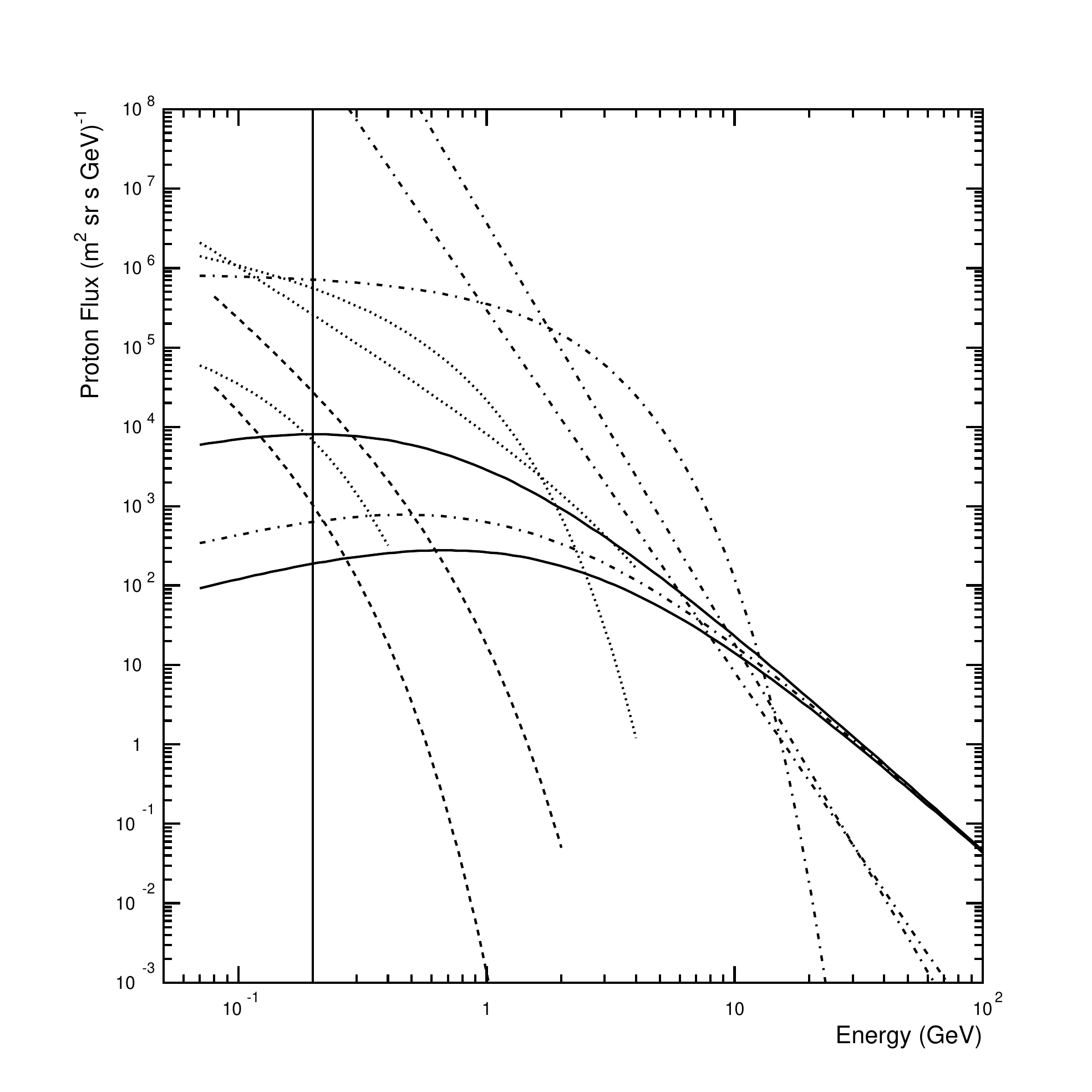}
\caption{Minimum and maximum proton energy spectra of galactic origin observed near Earth (top and bottom continuous lines, respectively). The bottom dot-dashed line represents the minimum proton flux expected during the LISA mission. Dashed lines (December 14, 2006; \citet{2011ApJ...742..102A}), dotted lines (May 7, 1978; \citeauthor{1984AdSpR...4b.117M}, \citeyear{1984AdSpR...4b.117M}) and top dot-dashed lines (February 23, 1956; \citeauthor{2008AdSpR..41..926V}, \citeyear{2008AdSpR..41..926V}) represent solar proton fluxes observed during the evolution of SEP events of different intensities. The vertical black line indicates the maximum energy of the NGRM proton spectrum measurement capability.}
\label{fig_min_and_max_proton_energy_spectra}
\end{figure}

The ESA NGRM is small ($<$1 dm$^{3}$), light ($<$ 1 kg) and characterized  by a low power consumption ($<$ 1 W).
This detector is optimized for low-energy particle measurements  in harsh radiation environments, typical of SEP events and of particles magnetically trapped in the Earth and other planetary magnetospheres.
The NGRM consists of two units. The proton unit allows for the measurement of the proton flux between 2 MeV and 200 MeV. In Figure~\ref{fig_min_and_max_proton_energy_spectra} proton energy spectra of particles of galactic and solar origin are compared. Protons with energies smaller than 200 MeV constitute a minor percentage of particles of galactic origin (continuous lines) at both solar minimum (upper curve) and maximum (lower curve).
As it was shown in the previous section, the bottom dot-dashed curve represents the minimum flux of galactic protons expected during the LISA mission.
In the same figure the proton energy spectra observed during the evolution of SEP events of different intensities appear softer than the GCR energy spectrum and, consequently, the particle flux below 200 MeV is representative of the total solar proton flux. As a result, the NGRM is an optimum monitor of the solar particle flux.
%fluxes during the evolution of SEP events of different intensities                                                                                   
%except at the onset of very intense events represented by the lowest of the  top dot-dashed curve set.                                            
Linear energy transfer of ions can be also measured with this detector.
The electron unit is meant to monitor electrons in the energy range 100 keV-7 MeV.

\begin{figure}[t!]
%\hspace{-2cm}                                                                                                                                   
\hspace{-0.3cm}
\centering
\includegraphics[width=9.cm,height=9.cm]{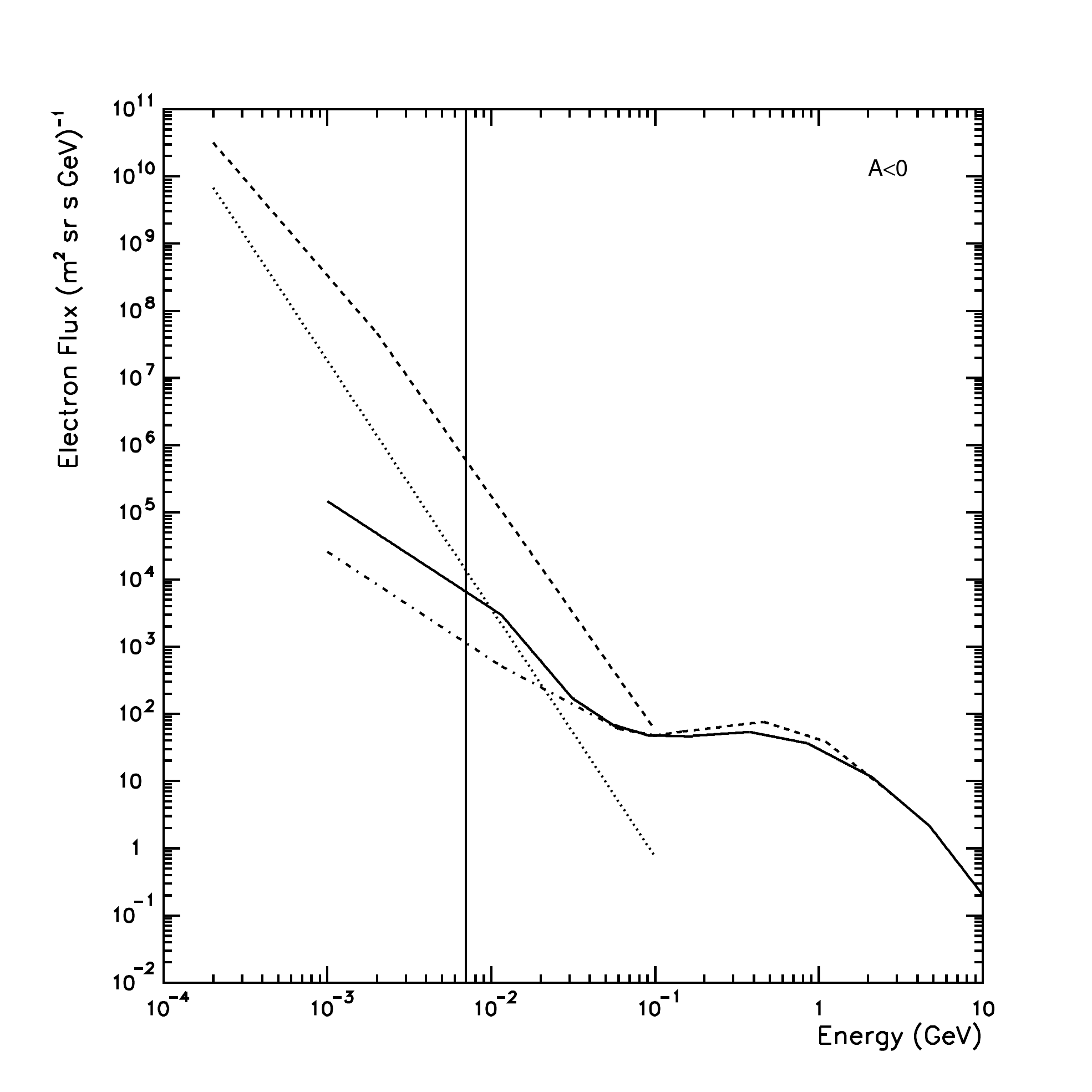}
\caption{Solar electron fluxes associated with impulsive (upper dashed line) and  gradual (dotted line) SEP events \citep{Grimani_2009}. The continuous line, the dot-dashed line and the bottom dashed line represent minimum and maximum interplanetary and galactic electron fluxes observed during a negative polarity period (A$<$0) of the GSMF (when the solar magnetic field lines enter the North pole of the Sun) during which the flux of negative particles is maximum with respect to periods of  positive GSMF polarity and similar conditions of solar activity \citep{Grimani_2009}. The vertical continuous line indicates the maximum electron energy measured by the NGRM. It can be observed that below 7 MeV the solar electron flux overcomes by several orders of magnitude the interplanetary and galactic components allowing for SEP event forecasting.}
\label{fig_solar_galactic_electron_fluxes}
\end{figure}

In Figure~\ref{fig_solar_galactic_electron_fluxes} solar (top dashed and dotted curves) and galactic  (solid, dot-dashed and bottom dashed lines) electron fluxes are compared \citep{{Grimani_2009},{gri09}}. A sudden increase of the  electron flux at MeV energies is a precursor of a solar energetic proton event due to particle velocity dispersion. Solar electron fluxes showing two different spectral indices indicate impulsive SEP events associated with proton acceleration below 50 MeV. Conversely, a single spectral index characterizes the solar electron flux associated with gradual events with proton acceleration above 50 MeV. In \citet{Grimani_2009} and references therein it was shown that the intensity of the solar electron flux in the MeV range is similar to the proton flux intensity in the 50 MeV range. The NGRM has excellent SEP event short-term forecasting and measurement capabilities. LISA will allow us to carry out the first multispacecraft SEP event observations at a few million kilometers with twin detectors. %equivalent to 1 degree %ver about one degree in heliolongituge  for the first time. 
The LISA data will be also compared to observations gathered near Earth at 50 million kilometers. %equivalent to 20 degrees. 
Conversely, the GCR flux short-term variations will not be nicely monitored with NGRM.  

The geometrical factor of the NGRM for GCR measurements is of about 0.01 cm$^2$, approximately hundred times smaller than that of the detector hosted on board LPF.
As a matter of fact, the NGRM will not allow us to monitor the dynamics of non-recurrent GCR short-term variations of typical durations of three-four days and grossly only recurrent GCR short-term variations that were found to last an average of 9 days with LPF \citep{paper3}. 

The classification of interplanetary processes associated with GCR short-term variations observed with LPF are reported in \citet{apj1,apj2}.
The evolution of each GCR short-term depression is unique and depends on the energy differential flux of GCR particles at the transit of subsequent interplanetary 
structures.
Consequently, the average monthly GCR flux observations are not representative of the GCR flux at shorter timescales.
Moreover, it is of fundamental importance to maintain separated the effects of cosmic-ray long-term modulations from short-term variations, due to a different dependence on the particle energies for the estimate of the LISA and LPF TM charging.
%are strictly correlated and both result energy dependent.                                                                                                               \
Only low statistical uncertainty data gathered hourly in space  above a few tens of MeV n$^{-1}$ would allow us to disentangle the role of long-term and short-term solar modulation of the GCR flux in the LISA sensitivity band.

The evolution of individual recurrent and transient GCR variations studied with data gathered on board the LPF S/C were compared to contemporaneous Earth neutron monitor observations. This comparison revealed  that very poor clues can be obtained on the physics of GCR flux short-term 
variations in space from  ground data affected by low-energy particle flux attenuation in the atmosphere and by the geomagnetic cut-off.
For instance, recurrent GCR flux depressions of intensities $<$ 5\% detected on board LPF were not observed on  neutron monitors by revealing that only particles with energies smaller than neutron monitor effective energies ($>$ 10 GeV; see for details \citeauthor{apj2}, \citeyear{apj2}) were involved in these processes. It is worthwhile to recall that particle count rates observed
with neutron monitors vary proportionally to the cosmic-ray flux at energies larger than the effective energy  \citep{gil17}, thus providing a direct measurement of the GCR flux at the top of the atmosphere above 10 GeV.
% and therefore, these data are not very useful to study the GCR flux short-term variations except during Forbush decreases that represent GCR flux depressions up to 
%about 10 GeV. 
While GCR recurrent short-term variations observed in space were associated with particle modulation below a few GeV within 1\% uncertainty, Forbush decreases (sudden drops of the  GCR flux observed at the transit of ICMEs, \citeauthor{1937PhRv...51.1108F}, \citeyear{{1937PhRv...51.1108F},{1954JGR....59..525F},{1958JGR....63..651F}}) were characterized by a GCR particle flux modulation at energies above 10 GeV.
Three Forbush decreases were observed with LPF and neutron monitors on July 20 and August 2, 2016 and on May 27, 2017.
The event depression amplitudes were found of 5.5\%, 9\% and 7\% in space and 
of less than  2-3\% in neutron monitor observations \citep{apj2}. In conclusion, also neutron monitor observations will not allow us to properly monitor the LISA TM charging during galactic cosmic-ray flux short-term variations. This may be useful, for instance, if TM glitches of unknown origin will be observed. As a result, a dedicated cosmic-ray detector with characteristics similar to those of the LPF PD would be more than recommended for LISA to improve the monitoring of the transit of interplanetary magnetic structures and high-speed streams.

%\begin{figure}[ht]
%\begin{center}
%\centering \epsfig{figure=protheNGRM.eps,width=12cm}
%\end{center}
%\caption{\label{gcr14} Minimum and maximum proton energy spectra of galactic origin (top and bottom continuous lines, respectively). The bottom 
%dot-dashed line represents the minimum proton flux expected during the LISA mission. Dashed lines (December 14, 2006; \citet{2011ApJ...742..102A}), 
%dotted lines (May 7, 1978; \citet{1984AdSpR...4b.117M}) and top dot-dashed lines (February 23, 1956; \cite{2008AdSpR..41..926V}) represent the solar proton fluxes 
%observed during the evolution of the indicated SEP events. The vertical black line indicates the energy upper limit to the NGRM proton 
%spectrum measurement capability.
%}
%\label{figure1}
%\end{figure}

%\begin{figure}[ht]
%\begin{center}
%\centering \epsfig{figure=intelenegpol_sol.eps,width=12cm}
%\end{center}
%\caption{\label{gcr14}
%Solar electron fluxes associated with impulsive (upper dashed line) and  gradual (dotted line) SEP events \cite{Grimani_2009}. The continuous line, 
%the dot-dashed line and the bottom dashed line represent minimum and maximum interplanetary and galactic electron fluxes observed during a negative 
%polarity period (A$<$0) of the GSMF (when the solar magnetic field lines enter the North pole of the Sun) during which  the flux of negative 
%particles is maximum with respect to periods of  positive GSMF polarity and similar conditions of solar activity. The vertical continuous line 
%indicates the maximum electron energy measured by the NGRM. It can be observed that below 7 MeV the solar electron flux overcomes by
%several orders of magnitude the interplanetary and galactic components allowing for  SEP event forecasting.
%}
%\label{figure1}
%\end{figure}

\section{Conclusions}
Spurious forces acting on the TMs of the future LISA and LISA-like interferometers for gravitational wave detection in space will be monitored and mitigated in order to improve the mission performance. Magnetometers and particle detectors will be hosted on board the LISA S/C to monitor the variations of the magnetic field and the TM charging process. Precious clues were obtained about the interplanetary medium impact in limiting the LISA performance with the LPF mission flown in 2016-2017 around the L1 Lagrange point to test the LISA technology.   
The LPF magnetometers measured a magnetic field varying from $700$ to $1200\,\mathrm{nT}$ near the TMs while the IMF values in L1 ranged between $1$ and $25\,\mathrm{nT}$. The on-board generated magnetic field components were observed to vary by about $150\,\mathrm{nT}$ on timescales of months and of $20-30\,\mathrm{nT}$ daily.
An empirical approach was considered in the present work to disentangle the IMF from the on-board originated magnetic field by processing the LPF platform magnetometer measurements.
% within a few \%. T 
An agreement within a few percent was found between the LPF estimated IMF components and those provided by the dedicated MFI instrument on board the Wind S/C also orbiting around L1. 
%(see Figures~\ref{fig_lpf__TD_TS1_components},~\ref{fig_lpf__TD_TS3_components} and~\ref{fig_lpf_IMF_comparison}). 
A similar approach could be considered for the future gravitational wave experiments carried out in space. 
Provided the availability of the whole housekeeping dataset, an empirical magnetic field noise model can be produced when a S/C is already in-flight, even in the case that platform magnetometers are immersed in a strong varying and direction dependent on-board generated magnetic field.
%{\red\bf The LPF data revealed that the passage of the interplanetary magnetic structures can be efficiently monitored on board space interferometers thus contributing to space weather science investigations and, possibly, to issue space weather alerts.}
%The LISA constellation will monitor the transit of ICME. 
Solar energetic particle event short-term forecasting and evolution monitoring will be also carried out with the NGRM on board the LISA S/C. An additional particle detector dedicated to galactic  cosmic-ray variation observations should be also included in the set of detectors meant for the mission diagnostics.   
LISA and the second generation LISA-like interferometers 
%such as the Big Bang Observer (BBO) \cite{2005PhRvD..72h3005C}, consisting of four LISA-like interferometers orbiting at $1\,\mathrm{AU}$ at large longitude intervals, will carry out similar observations but at much wider longitude ranges by providing 
will also naturally play the role of multipoint observatories for SEP events and interplanetary large-scale structure monitoring with magnetometers used in combination with dedicated GCR detectors.

%    'f1'           '334.93'       '11.48'        '[nT kg^(-1)]'
%    'o1'           '10.847'       '13.63'        '[nT kg^(-1)]'
%    's1'           '61.584'       '14.81'        '[nT kg^(-1)]'
%
%    'f1'           '73.462'       '12.86'        '[nT kg^(-1)]'
%    'o1'           '-64.988'      '10.76'        '[nT kg^(-1)]'
%    's1'           '-13.16'       '13'           '[nT kg^(-1)]'
%
%    'f1'           '342.99'       '12.24'        '[nT kg^(-1)]'
%    'o1'           '5.4586'       '11.49'        '[nT kg^(-1)]'
%    's1'           '-30.986'      '13.44'        '[nT kg^(-1)]'
%
%    'f1'           '81.605'       '9.199'        '[nT kg^(-1)]'
%    'o1'           '-15.273'      '7.86'         '[nT kg^(-1)]'
%    's1'           '-20.082'      '9.369'        '[nT kg^(-1)]'

%\end{document}

%
% ****** End of file template.aps ******
%\bibitem{gaiathrusters2014}
%J. Jarrige, P. Thobois, C. Blanchard, P.-Q. Elias, D. Packan, L. Fallerini, and G. Noci.  ``Thrust Measurements of the Gaia Mission Flight-Model Cold Gas Thrusters'', Journal of Propulsion and Power, Volume {\bf 30}, No. 4, pp. 934-943 (2014). 

% % %
\begin{acknowledgements}
      %Part of this work was supported by the German
      %\emph{Deut\-sche For\-schungs\-ge\-mein\-schaft, DFG\/} project
      %number Ts~17/2--1.
      The LISA Pathfinder mission is part of the space-science program of the European Space Agency.
      The Italian contribution has been supported by the Agenzia Spaziale Italiana (ASI) and the Istituto Nazionale di Fisica Nucleare (INFN).
      The LISA Pathfinder magnetometer data are available at the LISA Pathfinder Legacy Catalog (\url{http://lpf.esac.esa.int/lpfsa/}).
      Data from Wind experiment were collected from NASA CDAWeb (\url{https://cdaweb.sci.gsfc.nasa.gov/index.html/}). 
      %Data from Wind and ACE experiments were obtained from the NASA-CDAWeb website.
\end{acknowledgements}

%%    This version assumes use of bibtex with the jswsc.bib file being present
%%    If your bib file has a different name you need to change the following line

%\bibliography{jswsc}
\bibliography{lisa_magEnv_jswsc_v3}
   
%\end{linenumbers}

\end{document}